\documentclass[a4paper,11pt]{article}

\usepackage[english]{babel}
\usepackage[utf8x]{inputenc}
\usepackage[T1]{fontenc}

\usepackage{natbib}
 
\usepackage[margin=2.5cm]{geometry}

\usepackage{amsmath}
\usepackage{enumerate}
\usepackage{natbib}
\usepackage{url} 

\usepackage{algorithm,algpseudocode}
\usepackage{pdfpages}
\usepackage{bm} 
\usepackage{caption}
\usepackage{subcaption}
\usepackage{url}
\usepackage{verbatim} 
\usepackage{booktabs}
\usepackage{xcolor}
\usepackage{soul}
\usepackage{marginnote}

\usepackage{tikz}
\usetikzlibrary{decorations.pathreplacing}

\usepackage{fourier} 
\usepackage{array}
\usepackage{makecell}

\usepackage{mathtools}
\usepackage{nccmath}
 
\usepackage[normalem]{ulem}
\usepackage{enumitem}

\usepackage[textwidth=20mm,textsize=scriptsize]{todonotes}
\usepackage{soul}
\usepackage{hyperref}

\date{}

\begin{document}  
\def\spacingset#1{\renewcommand{\baselinestretch}%
{#1}\small\normalsize} \spacingset{1}

\title{Accelerating delayed-acceptance Markov chain Monte Carlo algorithms}

\author{Samuel Wiqvist$^{\ast}$, Umberto Picchini$^{\diamond\ast}$, Julie Lyng Forman$^{\dagger}$, Kresten Lindorff-Larsen$^{\ddagger}$, \\ Wouter Boomsma$^{\star}$ }
\date{
$^{\ast}$Centre for Mathematical Sciences, Lund University, Sweden
\\
$^{\diamond}$Department of Mathematical Sciences, Chalmers University of Technology and the University of Gothenburg, Sweden
\\
$^{\dagger}$Dept. Public Health, section of Biostatistics, University of Copenhagen, Denmark
\\
$^{\ddagger}$The Linderstrøm-Lang Centre for Protein Science, Department of Biology, University of Copenhagen, Denmark \\
$^{\star}$Department of Computer Science, University of Copenhagen, Denmark
\vskip 0.8cm
}

\maketitle

\begin{abstract}
Delayed-acceptance Markov chain Monte Carlo (DA-MCMC) samples from a probability distribution via a two-stages version of the Metropolis-Hastings algorithm, by combining the target distribution with a ``surrogate'' (i.e. an approximate and computationally cheaper version) of said distribution. DA-MCMC accelerates MCMC sampling in complex applications, while still targeting the exact distribution. We design a computationally faster, albeit approximate, DA-MCMC algorithm.  We consider parameter inference in a Bayesian setting where a surrogate likelihood function is introduced in the delayed-acceptance scheme. When the evaluation of the likelihood function is computationally intensive, our scheme produces a 2-4 times speed-up, compared to standard DA-MCMC. However, the acceleration is highly problem dependent. Inference results for the standard delayed-acceptance algorithm and our approximated version are similar, indicating that our algorithm can return reliable Bayesian inference. As a computationally intensive case study, we introduce a novel stochastic differential equation model for protein folding data.
\end{abstract}

\noindent%
{\it Keywords:}  Bayesian inference, Gaussian process, pseudo marginal MCMC, protein folding, stochastic differential equation
\vfill

\section{Introduction}

We introduce a new strategy to accelerate Markov chain Monte Carlo (MCMC) sampling when the evaluation of the target distribution is computationally expensive. We build on the ``delayed-acceptance'' (DA) strategy developed in \cite{christen2005markov} where a fast, ``two-stages'' DA-MCMC algorithm is proposed while still targeting the desired distribution exactly. We produce an approximated and accelerated delayed-acceptance MCMC algorithm (ADA-MCMC), where in exchange of exactness we obtain results even more rapidly than the standard DA-MCMC. In a computationally intensive case study, the run-time for ADA-MCMC is 2--4 times faster than for standard DA-MCMC. 

The methodology we consider is general, as our novel method pertains sampling from arbitrary distributions. However, in the interest of our applications, we will focus on Bayesian inference, and then suggest how to implement our ideas for general problems. In Bayesian inference we aim at sampling from the posterior distribution $p(\theta|y)\propto p(y|\theta)p(\theta)$, where $\theta$ are model parameters, $y$ denotes data, $p(y|\theta)$ is the likelihood function, and $p(\theta)$ is the prior distribution of $\theta$. We assume that the point-wise evaluation of the likelihood $p(y|\theta)$ (or an approximation thereof) is computationally intensive, because the underlying probabilistic model is complex and/or the data $y$ is large. For those situations, DA-MCMC algorithms turn particularly useful. In the approach originally outlined in \cite{christen2005markov} a DA strategy decomposes an MCMC move into two stages. At the first stage a proposal can either be rejected, according to a ``surrogate of the posterior'' (one that is computationally cheap to evaluate and chosen to approximate the desired posterior), or be sent to the second stage. If the proposal is not rejected at the first stage, at the second stage an acceptance probability is used that corrects for the discrepancy between the approximate surrogate and the desired posterior, and at this stage the proposal can finally be accepted or rejected. The advantage of using DA-MCMC is that the computationally expensive posterior only appears in the second stage, whereas the surrogate posterior in the first stage is cheap to evaluate. Therefore, in the first stage the surrogate posterior rapidly screens proposals, and rejects those that are unlikely to be accepted at the second stage, if the surrogate model is reliable. When considering a Bayesian approach, we build a surrogate of the computationally expensive likelihood function, while we assume the cost of evaluating the prior to be negligible. Therefore the expensive likelihood appears only in the second stage. Some implementations of the DA approach in Bayesian inference can be found e.g. in \cite{golightly2015delayed}, \cite{sherlock2017adaptive}, and \cite{banterle2015accelerating}, and similar approaches based on approximate Bayesian computation (ABC) can be found in \cite{picchini2014inference}, \cite{picchini2016accelerating}, and \cite{everitt2017delayed}. 

In this work, the sequence of computations pertaining the second stage of DA-MCMC are arranged so to find further opportunities to avoid the evaluation of the expensive likelihood. This leads to our accelerated and approximated ADA-MCMC. The computational benefit of using ADA-MCMC is that, unlike DA-MCMC, once a parameter proposal reaches the second stage, the expensive likelihood is not necessarily evaluated, but this comes at the price of introducing an approximation in the sampling procedure. 
We test and compare delayed-acceptance algorithms, particle marginal methods for exact Bayesian inference, and Markov-chain-within-Metropolis on two case studies: The stochastic Ricker model, and a novel state-space model for protein folding data, with dynamics expressed via a stochastic differential equation (SDE).
Therefore, in this work we contribute with: (i) a novel, approximate and accelerated delayed-acceptance MCMC algorithm, and (ii) a novel double-well potential state-space model for protein folding data. For practical applications, we use Gaussian processes to specify surrogates of the likelihood function, though this is not an essential component of our approach and other surrogates of the likelihood can be considered.
We found that the acceleration produced by ADA-MCMC, compared to DA-MCMC, is dependent on the specific application. If the exact or approximate likelihood function used in the second stage of the algorithm is not computationally intensive to evaluate, then our method produces negligible benefits. Therefore, the use of our ADA-MCMC, just as the standard DA-MCMC, is beneficial when each evaluation of the likelihood has a non-negligible impact on the total computational budget. Then, the time savings due to ADA-MCMC are proportional to the number of MCMC iterations where the evaluation of the likelihood at the second stage is avoided. In terms of inference quality, we find that ADA-MCMC returns results that are very close to DA-MCMC, so our approximations do not seem to harm the accuracy of the resulting inference.

The outline of this paper is as follows: The delayed-acceptance (DA) scheme and our novel accelerated DA algorithm are introduced in a general framework in Section \ref{sec:damcmc}. The Gaussian process (GP) surrogate model is introduced in Section \ref{sec:gpmodel}. The DA-GP-MCMC algorithm and the accelerated version ADA-GPMCMC are introduced in Section \ref{sec:dagpmcmc}. A simulation study for the stochastic Ricker model is in Section \ref{sec:rickermodel}. The protein folding data and the novel double-well potential stochastic differential equation model are introduced in Section \ref{sec:dwpsde}. A discussion in Section \ref{sec:discussion} closes our work. Further supplementary material is available, outlining: particle Markov chain Monte Carlo methods for state-space models, implementation guidelines for the algorithms, a further simulation study, and diagnostic analyses. The code used to generate results can be found at
\href{https://github.com/SamuelWiqvist/adamcmcpaper}{https://github.com/SamuelWiqvist/adamcmcpaper} and in the supplementary material.

\section{Delayed-acceptance MCMC}
\label{sec:damcmc}

We first introduce the delayed-acceptance (DA-MCMC) scheme due to \cite{christen2005markov} in full generality, then we specialize it for Bayesian inference. Our accelerated delayed-acceptance (ADA-MCMC) algorithm is introduced in section \ref{sec:ada-mcmc}. 
We are interested in sampling from some distribution $p(x)$ using Metropolis-Hastings \citep{hastings1970monte}. Metropolis-Hastings proceeds by evaluating random moves produced by a Markov kernel from the current value of $x$ to a new $x^\star$. The sequence of accepted moves forms a Markov chain having $p(x)$ as stationary distribution. Now, assume that the point-wise evaluation of $p(x)$ is computationally expensive.  The main idea behind a DA-MCMC approach is to delay (or avoid as much as possible) the evaluation of the computationally expensive $p(x)$, by first trying to early-reject the proposal $x^\star$ using some surrogate (cheap to evaluate) deterministic or stochastic model $\tilde{p}(x)$. 
To enable early-rejections while still targeting the distribution $p(x)$, a two-stages acceptance scheme is introduced in \cite{christen2005markov}. Say that we are at the $r$th iteration of the Metropolis-Hastings algorithm, and denote with $x^{r-1}$ the state of the chain produced at the previous iteration. At the ``first stage'' of DA-MCMC we evaluate the acceptance probability (though at this stage we do not really accept any proposal as explained below)
\begin{equation} \label{eq:da_stage_1}
\alpha_{1} = \min \biggl( 1, \frac{\tilde{p}(x^{\star})}{\tilde{p}(x^{r-1})} \cdot \frac{g(x^{r-1} | x^{\star}) }{g(x^{\star} | x^{r-1})}  \biggr),
\end{equation}
where $g(x | y)$ is the transition kernel used to generate proposals, i.e. at the $r$th iteration $x^\star\sim g( x | x^{r-1})$. If the proposal $x^\star$ ``survives'' the first stage (i.e. if it is not rejected) it is then promoted to the second stage where it is accepted with probability $\alpha_2$,
\begin{align} \label{eq:da_stage_2}
\alpha_{2} = \min \biggl( 1, \frac{p(x^{\star})}{p(x^{r-1})} \cdot \frac{\tilde{p}(x^{r-1}) }{ {\tilde{p}}(x^\star) } \biggr).
\end{align}
Therefore $x^\star$ can only be accepted at the second stage, while it can be rejected both at the first and second stage. A computational speed-up is obtained when $x^\star$ is early-rejected at the first stage, as there the expensive $p(x^\star)$ is not evaluated. Hence, to obtain a significant speed-up it is important to early-reject ``bad'' proposals that would likely be rejected at the second stage. The probability $\alpha_2$ corrects for the approximation introduced in the first stage and the resulting Markov chain has the correct stationary distribution $p(x)$. This result holds if $g$ is $p$-irreducible and reversible, and if $g(x | y) > 0 $ implies $\tilde{p}(x) > 0$.
From  \eqref{eq:da_stage_2} it is evident how the surrogate model acts as a proposal distribution. See \cite{franks2017importance} for a comparison in terms of asymptotic variances of Monte Carlo estimators  provided via importance sampling, pseudo-marginal and delayed-acceptance methods.

In a Bayesian framework we are interested in sampling from the posterior $p(\theta |y)\propto p(y | \theta)p(\theta)$. Furthermore, for the cases of interest to us, the log-likelihood function (or an approximation thereof) $\ell(\theta):=\log p(y|\theta)$, is computationally expensive while the prior distribution is assumed cheap to evaluate.
By introducing a deterministic or stochastic surrogate likelihood  $\tilde{L}(\theta) := \exp(\tilde{\ell}(\theta))$, DA has first stage acceptance probability $\alpha_1$, where
\begin{align*} 
\alpha_{1} = \min \biggl( 1, \frac{\tilde{L}(\theta^{\star})}{\tilde{L}(\theta^{r-1})} \cdot \frac{p(\theta^{\star})}{p(\theta^{r-1})} \cdot \frac{ g(\theta^{r-1}|\theta^{\star}) }{ g(\theta^{\star} |\theta^{r-1}) } \biggr),
\end{align*}
with transition kernel $g$. Similarly, by setting ${L}(\theta) := \exp({\ell}(\theta))$, the second stage acceptance probability is
\begin{align*} 
\alpha_{2} = \min \biggl( 1, \frac{L(\theta^{\star})}{L(\theta^{r-1})} \cdot \frac{\tilde{L}(\theta^{r-1}) }{ \tilde{L}(\theta^{\star}) } \biggr).
\end{align*}
An extension of the DA-MCMC scheme due to \cite{sherlock2017adaptive} is to generate a proposal $\theta^{\star}$ from a different transition kernel $\tilde{g}( \cdot | \theta^{r-1})$, and with a small but positive probability $\beta_{MH}\in (0,1)$ allow the evaluation of the proposal $\theta^{\star}$ in an ordinary Metropolis-Hastings algorithm, with acceptance probability denoted $\alpha_{MH}$,   
\begin{align}  \label{eq:regular-acceptance-probability}
\alpha_{MH} = \min \biggl( 1, \frac{L(\theta^{\star})}{L(\theta^{r-1})} \cdot \frac{p(\theta^{\star})}{p(\theta^{r-1})} \cdot \frac{ \tilde{g}(\theta^{r-1}| \theta^{\star}) }{ \tilde{g}(\theta^{\star} | \theta^{r-1}) } \biggr).
\end{align}
In this case the proposal can be immediately accepted or rejected as in a regular MCMC. The transition kernel $g$ should have a somewhat larger variance than $\tilde{g}$. With probability $1-\beta_{MH}$ a proposal is instead evaluated using the two-stages 
DA-MCMC algorithm.
When considering this ``extended version'' of DA-MCMC (where 
$\beta_{MH}$ is introduced) it is preferable to use a small $\beta_{MH}$ in order not to lose too much of the acceleration implied by a DA approach. Our experience also indicates that this extension can be critical to better explore the tails of the posterior distribution, compared to a standard DA-MCMC that uses $\beta_{MH}=0$. This "mixture" of the two Metropolis-Hastings kernels (i.e. the acceptance kernel for the DA scheme, and the acceptance kernel in \eqref{eq:regular-acceptance-probability}) produces a valid MCMC algorithm, since both kernels in the standard cases target the correct posterior \citep{rosenthal2007coupling}.

\subsection{Accelerated delayed-acceptance MCMC}\label{sec:ada-mcmc}

There have been a number of  attempts at accelerating the original DA-MCMC of \cite{christen2005markov}. For example, in a Bayesian framework, \cite{banterle2015accelerating} propose to break down the posterior into the product of $d$ chunks. The Metropolis-Hastings acceptance ratio becomes the product of $d$ acceptance ratios, each of which can be sequentially evaluated against one of $d$ independent uniform variates. The acceleration is given by the possibility to ``early-reject'' a proposal, as soon as one of those acceptance ratios leads to a rejection (in the same spirit of \citealp{solonen2012efficient}). However, an acceptance  requires instead the scanning of all $d$ components, i.e. the full posterior. \cite{quiroz2017speeding} never use the full data set in the second stage of DA and instead construct an approximated likelihood from subsamples of the data, which is particularly relevant for Big Data problems (see references therein and \citealp{angelino2016patterns}). Remarkably, \cite{quiroz2017speeding} prove that even when the full likelihood is approximated using data subsamples, the resulting chain has the correct stationary distribution. However, they assume data to be conditionally independent, a strong condition which does not apply to case studies considered in the present work.  

We now introduce the novel, accelerated DA-MCMC algorithm, shortly ADA-MCMC. The main idea behind ADA-MCMC is that, under some assumptions on how the likelihood function and the surrogate model relate, it is possible to arrange the computations in the second stage to obtain an acceleration in the computations. This is implied by the possibility to avoid the evaluation of the expensive likelihood in the second stage, in some specific circumstances. However, this also implies that ADA-MCMC is an approximated procedure, since a proposal can sometimes be accepted according to the surrogate model.
We introduce ADA-MCMC in a Bayesian setting where the surrogate model pertains the 
likelihood function. However, the idea can straightforwardly be adapted to the case where a surrogate model of a generic distribution $p(x)$ is used, as in Equations \eqref{eq:da_stage_1}-\eqref{eq:da_stage_2}. The more general setting is briefly described later in this section.  
As previously mentioned, at the $r$th iteration the DA algorithm is governed by the values of the likelihood function $L(\theta^\star)$ and $L(\theta^{r-1})$, and the values of the surrogate model  $\tilde{L}(\theta^\star)$ and $\tilde{L}(\theta^{r-1})$. These four values can be considered arranged in four mutually exclusive scenarios:
\begin{align*}
\text{case 1)} \,\, \tilde{L}(\theta^\star) > \tilde{L}(\theta^{r-1}) \,\, \textit{and} \,\, L(\theta^\star) > L(\theta^{r-1}), \\
\text{case 2)} \,\, \tilde{L}(\theta^\star) < \tilde{L}(\theta^{r-1}) \,\, \textit{and} \,\, L(\theta^\star) < L(\theta^{r-1}), \\
\text{case 3)} \,\, \tilde{L}(\theta^\star) > \tilde{L}(\theta^{r-1}) \,\, \textit{and} \,\, L(\theta^\star) < L(\theta^{r-1}), \\
\text{case 4)} \,\, \tilde{L}(\theta^\star) < \tilde{L}(\theta^{r-1})  \,\, \textit{and} \,\, L(\theta^\star) > L(\theta^{r-1}).
\end{align*}
We study each case separately to investigate any opportunity for accelerating the computations in the second stage of DA-MCMC, under the assumption that the relations between the evaluations of $\tilde{L}$ and $L$ hold. Afterwards, we suggest ways to determine approximately which of the four possibilities we should assume to hold, for any new proposal $\theta^\star$, without evaluating the expensive likelihood $L(\theta^\star)$.  

\paragraph{Case 1)} Under the assumption that $\tilde{L}(\theta^\star) > \tilde{L}(\theta^{r-1})$ \textit{and}  $L(\theta^\star) > L(\theta^{r-1})$ it is clear that $\frac{\tilde{L}(\theta^{r-1})}{\tilde{L}(\theta^\star)} < 1$ \textit{and} $\frac{L(\theta^{r-1})}{L(\theta^\star)} < 1$. It also holds that 
\begin{align} 
	\frac{\tilde{L}(\theta^{r-1})}{\tilde{L}(\theta^\star)} < \frac{L(\theta^{\star})}{L(\theta^{r-1})} \cdot \frac{\tilde{L}(\theta^{r-1}) }{ \tilde{L}(\theta^{\star})}.\label{eq:ratios-case1}
\end{align}
Hence, the acceptance region for the second stage can be split in two parts, where one part is ``governed'' by $\frac{\tilde{L}(\theta^{r-1})}{\tilde{L}(\theta^\star)}$ only. To clarify, at the second stage of the standard DA-MCMC, acceptance of a proposed $\theta^\star$ takes place if $u< \frac{L(\theta^{\star})}{L(\theta^{r-1})} \cdot \frac{\tilde{L}(\theta^{r-1}) }{ \tilde{L}(\theta^{\star})}$ where $u\sim U(0,1)$ is uniformly distributed in [0,1], hence, the acceptance region is $\bigl[0,\frac{L(\theta^{\star})}{L(\theta^{r-1})} \cdot \frac{\tilde{L}(\theta^{r-1}) }{ \tilde{L}(\theta^{\star})}\bigr]$. However, because of \eqref{eq:ratios-case1} we are allowed to further decompose the acceptance region, as presented below: 

\begin{center}

\begin{tikzpicture}

\draw (-0.3,0) --(11,0);
\draw (0,0.15) --(0,-0.15);
\draw (4,0.15) --(4,-0.15);
\draw (7,0.15) --(7,-0.15);
\draw (10,0.15) --(10,-0.15);

\node [below] at (0,-0.15) {0};
\node [below] at (10,-0.15) {1};
\node [below] at (10.75,0) {$u$};

\node [below] at (4,-0.15) {$\frac{\tilde{L}(\theta^{r-1})}{\tilde{L}(\theta^\star)}$};

\node [below] at (7,-0.15) {$\frac{L(\theta^{\star})}{L(\theta^{r-1})} \cdot \frac{\tilde{L}(\theta^{r-1}) }{ \tilde{L}(\theta^{\star})}$};

\draw [thick, decorate, decoration={brace,amplitude=8pt}](0,0.2) -- (4,0.2) node[midway,right,xshift=7pt] {};

\draw [thick, decorate, decoration={brace,amplitude=8pt}](4,0.2) -- (7,0.2) node[midway,right,xshift=7pt] {};

\draw [thick, decorate, decoration={brace,amplitude=8pt}](7,0.2) -- (10,0.2) node[midway,right,xshift=7pt] {};

\node[align=center, above] at (2,0.5)%
{early-accept};
\node[align=center, above] at (5.5,0.5)%
{accept};
\node[align=center, above] at (8.5,0.5)%
{reject};

\end{tikzpicture}

\end{center}
Hence, if a proposal $\theta^\star$ has survived the first stage and we assume that we are in case 1, we can first check whether we can ``early-accept'' the proposal (i.e. without evaluating the expensive likelihood), that is, check if 
\begin{align} \label{eq:case_1_a_2_1}
u < \frac{\tilde{L}(\theta^{r-1})}{\tilde{L}(\theta^\star)},
\end{align}
and if this is the case $\theta^\star$ is (early)-accepted and stored, and we can move to the next iteration of ADA-MCMC. If $\theta^\star$ is not early-accepted, we can look into the remaining part of the [0,1] segment to determine if the proposal can be accepted or rejected. Hence, when early-acceptance is denied, the expensive likelihood $L(\theta^{\star})$ is evaluated and the proposal is accepted and stored if 
\begin{align} \label{eq:case_1_a_2_2}
u <\frac{L(\theta^{\star})}{L(\theta^{r-1})} \cdot \frac{\tilde{L}(\theta^{r-1}) }{ \tilde{L}(\theta^{\star}) },  
\end{align}
and rejected otherwise, and we can move to the next iteration of ADA-MCMC. 
Since the acceptance region for the second stage is split in two parts (early-acceptance and acceptance), the \textit{same} random number $u$ is used in  \eqref{eq:case_1_a_2_1} and \eqref{eq:case_1_a_2_2}. By splitting  the region it is possible to early-accept proposals without evaluating $L(\theta^\star)$, and thereby obtaining a speed-up.

\paragraph{Case 2)} If this case holds, then $\frac{\tilde{L}(\theta^{r-1})}{\tilde{L}(\theta^{\star})}>1$ and $\frac{L(\theta^{\star})}{L(\theta^{r-1})}<1$. Hence, it is not possible to obtain any early-accept or early-reject opportunity in this case. 

\paragraph{Case 3)} If this case holds, then $\frac{\tilde{L}(\theta^{r-1})}{\tilde{L}(\theta^{\star})}<1$ and $\frac{L(\theta^{\star})}{L(\theta^{r-1})}<1$. Hence, it also holds that 
\begin{align*} 
	\frac{\tilde{L}(\theta^{r-1})}{\tilde{L}(\theta^\star)} > \frac{L(\theta^{\star})}{L(\theta^{r-1})} \cdot \frac{\tilde{L}(\theta^{r-1}) }{ \tilde{L}(\theta^{\star})}.
\end{align*}
The rejection region is $\bigl[\frac{L(\theta^{\star})}{L(\theta^{r-1})} \cdot \frac{\tilde{L}(\theta^{r-1}) }{ \tilde{L}(\theta^{\star})},1\bigr]$ and this can be split in two parts, where one part is only governed by $\frac{\tilde{L}(\theta^{r-1})}{\tilde{L}(\theta^\star)}$, see below: 

\begin{center}

\begin{tikzpicture}

\draw (-0.3,0) --(11,0);
\draw (0,0.15) --(0,-0.15);
\draw (4,0.15) --(4,-0.15);
\draw (7,0.15) --(7,-0.15);
\draw (10,0.15) --(10,-0.15);

\node [below] at (0,-0.15) {0};
\node [below] at (10,-0.15) {1};
\node [below] at (10.75,0) {$u$};

\node [below] at (4,-0.15) {$\frac{L(\theta^{\star})}{L(\theta^{r-1})} \cdot \frac{\tilde{L}(\theta^{r-1}) }{ \tilde{L}(\theta^{\star})}$};

\node [below] at (7,-0.15) {$\frac{\tilde{L}(\theta^{r-1})}{\tilde{L}(\theta^\star)}$};

\draw [thick, decorate, decoration={brace,amplitude=8pt}](0,0.2) -- (4,0.2) node[midway,right,xshift=7pt] {};

\draw [thick, decorate, decoration={brace,amplitude=8pt}](4,0.2) -- (7,0.2) node[midway,right,xshift=7pt] {};

\draw [thick, decorate, decoration={brace,amplitude=8pt}](7,0.2) -- (10,0.2) node[midway,right,xshift=7pt] {};

\node[align=center, above] at (2,0.5)%
{accept};
\node[align=center, above] at (5.5,0.5)%
{reject};
\node[align=center, above] at (8.5,0.5)%
{early-reject};

\end{tikzpicture}

\end{center}
By simulating a $u\sim U(0,1)$, we can first check if the proposal can be early-rejected. This happens if $u > \frac{\tilde{L}(\theta^{r-1})}{\tilde{L}(\theta^\star)}$. If the proposal is not early-rejected, it is accepted if 
\begin{align*}
u < \frac{L(\theta^{\star})}{L(\theta^{r-1})} \cdot \frac{\tilde{L}(\theta^{r-1}) }{ \tilde{L}(\theta^{\star}) }, 
\end{align*}
and rejected otherwise. Hence, in case 3 there is a chance to early-reject $\theta^\star$ without evaluating $L(\theta^\star)$.

\paragraph{Case 4)} Under the assumption we have that  $\frac{\tilde{L}(\theta^{r-1})}{\tilde{L}(\theta^{\star})}>1$ and $\frac{L(\theta^{\star})}{L(\theta^{r-1})}>1$, and we can immediately accept the proposal without evaluating $L(\theta^\star)$, since $\frac{L(\theta^{\star})}{L(\theta^{r-1})} \cdot \frac{\tilde{L}(\theta^{r-1}) }{ \tilde{L}(\theta^{\star})} > 1$.\\

Clearly, assuming a specific case to be the ``right one'', for proposal $\theta^\star$, is a decision subject to probabilistic error. This is why ADA-MCMC is an approximate version of DA-MCMC. Of course, the crucial problem is to determine which of the four cases to assume to hold for the proposed $\theta^\star$. One method is to consider a pre-run of some MCMC algorithm, to estimate the probability $p_j$ for each of the four different cases, where $p_j$ is the true but unknown probability that case $j$ holds, $j=1,...,4$. This is of course a possibly computationally heavy procedure, however, for the specific algorithms we study in Section \ref{sec:dagpmcmc}, such a pre-run is necessary to construct the surrogate model for the log-likelihood, hence the estimation of the $p_j$ comes as a simple by-product of the inference procedure. Then, once the estimates $\hat{p}_j$ are obtained, for a new $\theta^\star$ one first checks if $\frac{\tilde{L}(\theta^{r-1})}{\tilde{L}(\theta^{\star})}<1$ or if $\frac{\tilde{L}(\theta^{r-1})}{\tilde{L}(\theta^{\star})}>1$. If $\frac{\tilde{L}(\theta^{r-1})}{\tilde{L}(\theta^{\star})}>1$ then we can either be in case 2 or 4. We toss a uniform $u$ and if $u<\hat{p}_2$ case 2 is selected with probability $\hat{p}_2$ (and otherwise case 4 is selected, since $\hat{p}_4 = 1-\hat{p}_2$). Correspondingly, if $\frac{\tilde{L}(\theta^{r-1})}{\tilde{L}(\theta^{\star})}<1$
then we can be either in case 1 or 3. We toss a uniform $u\sim U(0,1)$, and if $u<\hat{p}_1$ case 1 is selected (otherwise case 3 is selected, since $\hat{p}_3 = 1-\hat{p}_1$). 
Another approach is to model the probabilities as a function of $\theta$. Hence, we are then interested in computing the probabilities $\hat{p}_1(\theta)$,  $\hat{p}_2(\theta)$,$\hat{p}_3(\theta)$, and $\hat{p}_4(\theta)$. For this task, we can for instance use logistic regression, or some other classification algorithm.   
The problem of the selection of cases 1--4 is discussed in detail in Section \ref{sec:adagpmcmc}. 

We stated early that ADA can also be used in a non-Bayesian setting, where we target a generic distribution $p(x)$ for some $x\in\mathcal{X}$. In that case we need to introduce a corresponding surrogate model $\tilde{p}(x)$. The $r$th iteration of ADA will then be governed by the four values $\tilde{p}(x^\star)$, $\tilde{p}(x^{r-1})$, $p(x^\star)$, and $p(x^{r-1})$, where $x^\star$ is a proposed value $x^*\in\mathcal{X}$. These can be arranged into four cases, similarly to what previously described: case 1)  $\tilde{p}(x^\star)$ > $\tilde{p}(x^{r-1})$ \textit{and} $p(x^\star)$ > $p(x^{r-1})$, 2)   $\tilde{p}(x^\star)$ < $\tilde{p}(x^{r-1})$  \textit{and} $p(x^\star)$ < $p(x^{r-1})$, 3) $\tilde{p}(x^\star)$ > $\tilde{p}(x^{r-1})$  \textit{and}  $p(x^\star)$ < $p(x^{r-1})$, and 4) $\tilde{p}(x^\star)$ < $\tilde{p}(x^{r-1})$  \textit{and} $p(x^\star)$ > $p(x^{r-1})$. 
Therefore, by adapting the methodology, possibilities for early-rejection and early-acceptance of a proposal $x^\star$ can straightforwardly be obtained regardless of whether we pursue a Bayesian analysis or not.

\section{Modeling the log-likelihood function using Gaussian processes} \label{sec:gpmodel}

We have outlined our methodology without reference to a specific choice for the surrogate likelihood. A possibility is to use Gaussian process regression to obtain a surrogate log-likelihood $\log\tilde{L}$. Gaussian processes (GPs) is a class of statistical models that can be used to  describe the uncertainty  about  an
unknown  function. In our case, the unknown function is the log-likelihood $\ell(\theta)=\log p(y|\theta)$. A GP has  the  property  that  the  joint  distribution  for  the  values  of  the
unknown function, at a finite collection of points, has a multivariate normal distribution. As such, each Gaussian process is fully specified by a mean function $m$, and a covariance function $k$  \citep{rasmussen2006gaussian}. We introduce a GP regression model, similar to the one used in \cite{drovandi2015accelerating}, as a computationally cheap proxy to the unknown log-likelihood $\ell(\theta)$. Our GP model uses covariates that are powers and interactions of the $d$ parameters of interest $\theta=(\theta_1,...,\theta_d)$ (see the supplementary material). The GP model assumes 
\begin{align*}  
\ell(\theta) \sim \mathcal{GP}(m_{\beta}(\theta), k_{\phi}(\theta, \theta^{\prime})),
\end{align*}
where $\eta= [\phi,\beta]$ are the auxiliary parameters for the mean and covariance function respectively. Since $\eta$ is in general unknown, this must be estimated by fitting the GP model to some ``training data''. In our case, training data is obtained by running a a number of preliminary MCMC iterations, and collect all generated parameter proposals and corresponding log-likelihood values. The GP regression considers the log-likelihood values as ``responses'' and the proposed parameters are used to construct the covariates. Once $\hat{\eta}$ is available, then for any new $\theta^\star$ we obtain a proxy to the unknown log-likelihood that is computationally much faster to evaluate than $\ell(\theta^\star)$.  
The training data we fit the GP model to is denoted $\mathcal{D}$, and how this data is collected is explained in Section \ref{sec:dagpmcmc}. Using the same assumptions for the Gaussian process model as in \cite{drovandi2015accelerating}, we have that the predictive distribution for the GP model is available in closed form. Therefore, for given $\mathcal{D}$ and $\hat{\eta}$ we can easily produce a draw from said distribution, which is Gaussian, and given by 
\begin{align}  
\ell(\theta^{\star}) |\mathcal{D}, \hat{\eta} \sim \mathcal{N}(\bar{\ell}(\theta^{\star}), \mathrm{Var} (\ell(\theta^{\star}))).
\label{eq:gp-predictive}
\end{align}
See the supplementary material for the definitions of $\bar{\ell}(\theta)$ and $\mathrm{Var} (\ell(\theta))$. 
It is computationally very rapid to produce draws from \eqref{eq:gp-predictive} at any new $\theta^{\star}$, which is why we use GP prediction as a surrogate of the log-likelihood within DA algorithms.
The derivation of \eqref{eq:gp-predictive}, and more details pertaining the GP model are found in the supplementary material.

\section{Delayed-acceptance Gaussian process Markov chain Monte Carlo} \label{sec:dagpmcmc}

We now make use of the fitted GP model discussed in Section \ref{sec:gpmodel} as a surrogate of the log-likelihood function, within DA-MCMC and ADA-MCMC.  By sampling a GP log-likelihood $\ell_{GP}(\theta^{\star}):=\ell(\theta^{\star}) |\mathcal{D}, \hat{\eta}$ from \eqref{eq:gp-predictive} for some $\theta^{\star}$,  we denote with $\hat{L}_{GP}(\theta^{\star}) = \exp(\ell_{GP}(\theta^{\star}))$ the GP prediction of the corresponding likelihood function. In addition to be computationally intensive to evaluate, the true likelihood $L(\theta)$ might also be unavailable in closed form. However, it is often possible to obtain Monte Carlo approximations returning non-negative unbiased estimates of $L$. We denote with $\hat{L}_{u}(\theta)$ such unbiased estimate. For our case studies, $\hat{L}_{u}(\theta)$ is obtained via sequential Monte Carlo (SMC, also known as particle filter, see \citealp{kantas2015particle} and \citealp{schon2018probabilistic} for reviews). A simple example of SMC algorithm (the bootstrap filter) and its use within particle-marginal methods \citep{andrieu2009pseudo} for inference in state-space models are presented in the supplementary materials. Two types of pseudo-marginal methods, particle MCMC (PMCMC) and Markov-chain-within-Metropolis (MCWM), are there described. In the supplementary material we give a brief technical presentation of PMCMC and MCWM.

Notice that MCMC algorithms based on GP-surrogates have already been considered, e.g. in \cite{meeds2014gps} and \cite{drovandi2015accelerating}. \cite{meeds2014gps} assume that the latent process has a Gaussian distribution with unknown moments, and these moments are estimated via simulations using ``synthetic likelihoods''. There, the discrepancy between the simulated (Gaussian) latent states and observed data is evaluated using a Gaussian ABC kernel, where ABC stands for ``approximate Bayesian computation'', see \cite{marin2012approximate} for a review. This computationally expensive setting is fitted to ``training data'', then used in place of the (unknown) likelihood into a pseudo-marginal MCMC algorithm. The work in \cite{drovandi2015accelerating} builds up on the ideas found in \cite{meeds2014gps}, with the difference that the former does not use synthetic likelihoods nor ABC to produce training data. Instead they use the MCWM algorithm to collect many log-likelihood evaluations at all proposed parameter values, then fit a GP regression model on these training data. Finally, they use the fitted GP regression in a pseudo-marginal algorithm, without ever resorting to expensive likelihood calculations. As opposed to \cite{drovandi2015accelerating}, we make use of both a surrogate of the likelihood and (with low frequency)  of the expensive likelihood approximated via a particle filter. We call DA-GP-MCMC a delayed acceptance MCMC algorithm using predictions from GP regression as a surrogate of the likelihood function. Similarly, we later introduce our accelerated version ADA-GP-MCMC.

The DA-GP-MCMC procedure is detailed in Algorithm \ref{alg:DAGPMCMC}. Using the notation in Section \ref{sec:damcmc}, we now have that the first stage acceptance probability for DA-GP-MCMC is 
\begin{align*}  
\alpha_{1} = \min \biggl( 1, \frac{\hat{L}_{GP}(\theta^{\star})}{\hat{L}_{GP}(\theta^{r-1})} \cdot \frac{p(\theta^{\star})}{p(\theta^{r-1})} \cdot \frac{ g(\theta^{r-1} |\theta^{\star}) }{ g(\theta^{\star} |\theta^{r-1}) } \biggr).
\end{align*}
The second stage acceptance probability is 
\begin{align*}  
\alpha_{2} = \min \biggl(1, \frac{\hat{L}_{u}(\theta^{\star})}{\hat{L}_{u}(\theta^{r-1}) }\cdot \frac{\hat{L}_{GP}(\theta^{r-1})}{\hat{L}_{GP}(\theta^{\star})} \biggr).
\end{align*}
As mentioned in Section \ref{sec:damcmc}, for our applications we found it beneficial to use the extended DA-MCMC introduced in \cite{sherlock2017adaptive}. However, this is in general not a requirement for using DA-MCMC. The DA-GP-MCMC algorithm is preceded by the following two steps, required to collect training data and fit the GP regression to these data: 

\paragraph{1. Collect training data using MCWM:} A MCWM algorithm is run 
to approximately target $p(\theta|y)$, where a bootstrap particle filter using $N$ particles is employed to obtain $\hat{L}_u(\theta)$, until the chain has reached apparent stationarity. When using MCWM we do not target the exact posterior for a finite number of particles $N$, however, this is not a concern to us. In fact, we use MCWM as in \cite{drovandi2015accelerating}, namely to ``harvest'' a large number of (approximate) log-likelihood function evaluations, in order to learn the dependence between loglikelihoods and corresponding parameters. Indeed, in this phase we store as training data $\mathcal{D}$ \textit{all} the proposed parameters $\theta^\star$ (regardless of whether these are accepted or rejected from MCWM) and their corresponding log-likelihoods $\ell_{u}(\theta^\star)$. Hence, all parameter proposals and corresponding log-likelihoods from MCWM (excluding some sufficiently long burnin period) are stored as training data $\mathcal{D} = \{ \theta^{*i}, \ell_{u}^{*i} \}$, (where here the superscript $i$ ranges from 1 to the number of iterations post-burnin). We also collect the generated Markov chain $\theta^{i}$ and their corresponding log-likelihood estimations in  $\tilde{\mathcal{D}} = \{ \theta^i, \ell_{u}^i \}$. Basically the difference between $\mathcal{D}$ and $\tilde{\mathcal{D}}$ is that parameters $\theta^i$ in the latter are the standard output of a Metropolis-Hastings procedure, i.e. $\tilde{\mathcal{D}}$ may contain ``repeated parameters'' (when rejections occur). Instead $\mathcal{D}$ contains all simulated proposals. We motivate the use for set $\tilde{\mathcal{D}}$ in Section \ref{sec:adagpmcmc}.   
\paragraph{2. Fit the GP model:} The Gaussian process model is fitted to  the training data $\mathcal{D}$ using the method described in Section \ref{sec:gpmodel}. 

\begin{algorithm}[H]
\scriptsize
\caption{DA-GP-MCMC algorithm}\label{alg:DAGPMCMC}
\begin{algorithmic}[1]
	
	\Require Number of iterations $R$, a GP model fitted to the training data, a starting value $\theta^0$ and corresponding $\hat{L}_{u}(\theta^0)$. 
	\Ensure The chain $\theta^{1:R}$.
    \For{$r = 1,...,R$} 
    \State Propose $\theta^{\star} \sim g( \, \cdot |\theta^{r-1})$. \Comment{Run two stages DA scheme}
	\State Sample from \eqref{eq:gp-predictive} to predict independently $\ell_{GP}(\theta^{\star})$ and $\ell_{GP}(\theta^{r-1})$. Define $\hat{L}_{GP}(\theta^{\star}):=\exp(\ell_{GP}(\theta^{\star}))$ and $\hat{L}_{GP}(\theta^{r-1}):=\exp(\ell_{GP}(\theta^{r-1}))$. 		
        \State Compute $\alpha_{1} = \min \big( 1, \frac{\hat{L}_{GP}(\theta^{\star})}{\hat{L}_{GP}(\theta^{r-1})} \cdot \frac{g(\theta^{r-1}|\theta^{\star})}{g(\theta^{\star} |\theta^{r-1})} \cdot \frac{p(\theta^{\star}) }{p(\theta^{r-1})}  \big)$.
	\State Draw $u \sim \mathcal{U}(0,1)$.
	\If{$u > \alpha_{1}$} \Comment{Early-reject}
	\State Set $\theta^{r} = \theta^{r-1}$.
	
	\Else
	\State Compute  $\hat{L}_{u}(\theta^{\star})$. \Comment{Second stage update scheme}
	\State Compute $\alpha_{2} = \min(1, \frac{\hat{L}_{u}(\theta^{\star})}{\hat{L}_{u}(\theta^{r-1})} \cdot \frac{\hat{L}_{GP}(\theta^{r-1}) }{  \hat{L}_{GP}(\theta^{\star}) } )$.  
	\State Draw $u \sim \mathcal{U}(0,1)$.
	\If{$u \le \alpha_{2}$} \Comment{Accept proposal}
	\State Set $\theta^{r} = \theta^{\star}$.
	\Else
	\State Set $\theta^{r} = \theta^{r-1}$. \Comment{Reject proposal}
	\EndIf
    \EndIf
	\EndFor
\end{algorithmic}
\end{algorithm}

\subsection{Accelerated delayed-acceptance Gaussian process MCMC}  \label{sec:adagpmcmc}

Our accelerated delayed-acceptance Gaussian process MCMC algorithm (ADA-GP-MCMC) is described in Algorithm \ref{alg:ADAGPMCMC_V2}. Same as for DA-GP-MCMC, also ADA-GP-MCMC is preceded by two phases (collection of training data and GP regression). After fitting the GP model, the training data is also used to produce a ``selection method'' for the four cases introduced in Section \ref{sec:ada-mcmc}.
As already mentioned in Section \ref{sec:ada-mcmc}, we can either select which case to use independently of the current proposal $\theta^{\star}$, or make the selection of cases a function of $\theta^{\star}$. We introduce three selection methods, where the first one selects which case to assume independently of $\theta^{\star}$, while the other two depend on the proposal.  

\paragraph{Biased coin:} In the most naive approach, selecting a case between 1 and 3, or between 2 and 4 can be viewed as the result of tossing a biased coin. Hence, we just compute the relative frequency of occurrence for cases 1, 2, 3 and 4 (see Section \ref{sec:ada-mcmc}) as observed in the training data. These are obtained as follows: using the fitted GP model we predict log-likelihoods $\ell_{GP}(\theta)\equiv\ell(\theta)|\mathcal{D},\hat{\eta}$ using \eqref{eq:gp-predictive} for all collected $\theta\in\Theta$ ($\Theta$ denotes the matrix of the $\theta$ proposals that belong to the training data $\mathcal{D}$). Then we obtain corresponding $\hat{L}_{GP}:=\exp(\ell_{GP}(\theta))$, for all $\theta\in\Theta$. Now, since all the corresponding $\hat{L}_{u}(\theta)$ are already available as training data, it is possible to compute said relative frequencies $\hat{p}_j$ of occurrence for each case $j$ ($j=1,..,4$). At iteration $r$ of the ADA-GP-MCMC algorithm, for  proposal $\theta^\star$, and supposing we have survived the first stage, 
then if $\hat{L}_{GP}(\theta^{\star}) > \hat{L}_{GP}(\theta^{r-1})$ we draw from the $\mathrm{Bernoulli}(\hat{p}_1)$ distribution and go for case 1 if the draw equals one, and go for case 3 otherwise. If instead $\hat{L}_{GP}(\theta^{\star}) < \hat{L}_{GP}(\theta^{r-1})$ we draw from $\mathrm{Bernoulli}(\hat{p}_2)$ and go for case 2 if the draw equals one, and go for case 4 otherwise. 

\paragraph{State-dependent selection:} The biased coin model does not take into account the specific value of the current proposal $\theta^{\star}$, that is, the same $\hat{p}_j$ are applied to all proposals during a run of ADA-GP-MCMC. We could instead estimate $\hat{p}_j(\theta)$ using logistic regression or a decision tree model. When using logistic regression, we have two regression models to estimate, one for cases 1 and 3, and one for cases 2 and 4.  By combining the training data $\mathcal{D}$, and the accepted proposals stored in $\tilde{\mathcal{D}}$, we have access to both the particle filter evaluations corresponding to all generated proposals, and to the ones for the accepted proposals. Using $\mathcal{D}$ and  $\tilde{\mathcal{D}}$ we can now classify which case each proposal \textit{should} belong to. This is done by computing GP predictions, independently for both sets of parameters stored in $\mathcal{D}$ and  $\tilde{\mathcal{D}}$. Note, after computing the GP predictions we have (i)  particle filter predictions and GP predictions for all proposals in ${\mathcal{D}}$, i.e. $\hat{L}_{u}(\theta^{\star})$ and $\hat{L}_{GP}(\theta^{\star})$, and (ii) particle filter predictions and GP predictions for all accepted proposals in $\tilde{\mathcal{D}}$, hence, $\hat{L}_{u}(\theta^{r-1})$ and $\hat{L}_{GP}(\theta^{r-1})$. We can now loop over the proposals in the training data and assign labels for which of the four cases each proposal belongs to. As an example, after labelling is performed, all proposals in the training data that are classified to belong to case 1 or 3 are denoted $\theta^{\star}_{1,3}$, and an associated indicator vector $y_{1,3}$, having 1 for proposals belonging to case 1 and 0 for proposals belonging to case 3, is created. We now fit a logistic regression model on $\{ \theta^{\star}_{1,3}, y_{1,3} \}$, where the $\theta^{\star}_{1,3}$ take the role of ``covariates'' and the $y_{1,3}$ are binary ``responses''. We denote with $\hat{p}_1(\theta)$ the resulting fitted probability of selecting case 1 (so that $\hat{p}_3(\theta)=1-\hat{p}_1(\theta)$). In a similar way, after labelling is performed, all proposals in the training data that are classified to belong to case 2 or 4 are denoted $\theta^{\star}_{2,4}$, with associated indicator vector $y_{2,4}$.  We fit a logistic regression model on $\{ \theta^{\star}_{2,4}, y_{2,4} \}$ to obtain $\hat{p}_2(\theta)$ (and $\hat{p}_4(\theta)=1-\hat{p}_2(\theta)$). 

All the above is preliminary to starting ADA-GP-MCMC. Then we proceed as described for the biased coin case, with minimal notation adjustment. Namely for a new proposal $\theta^{\star}$, if $\hat{L}_{GP}(\theta^{\star}) > \hat{L}_{GP}(\theta^{r-1})$ we decide between case 1 and 3 by drawing from $\mathrm{Bernoulli}(\hat{p}_1(\theta^\star))$. If instead $\hat{L}_{GP}(\theta^{\star}) < \hat{L}_{GP}(\theta^{r-1})$ we draw from $\mathrm{Bernoulli}(\hat{p}_2(\theta^\star))$ to decide between case 2 and 4.
Alternatively, in place of a logistic regression model we can use decision trees, but still employ the same ideas as for logistic regression. Decision trees can perform better at modeling non-linear dependencies in the data. Importantly, a decision tree does not produce an estimation of the probabilities for each case (hence, we do not obtain a direct estimation of $\hat{p}_j(\theta)$), instead a classification decision is computed, which will directly select which case to assume for the given proposal $\theta^{\star}$. 
We obtained the best results with the decision tree model. We have found beneficial to include, as a covariate in the decision tree model, the ratio between the GP-based log-likelihood  estimates at the current proposal and the previous log-likelihood estimate. 

In conclusion, we have introduced three selection methods. In Algorithm  \ref{alg:ADAGPMCMC_V2} the selection methods are denoted  $s_{13}(\cdot)$ (for selection between case 1 and 3) and $s_{24}(\cdot)$ (for selecting between case 2 and 4), to highlight the fact that different selection methods are available.   
In the supplementary material we describe how to test the fit of the GP model and the performance of the selection method.

\begin{algorithm}[H]
\scriptsize
\caption{ADA-GP-MCMC algorithm}\label{alg:ADAGPMCMC_V2}
\begin{algorithmic}[1]
	
	\Require Number of iterations $R$, a GP model fitted to the training data, model $s_{13}()$ to select between case 1 and 3, model $s_{24}()$ to select between case 2 and 4, a starting value $\theta^0$ and corresponding $\hat{L}_{u}(\theta^0)$. 
    \For{$r = 1,...,R$} 
    \State Propose $\theta^{\star} \sim g( \, \cdot |\theta^{r-1})$. \Comment{Run A-DA scheme}
    \State Sample from the predictive distribution of the GP model  to obtain independently $\ell_{GP}(\theta^{\star})$ and $\ell_{GP}(\theta^{r-1})$. Define $\hat{L}_{GP}(\theta^{\star}):=\exp(\ell_{GP}(\theta^{\star}))$ and $\hat{L}_{GP}(\theta^{r-1}):=\exp(\ell_{GP}(\theta^{r-1}))$.
    \State Compute $\alpha_{1} = \min \big( 1, \frac{\hat{L}_{GP}(\theta^{\star})}{\hat{L}_{GP}(\theta^{r-1})} \cdot \frac{g(\theta^{r-1}|\theta^{\star})}{g(\theta^{\star} |\theta^{r-1})} \cdot \frac{p(\theta^{\star}) }{p(\theta^{r-1})}  \big)$.
    \State Draw $u \sim \mathcal{U}(0,1)$. 
    \If{$u < \alpha_{1}$} \Comment{Run second stage of the A-DA scheme}
    \If{$\hat{L}_{GP}(\theta^{\star}) > \hat{L}_{GP}(\theta^{r-1})$} 
     \State Select case 1 or 3 according to the model $s_{13}(\theta^{\star})$.
     \State Run the accelerated delayed-acceptance scheme for the selected case.
    \Else
     \State Select case 2 or 4 according to the model $s_{24}(\theta^{\star})$.
     \State Run the accelerated delayed-acceptance scheme for the selected case.
    \EndIf
    \Else \Comment{Early-reject}
    \State Set $\theta^{r} = \theta^{r-1}$. 
    \EndIf
	\EndFor
\end{algorithmic}
\end{algorithm}

\section{Case studies}\label{sec:case-studies}

In Section \ref{sec:rickermodel} we consider the Ricker model, which has been used numerous times as a toy model to compare inference methods (e.g. \cite{fearnhead2012constructing}, \cite{fasiolo2016comparison} to name a few). In Section \ref{sec:dwpsde} we consider a novel double-well potential stochastic differential equation (DW-SDE) model for protein folding data, which is a considerably more complex case study.
An additional simulation study for the DW-SDE model,  diagnostics and further methodological sections are presented in the supplementary material. The code can be found at
\href{https://github.com/SamuelWiqvist/adamcmcpaper}{https://github.com/SamuelWiqvist/adamcmcpaper}.

\subsection{Ricker Model} \label{sec:rickermodel}

The Ricker model is used in ecology to describe how the size of a population varies in time and follows
\begin{align} \label{eq:ricker_model} 
\begin{cases}
y_{t+1}\sim \mathcal{P}(\phi x_{t+1}),\\ 
x_{t+1}= r x_t e^{-x_t + \epsilon_t}, \, \epsilon_t \overset{i.i.d.}{\sim}  \mathcal{N}(0,\sigma^2), 
\end{cases}
\end{align}
where $\mathcal{P}(\lambda)$ is the  Poisson distribution with mean $\lambda$.
The $\{x_t\}$ process is a latent (i.e. unobservable) Markov process and realizations from the observable process $\{y_t\}$ are conditionally independent given the latent states, since the $\epsilon_t$ are assumed independent. Even though the model is fairly simple its dynamics are highly non-linear and close to chaotic for some choice of the parameter values \citep{wood2010statistical}. The likelihood function is also both analytically and numerically intractable, if evaluated at parameters very incompatible with the observed data, see \cite{fasiolo2016comparison} for a review of inference methods applied  to the Ricker model.  

We are interested in $\theta = [\log r, \log \phi, \log \sigma]$, and we use PMCMC, MCWM, DA-GP-MCMC, and ADA-GP-MCMC for this task. That is, MCWM is not only used to provide the training data for fitting a GP regression, but also to provide inference results, in the interest of comparison between methods. PMCMC is used to provide exact Bayesian inference. 
A data set containing $T=50$ observations, generated from the model with ground-truth parameters $\theta_{true} = [3.80, 2.30, -1.20]$ at integer sampling times $t\in[1,2,...,T]$, and the starting value $x_0$ for the latent state was deterministically set to $x_0=7$ and considered as a known constant throughout.      

Results obtained with PMCMC and MCWM are produced using in total 52,000 iterations (including a burnin period of 2,000 iterations), and $N = 1,000$ particles (the standard deviation of the log-likelihood obtained from the particle filter is about $0.5$). The proposal distribution was adaptively tuned using the generalized AM algorithm (\citealp{andrieu2008tutorial}, \citealp{mueller2010exploring}), which is set to target an acceptance rate of 40\%.  For DA-GP-MCMC algorithm, we used the last 2,000 iterations of a previous MCWM run to obtain training data. Prior to fitting the GP model we removed the 10\% of the cases having the lowest log-likelihood values from the training data, as these cases badly affected the GP predictions.  After fitting the GP model, we use the ``extended'' version of the DA algorithm discussed in section \ref{sec:damcmc} and  set $\beta_{MH}=0.15$ (that is a 15\% probability to skip the delayed-acceptance step and execute a regular Metropolis-Hastings step), $N = 1,000$, and ran DA-GP-MCMC for further 50,000 iterations. The Gaussian kernels for the Metropolis random walks, $g$ and $\tilde{g}$, were kept fixed during the entire run of the DA-GP-MCMC algorithm: specifically, $\tilde{g}$ used the covariance matrix $\Sigma$ returned by the final iteration of the MCWM algorithm that was used to collect training data, and $g$ was set to a kernel having slightly larger terms in the covariance, i.e. we used a covariance $a^2\Sigma$ with $a>1$. An important modification of DA-GP-MCMC as described in Algorithm \ref{alg:DAGPMCMC}, is that in our case studies the value $\hat{L}_{u}(\theta^{r-1})$ at the denominator of $\alpha_2$ is ``refreshed''. Hence, we employ a MCWM updating procedure in the second stage. This is to obtain a reasonable high acceptance rate and to avoid problems with stickiness. The same modification was used for ADA-GP-MCMC. At the second-stage of the $r$-th iteration of ADA-GP-MCMC, a decision tree model was used to select a case from the four ones discussed in sections \ref{sec:ada-mcmc} and \ref{sec:adagpmcmc}.

\begin{figure}[ht] 
\begin{center}
	\begin{subfigure}[b]{.30\textwidth}
      \includegraphics[width=5cm,height=5cm]{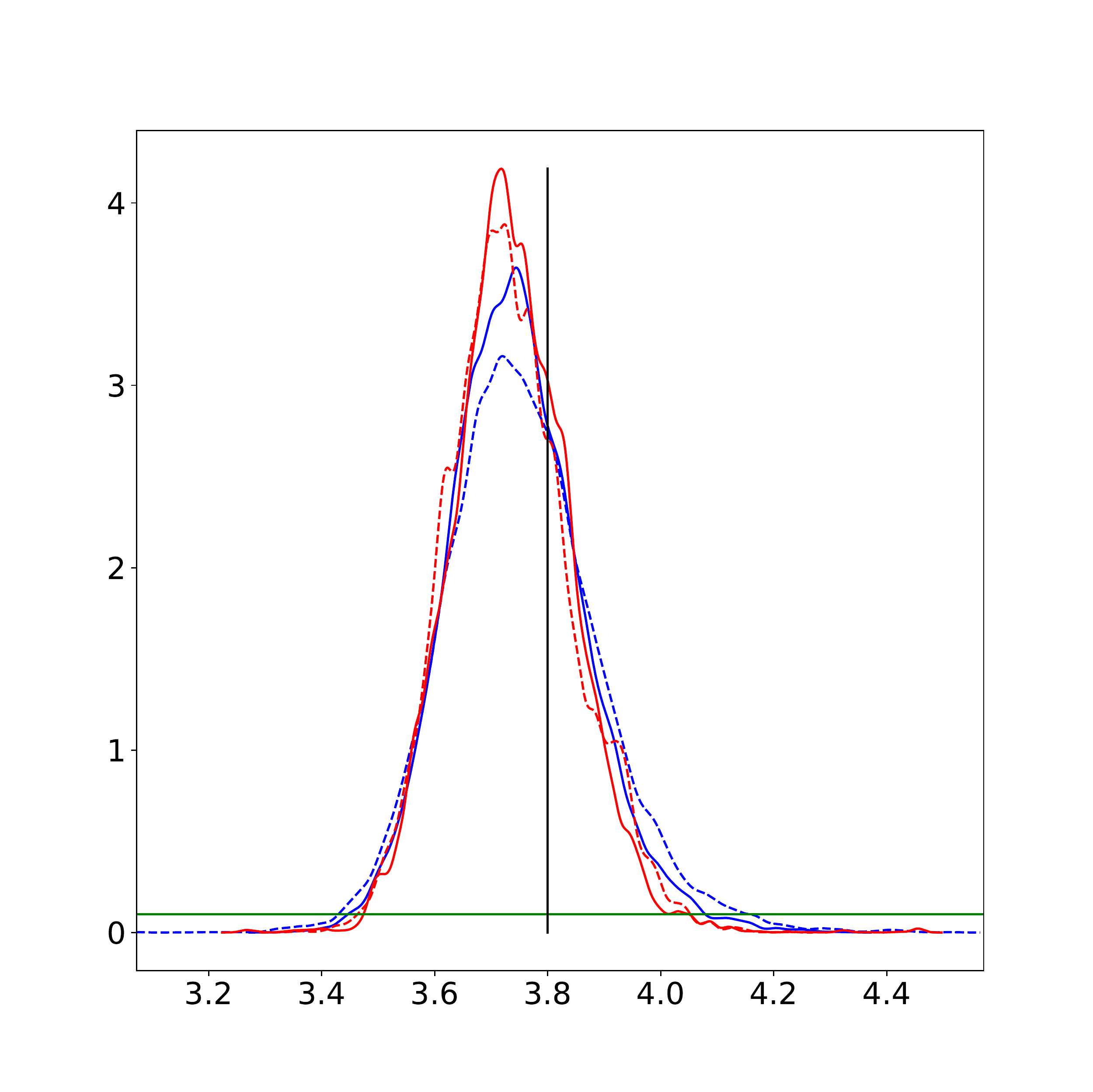}
      \caption{$\log r$.}
	\end{subfigure}
	\begin{subfigure}[b]{.30\textwidth}
	\includegraphics[width=5cm,height=5cm]{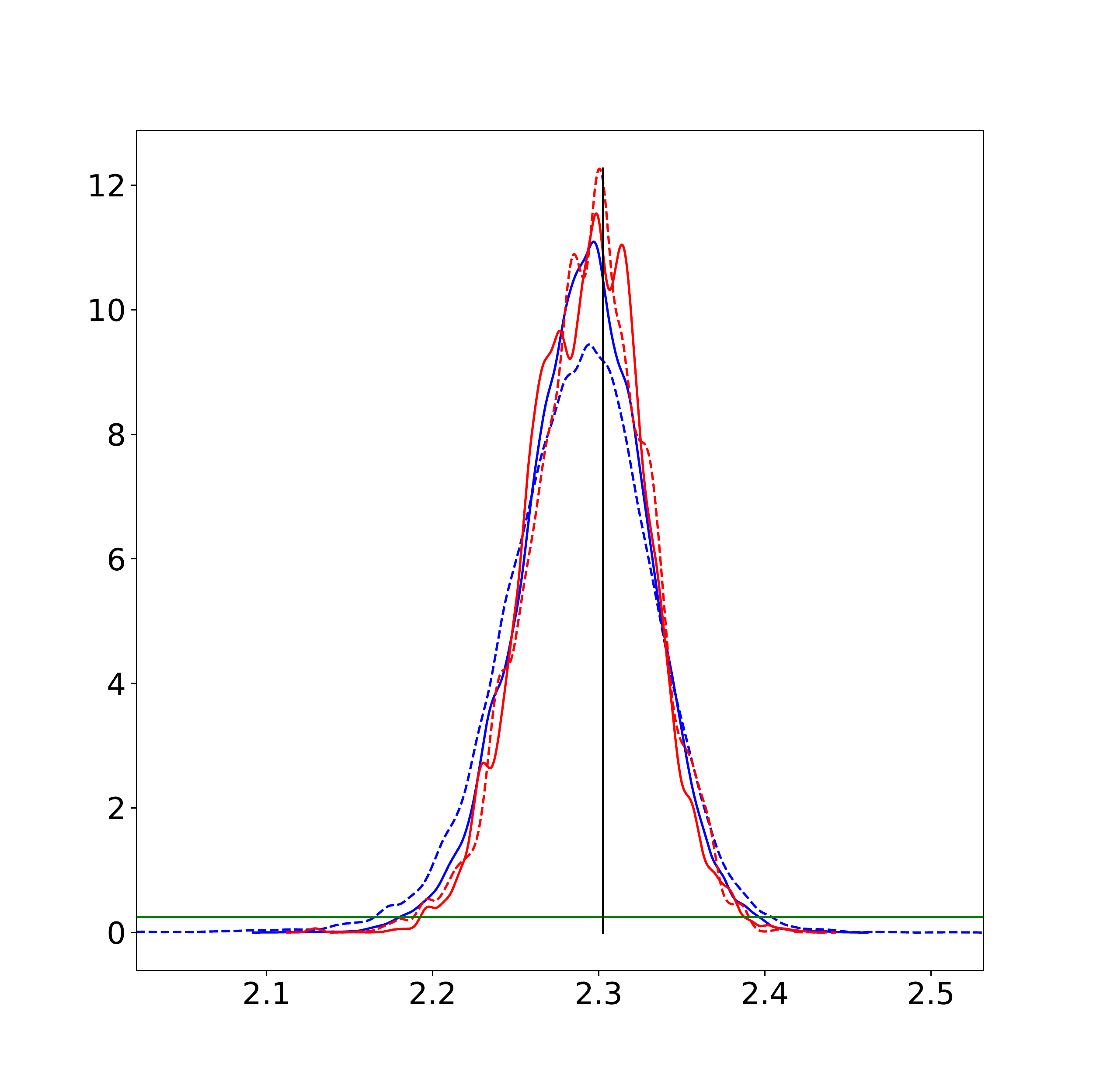}
	\caption{$\log \phi$.}
	\end{subfigure}
    \begin{subfigure}[b]{.30\textwidth}
	\includegraphics[width=5cm,height=5cm]{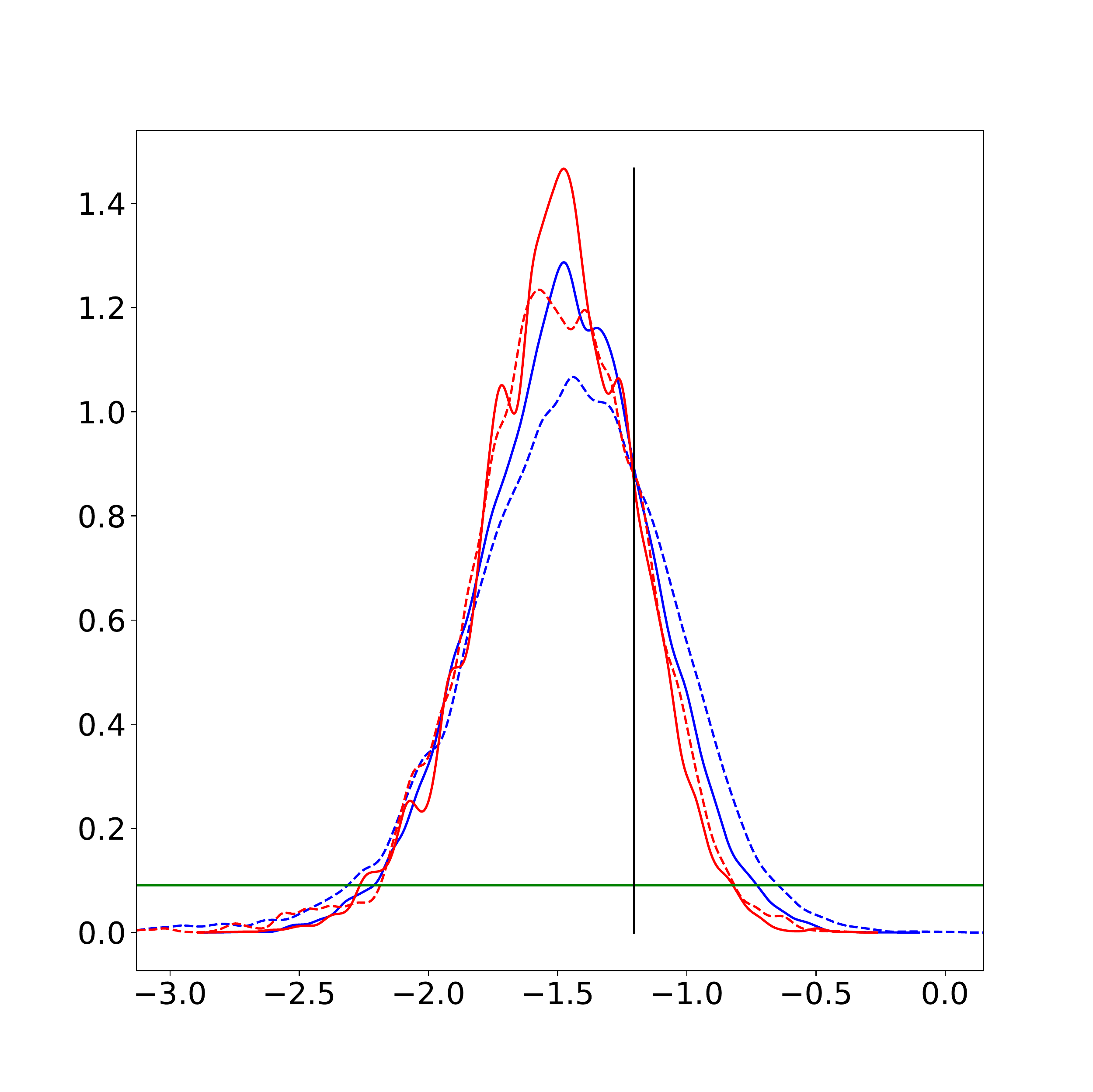}
	\caption{$\log \sigma$.}
	\end{subfigure}
\end{center}
\caption{\footnotesize{Marginal posteriors for the Ricker model: PMCMC (blue solid line), MCWM (blue dashed line), DA-GP-MCMC (red solid line), and ADA-GP-MCMC (red dashed line). Priors distributions are denoted with green lines (these look ``cut'' as we zoom on the bulk of the posterior), and the true parameter values are marked with black vertical lines.}}
\label{fig:posterior_ricker_model}
\end{figure}

Wide uniform priors were employed for all unknown parameters; $p(\log r) \sim \mathcal{U}(0,10)$, $p(\log \phi) \sim \mathcal{U}(0,4)$ and $p(\log \sigma) \sim \mathcal{U}(-10,1)$. The starting values were also deliberately set far away from the true parameter values: $\log r_{0} = 1.10$, $\log \phi_0 = 1.10$, and $\log \sigma_0 = 2.30$.      
Results are presented in Table \ref{tab:param_est_ricker} and Figure \ref{fig:posterior_ricker_model}. We can conclude that all parameters are well inferred. The results for the different algorithms are also similar. The parameter with the highest estimation uncertainty is $\sigma$, which is in not surprising since $\sigma$ is the parameter that governs the noise in the model, and this is often the hardest parameter to estimate from discretely observed measurements. Notice that results produced by ADA-GP-MCMC are essentially identical to those from DA-GP-MCMC. We find this very encouraging since the most relevant way to judge inference results from the accelerated ADA procedure is to compare those to the standard DA algorithm rather than, say, PMCMC.  

\begin{table}[ht]
    \small
   	\centering 
   	\caption{\footnotesize{Ricker model: Posterior means (2.5th and 97.5th quantiles) for PMCMC, MCWM, DA-GP-MCMC, and ADA-GP-MCMC.}}
   	\begin{tabular}{ lccccc  }
   		& True value & PMCMC & MCWM & DA-GP-MCMC & ADA-GP-MCMC    \\
   		\midrule 
   		$\log r$        &  3.80  & 3.75   [3.53, 4.00]   & 3.75  [3.51, 4.05] & 3.74   [3.54, 3.96] & 3.73   [3.54, 3.97] \\
   		$\log \phi$    	& 2.30  & 2.29   [2.21, 2.36]   & 2.29  [2.20, 2.37] & 2.29  [2.23, 2.36] & 2.29  [2.22, 2.36] \\
   		$\log \sigma$  	& -1.58 & -1.47   [-2.13, -0.85] & -1.46  [-2.3, -0.75] & -1.5  [-2.12, -0.95] & -1.51  [-2.16, -0.92] \\
   		\bottomrule
   	\end{tabular}
   	\label{tab:param_est_ricker}
\end{table}

\begin{table}[ht] 
\small
\centering 
\caption{\footnotesize{Ricker model: Efficiency of PMCMC, MCWM, DA-GP-MCMC, and ADA-GP-MCMC. Timings for (A)DA-GP-MCMC do not include the training data collection and the fitting of the GP model.} }
\begin{tabular}{ l c c   c  c c  }
 	& \makecell{Seconds per\\1000 iter.} &  \makecell{Acceptance \\ rate (\%)} & $\min$ ESS/sec & \makecell{Skip DA run \\ MH update (\%) } & \makecell{Early-\\rejections (\%) }   \\
 	\midrule
 PMCMC  		& 20.26 & 40.21 & 2.53 & NA  & NA \\
 MCWM 			& 39.83 & 39.70 & 1.26 & NA  & NA \\
 DA-GP-MCMC 	& 10.32 &  7.66 & 1.99 & 14.75  & 81.05  \\
 ADA-GP-MCMC 	& 9.46 & 7.89 & 1.75 & 15.02 & 80.49 \\
	\bottomrule
\end{tabular}
\label{tab:alg_prop_ricker_model}
\end{table}

\begin{table}[ht] 
\small
\centering 
\caption{\footnotesize{Ricker model: Estimated probabilities for the different cases and percentage of times the assumption for the different cases in the ADA-GP-MCMC algorithm holds.  }}
\begin{tabular}{ l c c   c  c }
& Case 1 & Case 2 & Case 3 & Case 4 \\
 	\midrule 
Est. probab. ($\hat{p}_1$,$\hat{p}_2$,$\hat{p}_3$,$\hat{p}_4$)  & 0.59 & 0.91 & 0.41 & 0.09 \\
Perc. assum. holds (\%) & 73.51 & 88.24 & 39.80 & 21.21 \\
	\bottomrule
\end{tabular}
\label{tab:ada_prop_ricker_model}
\end{table}

Properties of the algorithms are presented in Table \ref{tab:alg_prop_ricker_model}. Before discussing these results, we emphasize that the benefits of our accelerated procedure are to be considered when the case study has a likelihood that is computationally very challenging, and this is not the case for the present example, see instead Section \ref{sec:dwpsde}.
The ADA-GP-MCMC algorithm is the fastest algorithm, though only marginally faster than DA-GP-MCMC (4.2 times faster than MCWM and 1.09 times faster than DA-GP-MCMC), while MCWM is the slowest one. Not surprisingly, PMCMC is almost twice as fast as MCWM, and this is because PMCMC only requires one evaluation of the particle filter per iteration, while the MCWM requires two evaluations. The four algorithms are, however, essentially equally efficient, as from the $\min$ ESS/sec values.

The estimated probabilities $\hat{p}_j$ for the four different cases characterizing ADA-GP-MCMC (recall that $\hat{p}_3=1-\hat{p}_1$ and $\hat{p}_4=1-\hat{p}_2$), and the percentage for each case to hold, i.e. the probability that the selected case indeed is the correct one, are presented in Table \ref{tab:ada_prop_ricker_model}.  
We notice that the probability for the different cases vary considerably, and also that the percentages that the assumption holds vary for the different cases. We also notice that the performance of the selection algorithm is much better for case 2 than for case 4: this is due to the unbalance of the two classes, meaning that in our training data case 2 occurs more frequently than case 4, and therefore it is more difficult to estimate the latter case accurately.

\subsection{Double-well potential stochastic differential equation model for protein folding data} \label{sec:dwpsde}

We now consider a computationally intensive case study concerning statistical inference for protein folding data. The challenges for this case study are: (a) the sample size is large, data being a long time-series (about $2.5 \times 10^4$ observations), (b) the non-linear dynamics, and (c) the presence of local perturbations. ``Protein folding'' is the last and crucial step in the
transformation of genetic information, encoded in DNA, into a functional protein
molecule. Studying the time-dynamics of real protein folding dynamics results in a very high dimensional problem, which is difficult to analyze using exact Bayesian methodology. Therefore, for reasons of simplification and tractability, the dynamics of a protein are often modelled as diffusions along a single
``reaction coordinate'', that  is  one-dimensional  diffusion  models  are  considered  to  model  a  projection of  the  actual  dynamics in  high-dimensional  space (\citealp{best2011diffusion}).

\begin{figure}[ht] 
\begin{center}
	\begin{subfigure}[b]{.4\textwidth}
      \includegraphics[width=6cm,height=4cm]{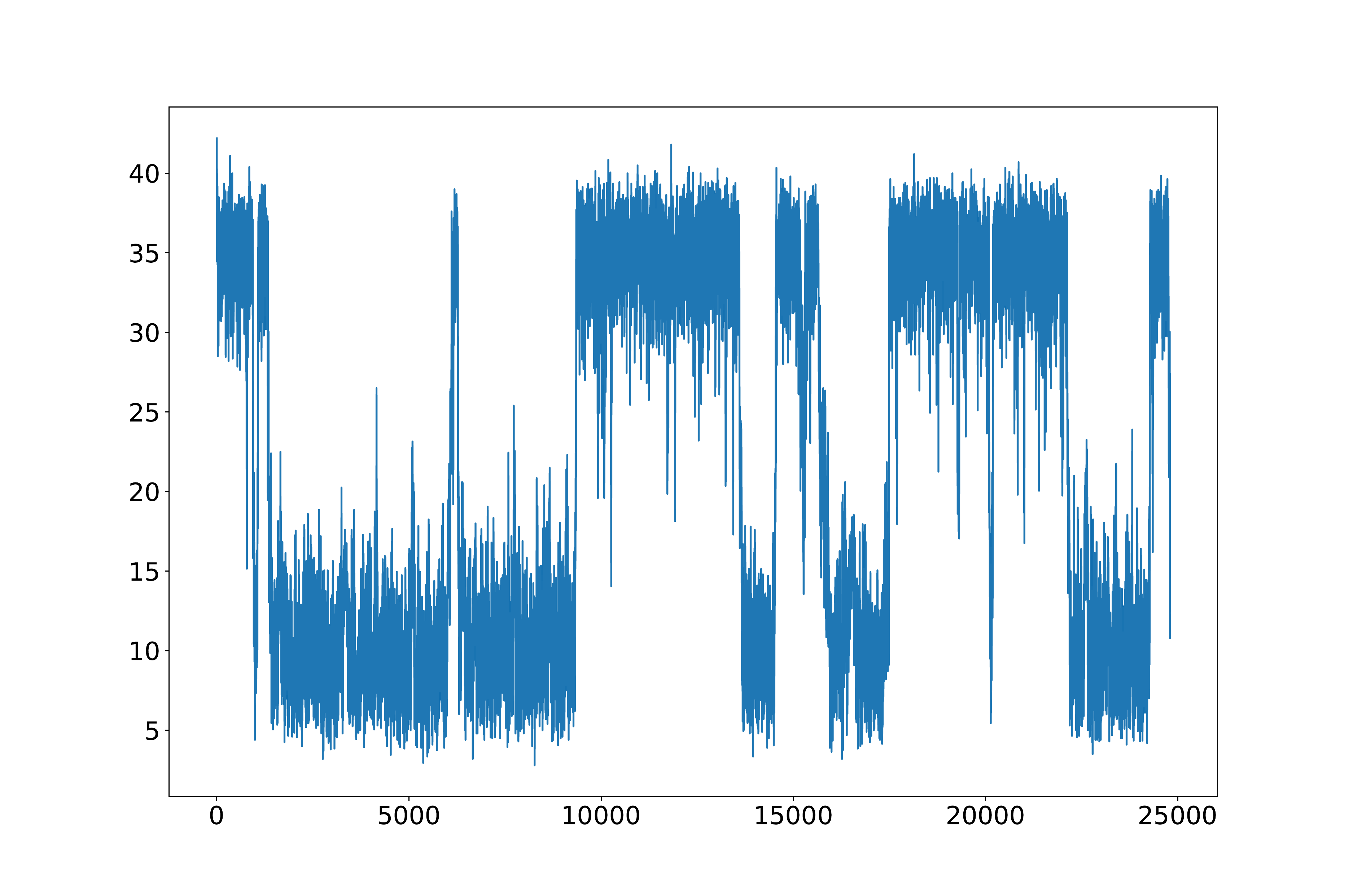}
      \caption{}
	\end{subfigure}
	\begin{subfigure}[b]{.4\textwidth}
      \includegraphics[width=6cm,height=4cm]{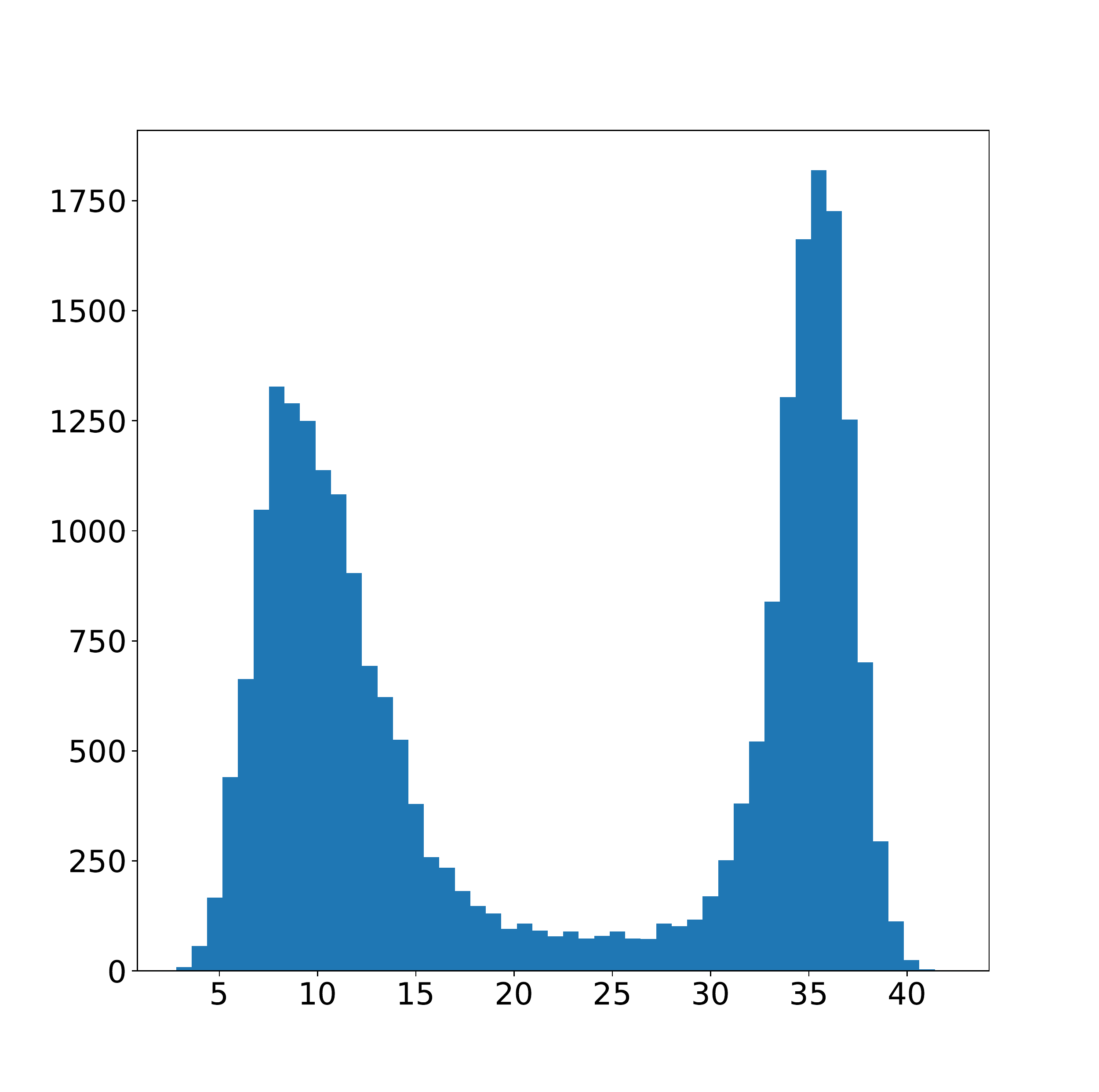}
      \caption{}
	\end{subfigure}
\end{center}
\caption{Data time course (left) and its marginal distribution (right). }
\label{fig:protein_data}
\end{figure}

The (reaction coordinate) data is in Figure \ref{fig:protein_data}. We notice that data have a marginal bimodal structure, with irregular change-points where the mean of the data shifts, and a local noisy structure. A class of models shown to be suitable for statistical modeling of protein folding (at least when these data result into a low-dimensional projection of the original data) is given by  stochastic differential equations (SDEs), see \cite{forman2014transformation} and \cite{picchini2016accelerating}. Monte Carlo inference methods are very  computationally  intensive for these models (in \citealp{picchini2016accelerating} data sub-sampling and special approximate Bayesian computation methods were used to accelerate the inference problem). We now introduce a novel double-well potential stochastic differential equation (DWP-SDE) model for protein folding data. This model is faster to simulate  than the one proposed in \cite{forman2014transformation} and \cite{picchini2016accelerating}. 
The DWP-SDE model is defined as 

\begin{align} \label{eq:structure_for_the_model}
\begin{cases}
z_t = x_t + y_t, \\
d x_t = - \nabla V(x_t)  \, dt + \sigma \, d W^{x}_t, \\
d y_t = - \kappa y_t \, dt + \sqrt{2 \kappa \gamma^2}  \, d W_t^{y}.
\end{cases}
\end{align}
Here $\{z_t\}$ is  the observable process, consisting in the sum of the solutions to the double-well potential SDE process $\{x_t\}$ and process $\{y_t\}$, the latter being unobservable and representing autocorrelated error. Here $\nabla V(\cdot)$ is the gradient of the double-well potential function $V(\cdot)$ with respect to $x_t$, further specified by six parameters introduced in \eqref{eq:DWP_V_2_1}. Finally $W_t^X$ and $W_t^Y$ are independent standard Wiener processes, that is their increments $dW_t^X$ and $dW_t^Y$ are independent, Gaussian distributed with zero mean and variance $dt$. 
We consider the following double-well potential function
\begin{equation} \label{eq:DWP_V_2_1}
V (x) = \frac{1}{2} \Big  \lvert \frac{1}{2} \lvert x - c \rvert ^{p_1} - d + g x \Big  \rvert ^{p_2} + \frac{1}{2} A x^2, 
\end{equation}
which is  based on the potential described in equation 1 in \cite{fang2015double}. The formulation in \eqref{eq:DWP_V_2_1} is fairly general, in the sense that many different potentials can be specified by varying its parameters. The parameters in  \eqref{eq:DWP_V_2_1} have the following interpretation: $c$ specifies the location for the potential (i.e. where the potential is centered); $d$ determines the spread of the potential; $A$  is an asymmetry parameter; $g$  compresses the two modes of the long term (stationary) density of process $\{X_t\}$; parameters $p_1$ and $p_2$ control the shape of the two modes (if the parameters  $p_1$ and $p_2$ are set to low values the long term probability distribution becomes more flat with less distinct modes); $\sigma$ governs the noise in the latent $\{X_t\}$ process. 
The error-model $Y_t$ is an Ornstein-Uhlenbeck  process specified by two parameters: $\kappa$ is the autocorrelation level, and $\gamma$ is the noise intensity. In principle, inference should be conducted for  $[\log \kappa, \log \gamma, \log A, \log c, \log d, \log g, \log p_1, \log p_2, \log \sigma]$. However, the model parameters $A$ and $g$ are ``stiff'', i.e. small changes in their values result in considerable changes in the output, and are therefore hard to estimate. Estimating all the parameters of the DWP-SDE model is also a complex task since a larger data set seems needed to capture the stationary distribution of the data. We will therefore consider the easier task of estimating  the parameters  $\theta = [\log \kappa, \log \gamma, \log c, \log d, \log p_1, \log p_2, \log \sigma]$. The remaining parameters, $A$ and $g$, will be fixed to arbitrary values, as discussed later. 

Simulating the $y_t$ process in \eqref{eq:structure_for_the_model} is easy since the transition density for the Ornstein-Uhlenbeck  process process is known. We have that 
\begin{align*}
y_{t+\Delta_t} | y_t = x \sim  \mathcal{N}(xe^{-\kappa\Delta_t}, \gamma^2(1-e^{-2\kappa\Delta_t}) ), 
\end{align*}
where $\Delta_t >0$.  The transition density for the $x_t$ process is not analytically known, and we use the Euler-Maruyama scheme to propagate the $x_t$ process, that is we use 
\begin{align*}
x_{t+\delta_t}  | x_t = x \approx x  - \nabla V(x) \delta_t + \sigma \epsilon_t,   
\end{align*}
where $\epsilon_t \sim \mathcal{N}(0, \delta_t^2)$, and $\delta_t >0$ is the stepsize for the Euler-Maruyama numerical integration scheme (typically $\delta_t\ll \Delta_t$).

Let us now consider the likelihood function for the $z_t$ process in \eqref{eq:structure_for_the_model}, for a set of discrete  observations $z = [z_1,\ldots, z_T]$ that we assume observed at integer sampling times $t\in[1,2,...,T]$. Corresponding (unobservable)  values for the $X_t$ process at the same sampling times are $[x_1,\ldots, x_T]$. In addition, we denote with $x$ the set $x= [x_0,x_1,...,x_T]$, which includes an arbitrary value $x_0$ from which simulations of the latent system are started. The likelihood function can be written as 

\begin{align*}
L(\theta) &= p(z|\theta) = p(z_1|\theta) \prod_{t = 2}^{T} p(z_t | z_1, \ldots, z_{t-1}, \theta), \\
&= \int p(z_1, \ldots, z_T | x_0, \ldots x_T, \theta)p(x_0, \ldots, x_T | \theta  ) d x_0 \cdots x_T, \\
&= \int p(z_1, \ldots, z_T | x_0, \ldots x_T, \theta) p(x_0) \prod_{t = 1}^{T}p(x_t  | x_{t-1}, \theta) d x_0 \cdots x_T.
\end{align*}
The last product in the integrand is due to the Markov property of $X_t$. Also, we have introduced a density $p(x_0)$, and if $x_0$ is deterministically fixed (as in our experiments) this density can be discarded. We cannot compute the likelihood function analytically (as the integral is typically intractable), but we can use sequential Monte Carlo (for example, the bootstrap filter in supplementary material) to compute an unbiased approximation $\hat{p}(z|\theta)$, which allows us to use PMCMC or MCWM for the inference. 
Furthermore, the $Z_t$ process is a transformation of the measurement noise that follows an Ornstein-Uhlenbeck process, and the density for $p(z_1, \ldots, z_T | x_0, \ldots x_T, \theta)$ is known \citep{picchini2016accelerating}. We have that 
\begin{align*}
p(z_1, \ldots, z_T | x_0, \ldots x_T, \theta) = \frac{1}{\gamma} \cdot \phi \Big( \frac{z_1 - x_1}{\gamma} \Big) \cdot \prod_{t=2}^{T} \frac{1}{\gamma\sqrt{1-e^{-2\kappa\Delta_t}}} \cdot \phi\Big(\frac{z_t - x_t - e^{-\kappa\Delta_t}(z_{t-1}-x_{t-1})}{\gamma\sqrt{1-e^{-2\kappa\Delta_t}}}\Big),
\end{align*}
where $\Delta_t = t_i - t_{i-1}$, and $\phi(\cdot)$ denotes the density function for the standard Gaussian distribution.

We now explain how an unbiased approximation to $p(z| \theta)$ is computed. To facilitate this  explanation we  introduce the following notation: let $x_{t-1}^{1:N}$ denote the set of $N$ particles we have at time $t-1$ before resampling is performed (see the bootstrap filter algorithm in the supplementary material). Let  $\tilde{x}_{t-1}^{1:N}$ denote the resampled particles that are used to propagate the latent system forward to time $t$ (using Euler-Maruyama). We approximate $p(z| \theta)$ unbiasedly with $\hat{p}(z| \theta)$ as
\begin{align*}
\hat{p}(z| \theta) = \hat{p}(z_1|\theta) \prod_{t=2}^{T} \hat{p}(z_t | z_1, \ldots, z_{t-1}, \theta) = \hat{p}(z_1|\theta) \biggl\{\prod_{t=2}^{T}\frac{1}{N}\sum_{n=1}^{N}w_{t}^{n}\biggr\},
\end{align*}
where the weights $w_{t}^{n}$ are 
\begin{align*}
w_{t}^{n} = \frac{1}{\gamma\sqrt{1-e^{-2\kappa\Delta_t}}} \cdot \phi \Big( \frac{z_t-x_t^{n}-e^{-\kappa\Delta_t}(z_{t-1}-\tilde{x}^{n}_{t-1})}{\gamma\sqrt{1-e^{-2\kappa\Delta_t}}} \Big), \qquad t\geq 2
\end{align*}
and
\begin{align*}
\hat{p}(z_1 | \theta) = \frac{1}{N}\sum_{n=1}^{N}w_{1}^{n}, \,\,\,\,\, \text{with} \,\,\,\,\, w_{1}^{n} = \frac{1}{\gamma} \cdot \phi \Big( \frac{z_1 - \tilde{x}^n_1}{\gamma} \Big).
\end{align*}

\subsubsection{Inference for protein folding data} \label{sec:real_data_old}
We now consider the data in Figure \ref{fig:protein_data}. We fixed $A$ and $g$ to $A = -0.0025$ and $g = 0$ as these parameters are difficult to identify, as already mentioned. Ideally, we should estimate $A$ and $g$, however, the data that we have access to seem to be not informative enough to infer all parameters simultaneously.
We set Gaussian priors as follows (notice these are not really motivated by biophysical considerations, we just set priors to be weakly informative): $p(\log \kappa) \sim \mathcal{N}(-0.7,0.8^2)$, $p(\log \gamma) \sim \mathcal{N}(-0.7,0.8^2)$, $p(\log c) \sim \mathcal{N}(3.34,0.173^2)$, $p(\log d) \sim \mathcal{N}(2.3,0.4^2)$, $p(\log p_1) \sim \mathcal{N}(0,0.5^2)$, $p(\log p_2) \sim \mathcal{N}(0,0.5^2)$, and $p(\log \sigma) \sim \mathcal{N}(0.69,0.5^2)$. The starting parameter values were set to $\exp(\theta_0) =[0.5, 2, 20, 15, 1.5, 1.5, 2.5]$. 

We use MCWM, DA-GP-MCMC, and ADA-GP-MCMC to estimate the unknown parameters. For each iteration of MCWM we compute 4 unbiased approximations of the likelihood function, one for each core of our computer, using $N=250$  particles for each of the 4 likelihoods. Taking the sample average of these likelihoods produces another unbiased estimate of the likelihood, but with a smaller variance than the individual ones (this is obviously true and also studied in detail in \citealp{drovandi2014pseudo}). However, given the length of the time-series, the obtained approximated likelihood is still fairly variable, and should we use PMCMC this would produce sticky chains. Therefore MCWM comes to our help for this example, as ``refreshing'' the denominator of the acceptance ratio helps escaping from sticky points, occurring when the likelihood approximation is overestimated. 

We used the following settings with MCWM: 20,000 iterations in total and a burnin of 10,000 iterations. The proposal distribution used the generalized AM algorithm, set to target an acceptance rate of 15\%. 
The training part for DA-GP-MCMC and  ADA-GP-MCMC was the output of an MCWM algorithm with the settings specified above. We fit a GP model to the output from the first  5,000 iterations of MCWM obtained after burnin. In a similar manner as for the Ricker model the two transition kernels $g$ and $\tilde{g}$ were based on the covariance matrices returned by the final iteration of the MCWM algorithm. A decision tree model, similar to the one used for the Ricker model, was used for the selection problem. Then we ran DA-GP-MCMC and ADA-GP-MCMC for 10,000  iterations, using $\beta_{MH}=0.15$. Same as with the Ricker model, a MCWM-style updating scheme was used in the second stage of both DA and ADA algorithms.

Marginal posteriors are in Figure \ref{fig:posterior_dwpsde_realdata}, and inference results are in Table \ref{tab:param_est_dwp_sde_realdata} and, same as for the Ricker model, we conclude that all three algorithms generate similar posterior inference. Algorithmic properties are in Table \ref{tab:alg_prop_dwp_sde_realdata}, and we conclude that in this case we obtain a higher speed-up compared to the Ricker model. Results are commented in detail in section \ref{sec:analyses_ada}. The estimated probabilities for the selection of the four different cases are in Table \ref{tab:prob_prop_dwp_model_realdata}, and we observe that case 4 is the least likely case. Similarly as for the Ricker model, and due to the same reasons, the performance of the selection algorithm is much better for case 2 than for case 4. 

\begin{figure}[h!] 
\begin{center}
	\begin{subfigure}[b]{.4\textwidth}
      \includegraphics[width=6cm,height=4cm]{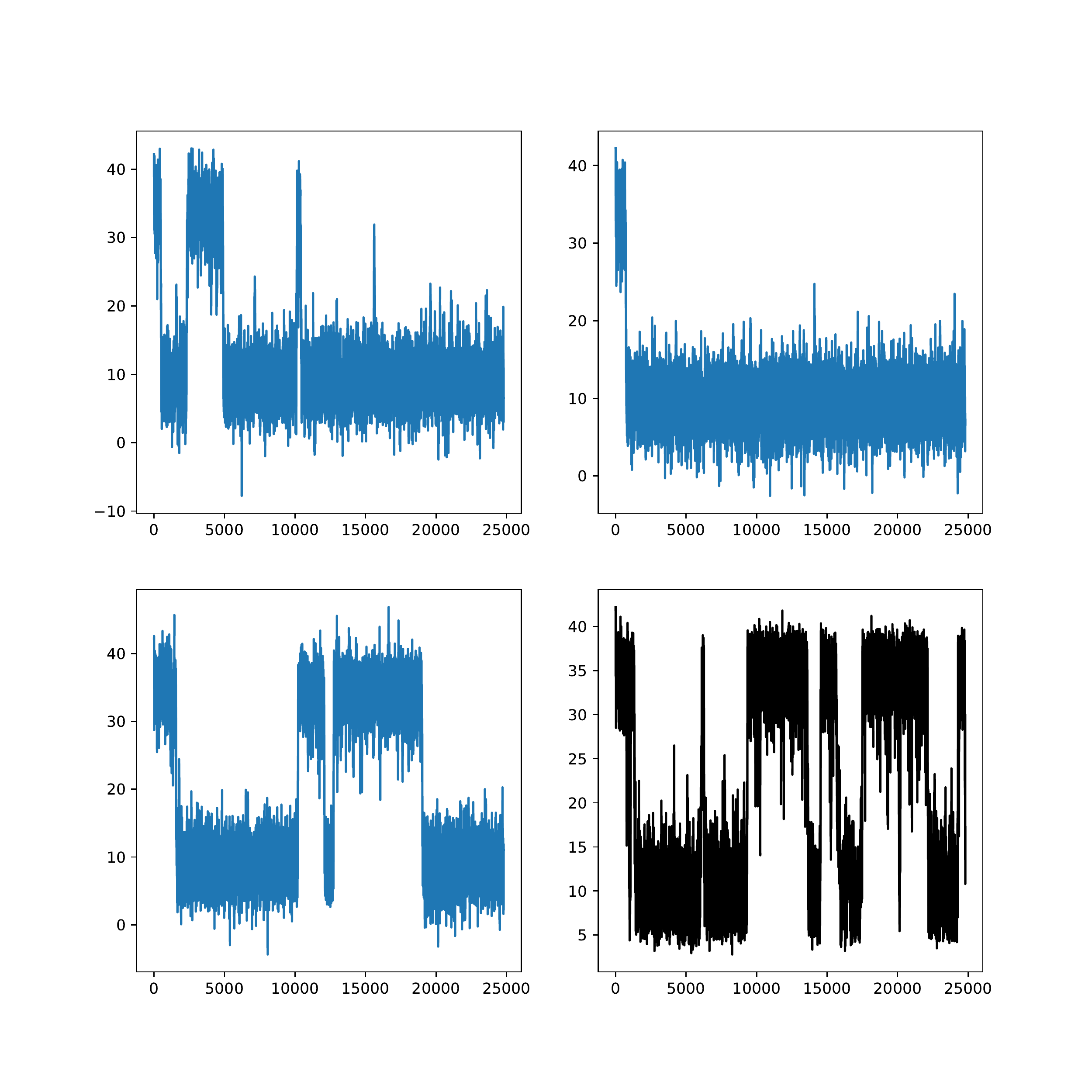}
      \caption{ }
	\end{subfigure}
	\begin{subfigure}[b]{.4\textwidth}
	\includegraphics[width=6cm,height=4cm]{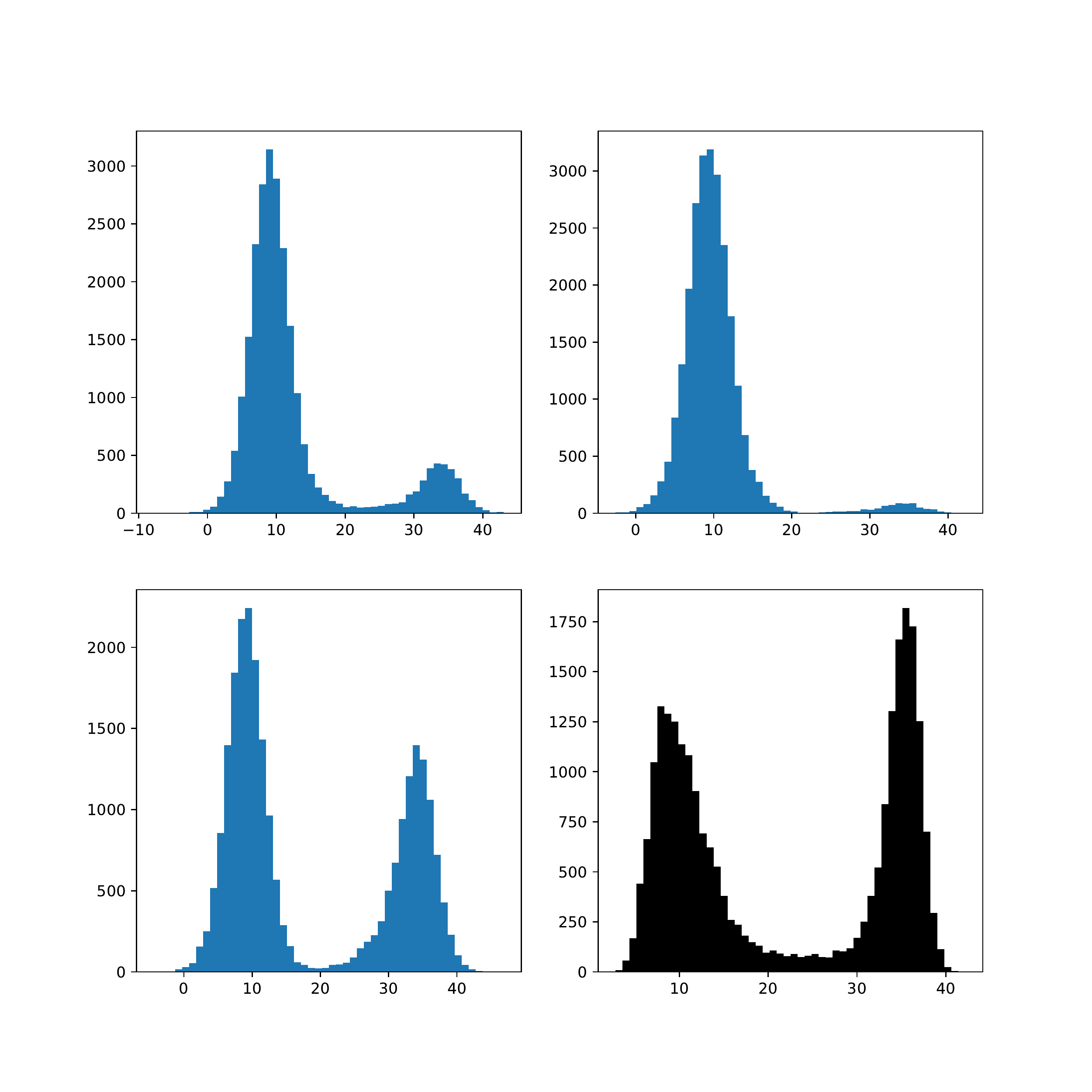}
	\caption{}
	\end{subfigure}
	\begin{subfigure}[b]{.4\textwidth}
	\includegraphics[width=6cm,height=4cm]{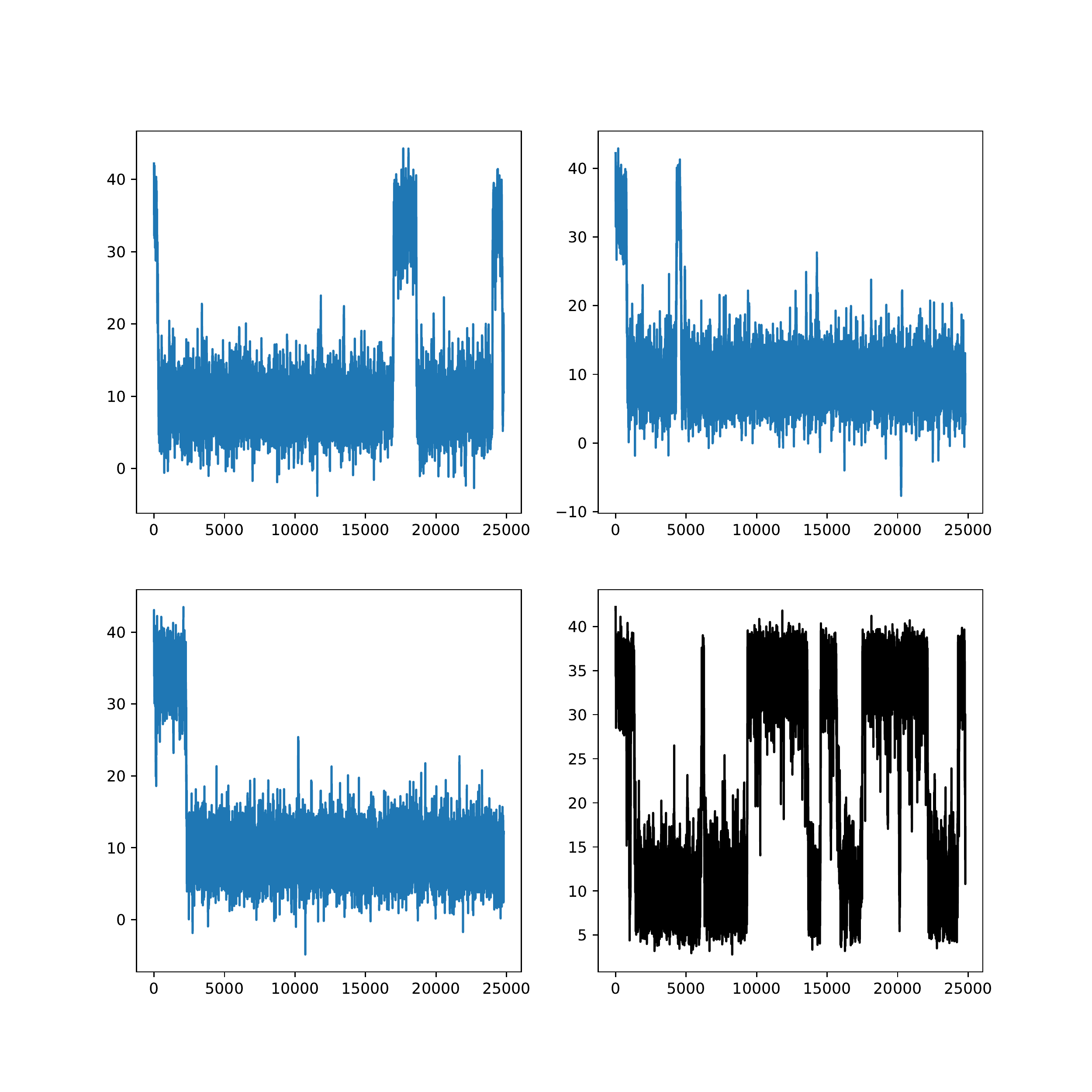}
	\caption{}
	\end{subfigure}
    \begin{subfigure}[b]{.4\textwidth}
	\includegraphics[width=6cm,height=4cm]{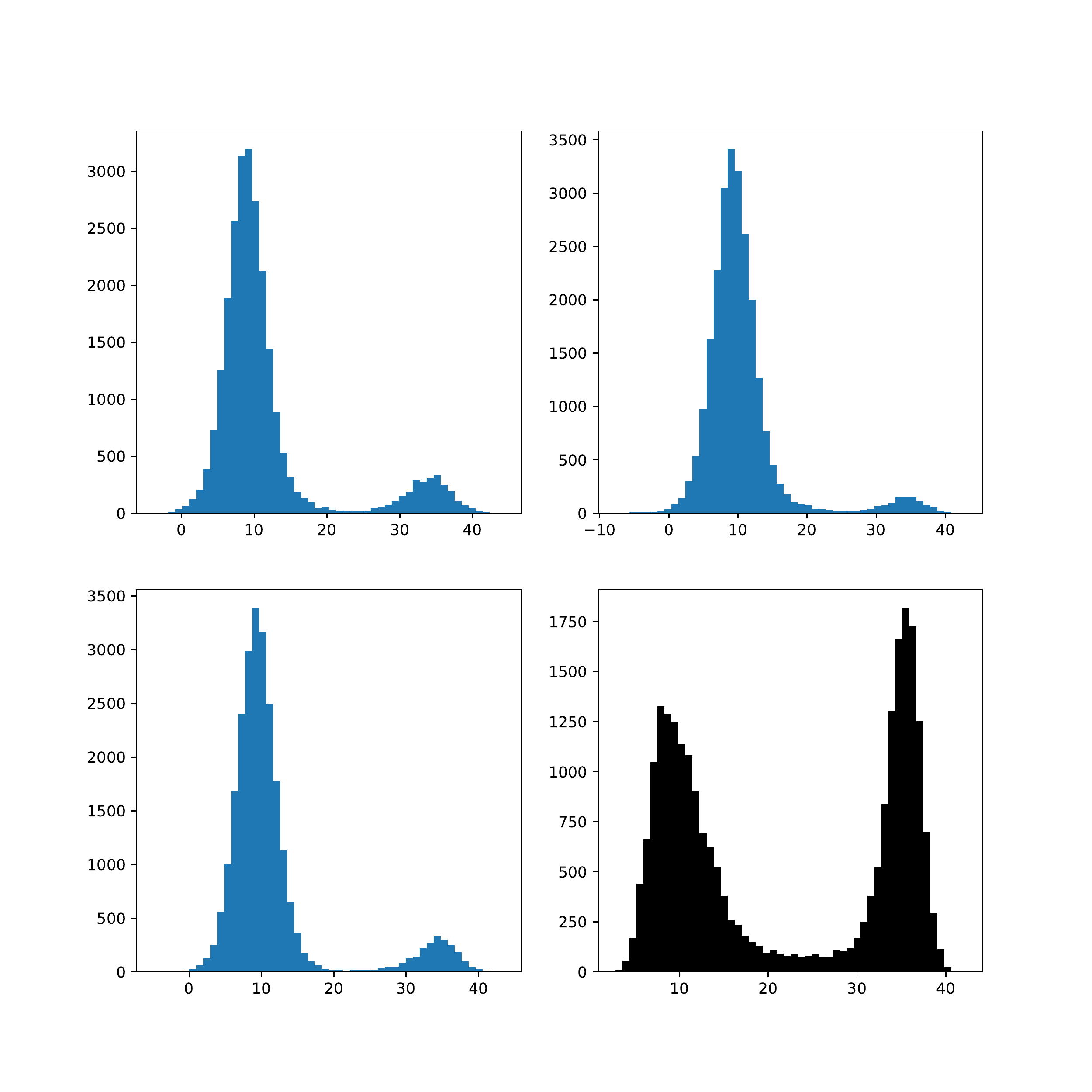}
	\caption{}
	\end{subfigure}
\end{center}
\caption{\footnotesize{Trajectories in blue are forward simulated from the DWP-SDE using draws from MCWM (a) and ADA-GP-MCMC (c). Black trajectories are data. Corresponding marginal distributions from MCWM (b) and ADA-GP-MCMC (d). }}
\label{fig:forwardsim_dwpsde_realdata}
\end{figure}

\begin{figure}[h!] 
\begin{center}
	\begin{subfigure}[b]{.25\textwidth}
      \includegraphics[width=4cm,height=4cm]{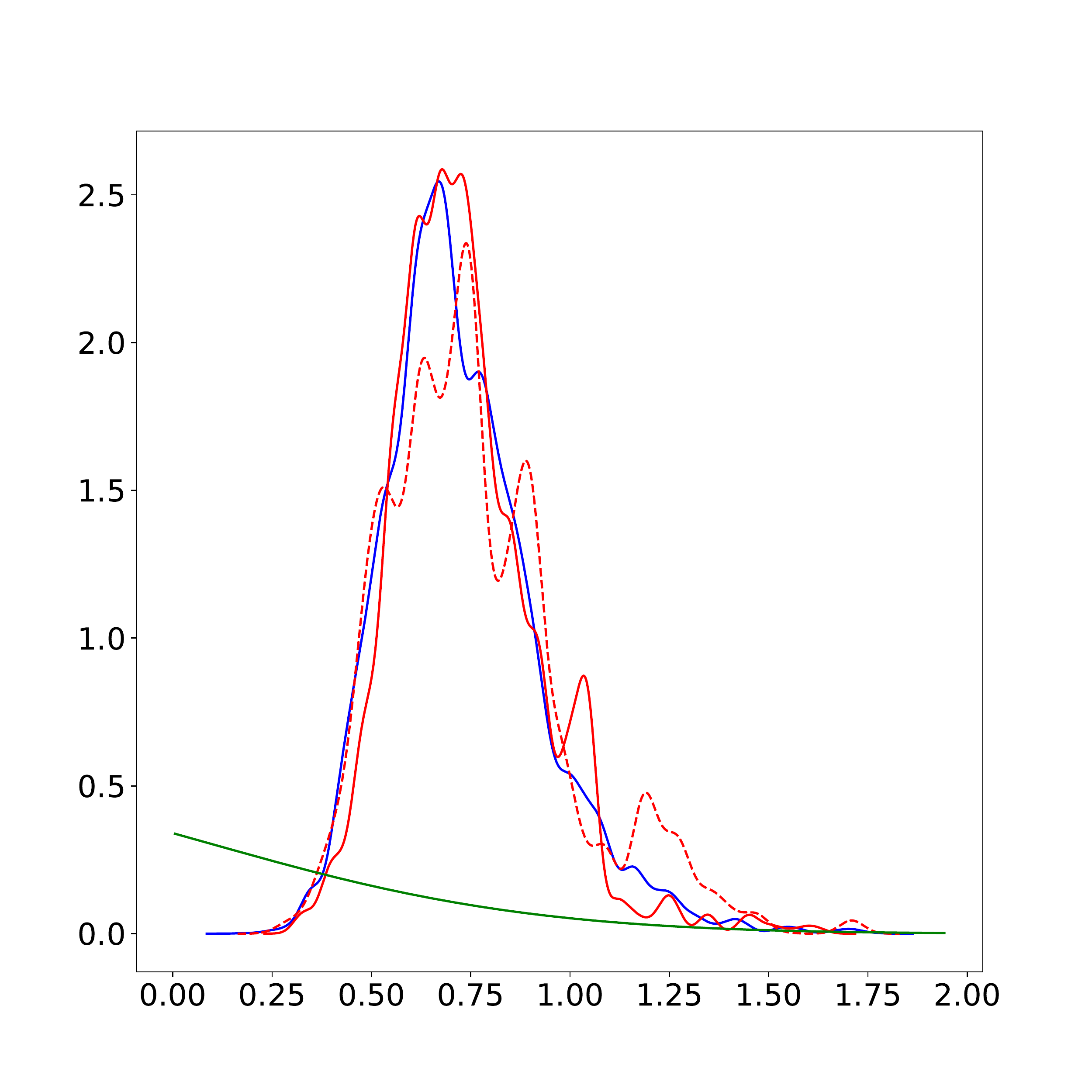}
      \caption{$\log \kappa$.}
	\end{subfigure}
	\begin{subfigure}[b]{.25\textwidth}
	\includegraphics[width=4cm,height=4cm]{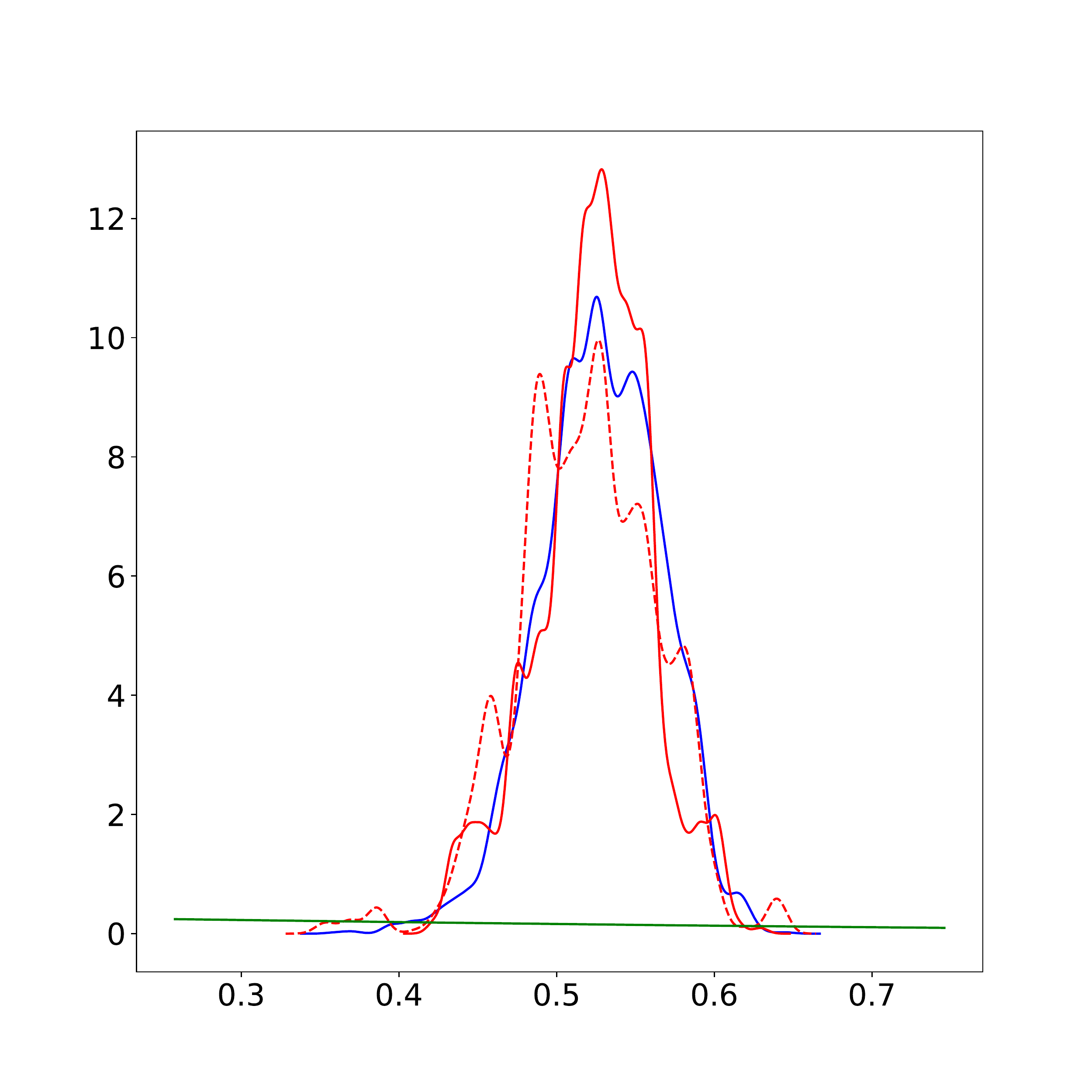}
	\caption{$\log \gamma$.}
	\end{subfigure}
    \begin{subfigure}[b]{.25\textwidth}
	\includegraphics[width=4cm,height=4cm]{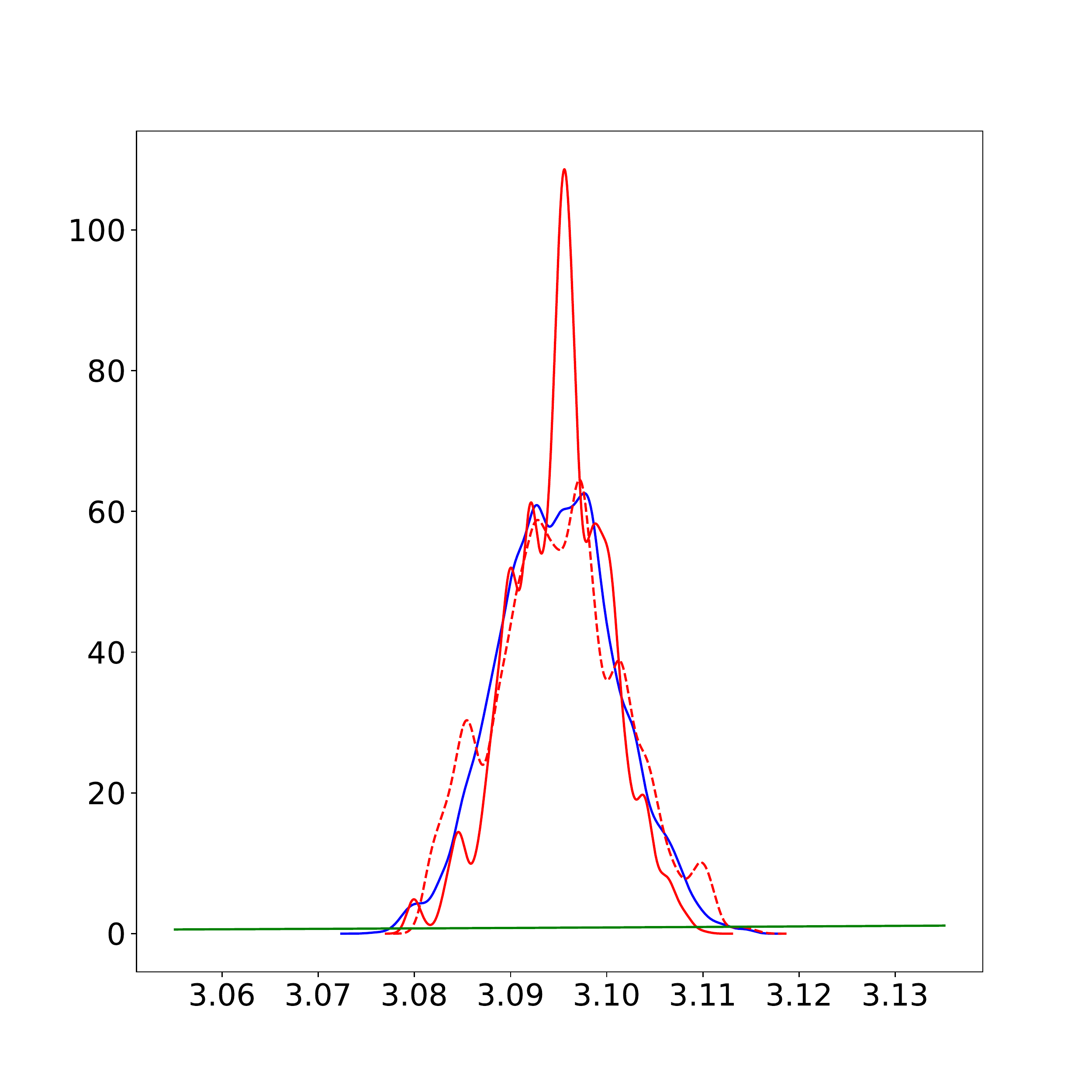}
	\caption{$\log c$.}
	\end{subfigure}
    \begin{subfigure}[b]{.25\textwidth}
      \includegraphics[width=4cm,height=4cm]{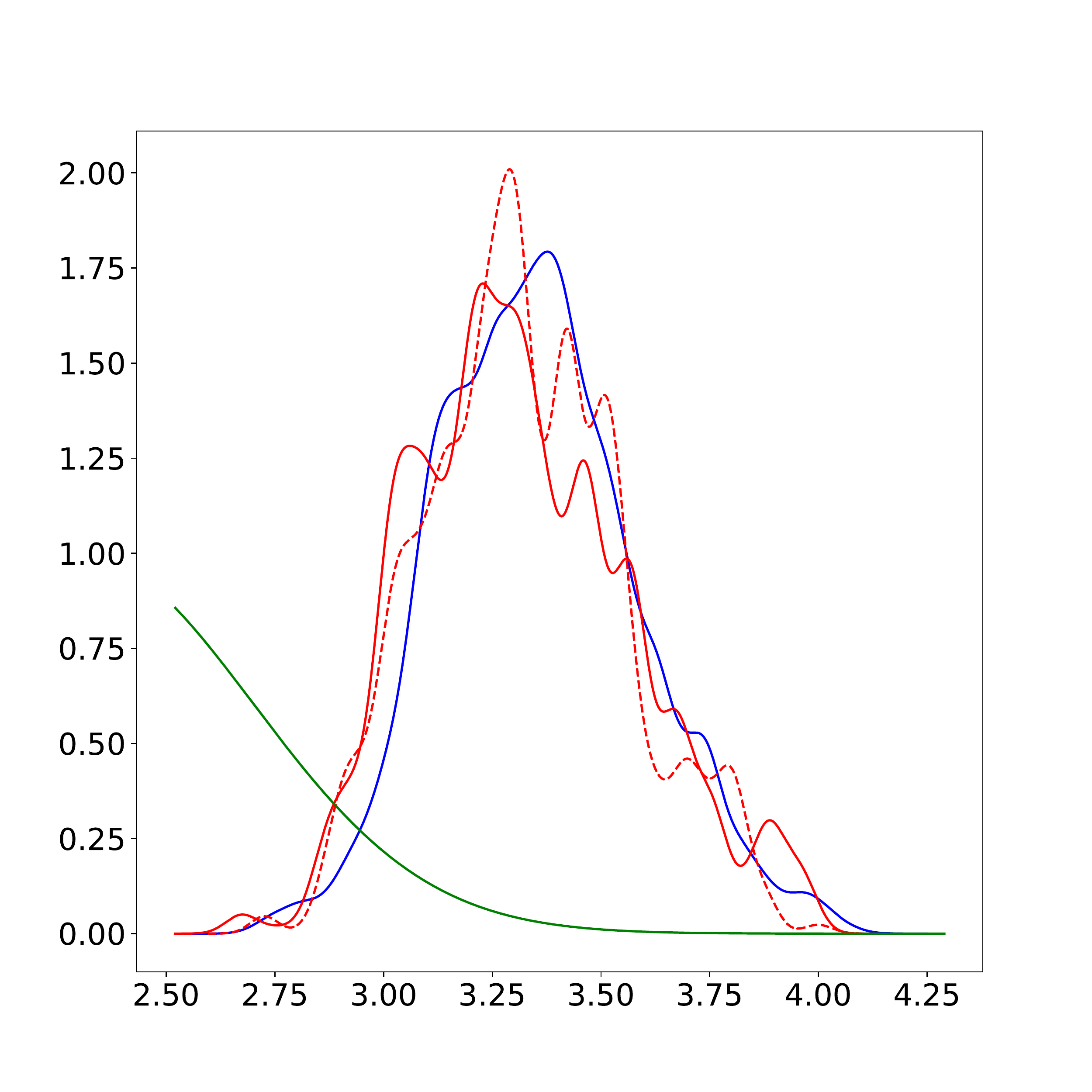}
      \caption{$\log d$.}
	\end{subfigure}
	\begin{subfigure}[b]{.25\textwidth}
	\includegraphics[width=4cm,height=4cm]{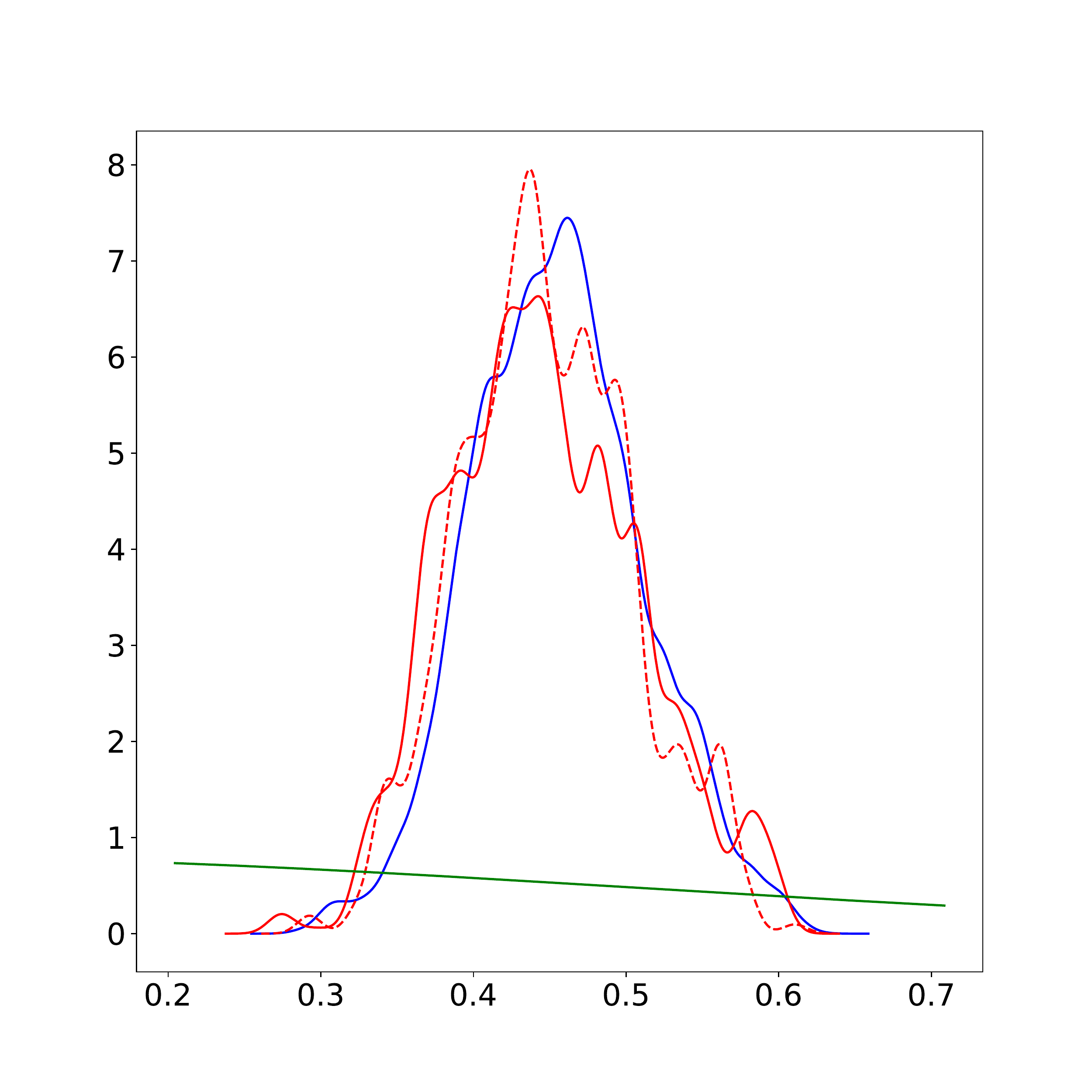}
	\caption{$\log p_1$.}
	\end{subfigure}
    \begin{subfigure}[b]{.25\textwidth}
	\includegraphics[width=4cm,height=4cm]{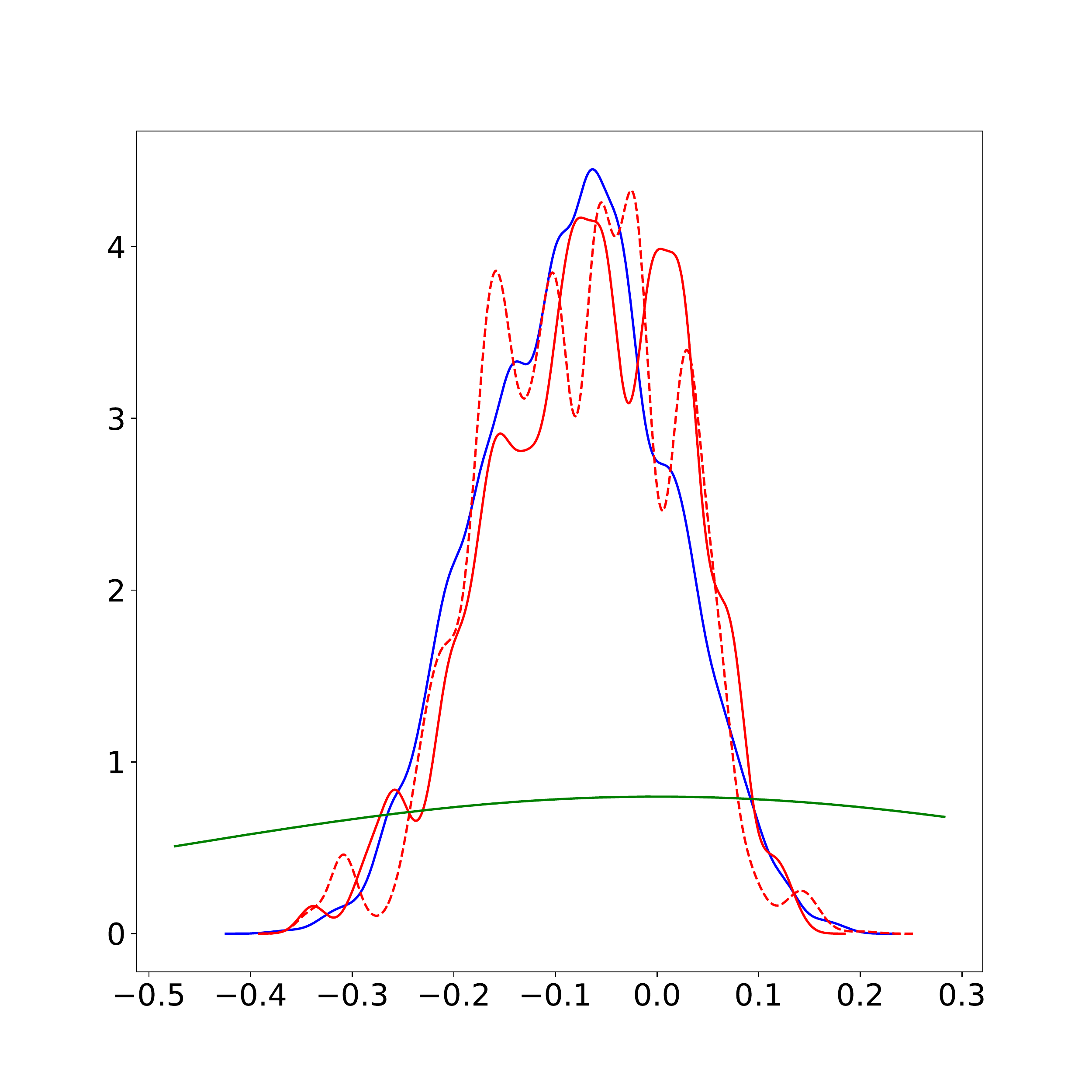}
	\caption{$\log p_2$.}
	\end{subfigure}
    \begin{subfigure}[b]{.25\textwidth}
	\includegraphics[width=4cm,height=4cm]{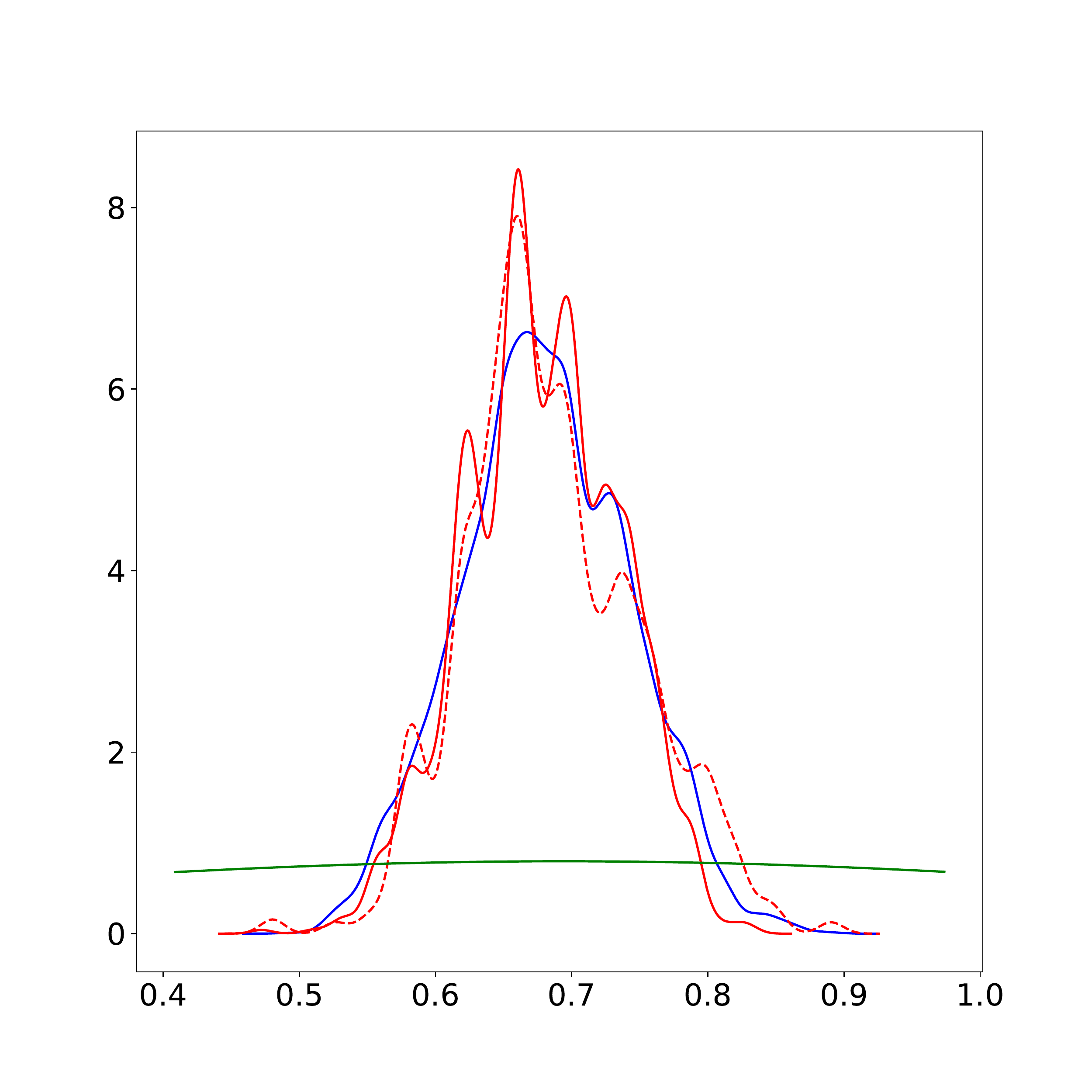}
	\caption{$\log \sigma$.}
	\end{subfigure}
\end{center}
\caption{\footnotesize{Marginal posteriors for the DWP-SDE model: MCWM (blue solid line), DA-GP-MCMC (red solid line), and ADA-GP-MCMC (red dashed line). Priors are denoted with green lines (these look ``cut'' as we zoom on the bulk of the posterior).}}
\label{fig:posterior_dwpsde_realdata}
\end{figure}

To further illustrate inference results, we randomly pick posterior draws from the high-density region of the posterior distribution, and conditionally to these we run forward simulations using the model in \eqref{eq:structure_for_the_model}. In Figure \ref{fig:forwardsim_dwpsde_realdata} we show three such forward simulations obtained from parameters sampled via MCMW and ADA-GP-MCMC. These look similar, which is not surprising since the posterior distribution that we obtain for the two methods also are similar. The number of regime switches appears underestimated compared to data. The forward simulations also show that we over-estimate the probability mass in the folded regime. This is likely due to not having estimated $A$ and $g$ from data. The values set for these two parameters are likely suboptimal, and (conditionally to those) the resulting inference for the remaining parameters is probably biased. We believe we require a longer dataset to be able to fit correctly all parameters, including $A$ and $g$.

\begin{table}[H]
\small
   	\centering 
   	\caption{\footnotesize{DWP-SDE model: Posterior means (2.5th and 97.5th quantiles) for MCWM, DA-GP-MCMC, and ADA-GP-MCMC.}}
   	\begin{tabular}{ l  c  c  c  }
   		 & MCWM & DA-GP-MCMC & ADA-GP-MCMC  \\
\midrule 
$\log \kappa$         & 0.73	 [0.42,1.19]   & 0.74   [0.45,1.12] & 0.76   [0.42,1.29] \\
$\log \gamma$    	  & 0.53    [0.45,0.59]   & 0.52  [0.44,0.6] & 0.52   [0.44,0.59] \\
$\log c$  			  & 3.09  	 [3.08,3.11]    & 3.1  [3.08,3.1] & 3.1   [3.08,3.11]\\ 
$\log d$  		  	  & 3.36  	 [2.94,3.84]     &3.32  [2.89,3.89] & 3.32  [2.91,3.81]\\ 
$\log p_1$  	  	  & 0.46  	 [0.35,0.57]    & 0.45  [0.34,0.58] & 0.45   [0.34,0.56]\\ 
$\log p_2$  	  	  & -0.08  	 [-0.26, 0.09]   & -0.07  [-0.26,0.08] & -0.08   [-0.25,0.07] \\ 
$\log \sigma$  	  	  & 0.68  	 [0.56,0.8]      & 0.68  [0.57,0.78] & 0.69  [0.57,0.82] \\                                    
   	\bottomrule
   	\end{tabular}
   	\label{tab:param_est_dwp_sde_realdata}
\end{table}

\begin{table}[H]
\small
\centering 
\caption{\footnotesize{DWP-SDE model: Efficiency of MCWM, DA-GP-MCMC, and ADA-GP-MCMC. Timings for (A)DA-GP-MCMC do not include the training data collection and the fitting of the GP model.}}
\begin{tabular}{ l c c   c  c c  }
 	& \makecell{Minutes per\\ 1000 iter.} &  \makecell{Acceptance \\ rate (\%)} & $\min$ ESS/min & \makecell{Second stage \\ direct (\%) } & \makecell{Early-\\rejections (\%) }   \\
 	\midrule
 MCWM 	& 75.88 & 18.5 & 0.39 & NA  & NA \\
 DA-GP-MCMC	& 24.81 & 3.96 & 0.69 & 15.27  & 68.95  \\
 ADA-GP-MCMC	& 15.37 & 3.34 & 0.94 & 14.52 & 69.21 \\
	\bottomrule
\end{tabular}
\label{tab:alg_prop_dwp_sde_realdata}
\end{table}

\begin{table}[H] 
\small
\centering 
\caption{\footnotesize{DWP-SDE model: Estimated probabilities for the different cases and percentage of times the assumption for the different cases in the ADA-GP-MCMC algorithm holds.}}
\begin{tabular}{ l c c   c  c }
& Case 1 & Case 2 & Case 3 & Case 4 \\
 	\midrule 
Est. prob. & 0.22 & 0.91 & 0.78 & 0.09 \\
Perc. assum. holds (\%) & 43.14 & 87.67 & 65.92 & 25.38 \\
	\bottomrule
\end{tabular}
\label{tab:prob_prop_dwp_model_realdata}
\end{table}

\subsection{Analysis of ADA-GP-MCMC} \label{sec:analyses_ada}

In the following we simplify the notation and refer to ADA-GP-MCMC and DA-GP-MCMC as ADA and DA. To analyze the runtime speed-up produced by ADA we execute multiple runs of both DA and ADA. We focus on four metrics measured over 1,000 MCMC iterations: runtimes for DA and ADA; the speed-up attained by ADA, expressed as how much faster ADA is in comparison with DA; the number of particle filter evaluations in the second stage for DA and ADA (notice, in the DA case this corresponds exactly to the number of times the second stage is reached); the reduction in the number of particle filter evaluations for ADA compared to DA. Since we are interested in analyzing the speed-up potential of ADA and not necessarily the inference results we set $\beta_{MH} = 0$, hence, we never skip the ADA/DA part of the algorithms. 

\begin{figure}[h!]
\begin{center}
	\begin{subfigure}[b]{.49\textwidth}
      \includegraphics[width=6cm,height=5cm]{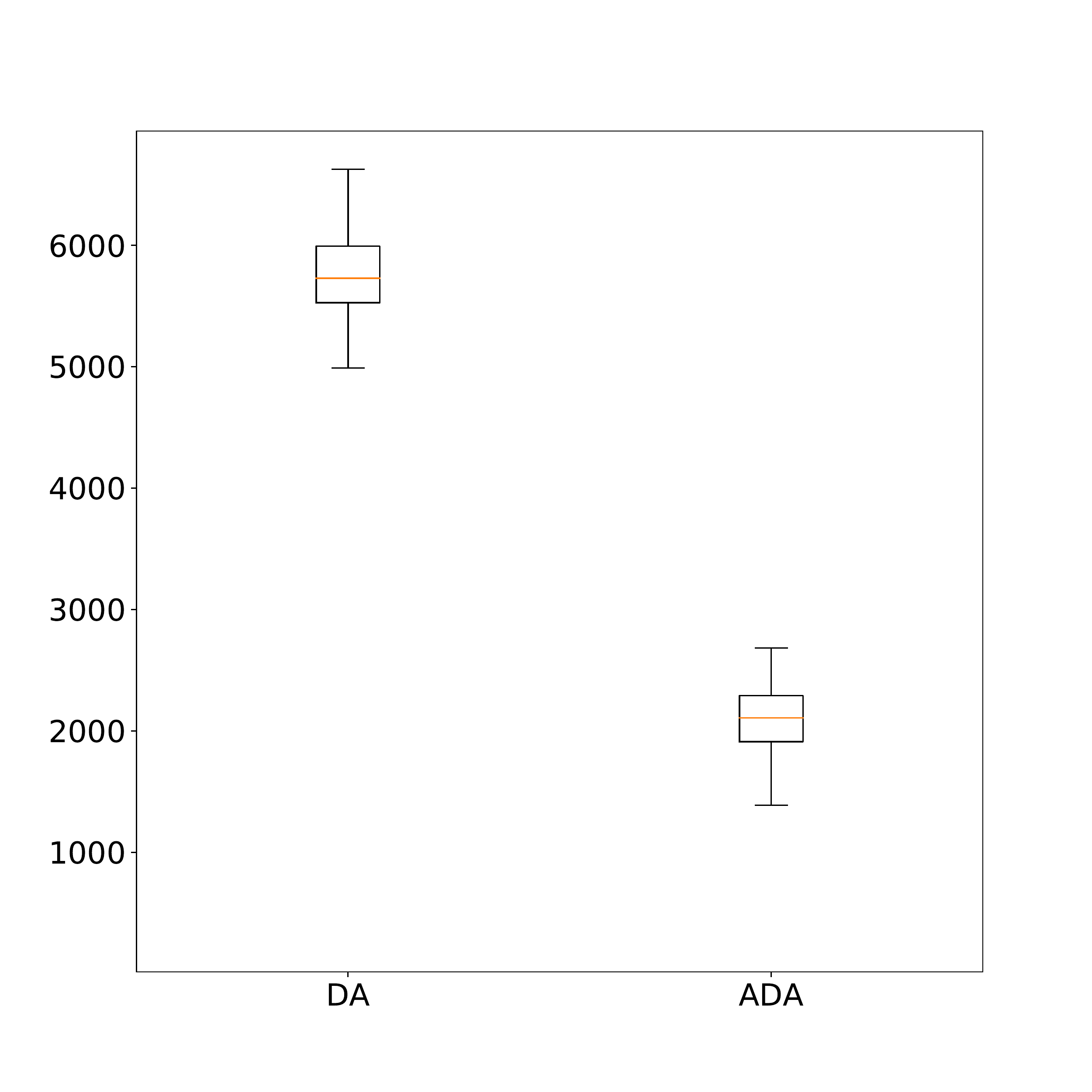}
      \caption{}
      \label{fig:speed_up_ana_dwp_real_data_runtime_diff}
	\end{subfigure}
	\begin{subfigure}[b]{.49\textwidth}
	\includegraphics[width=6cm,height=5cm]{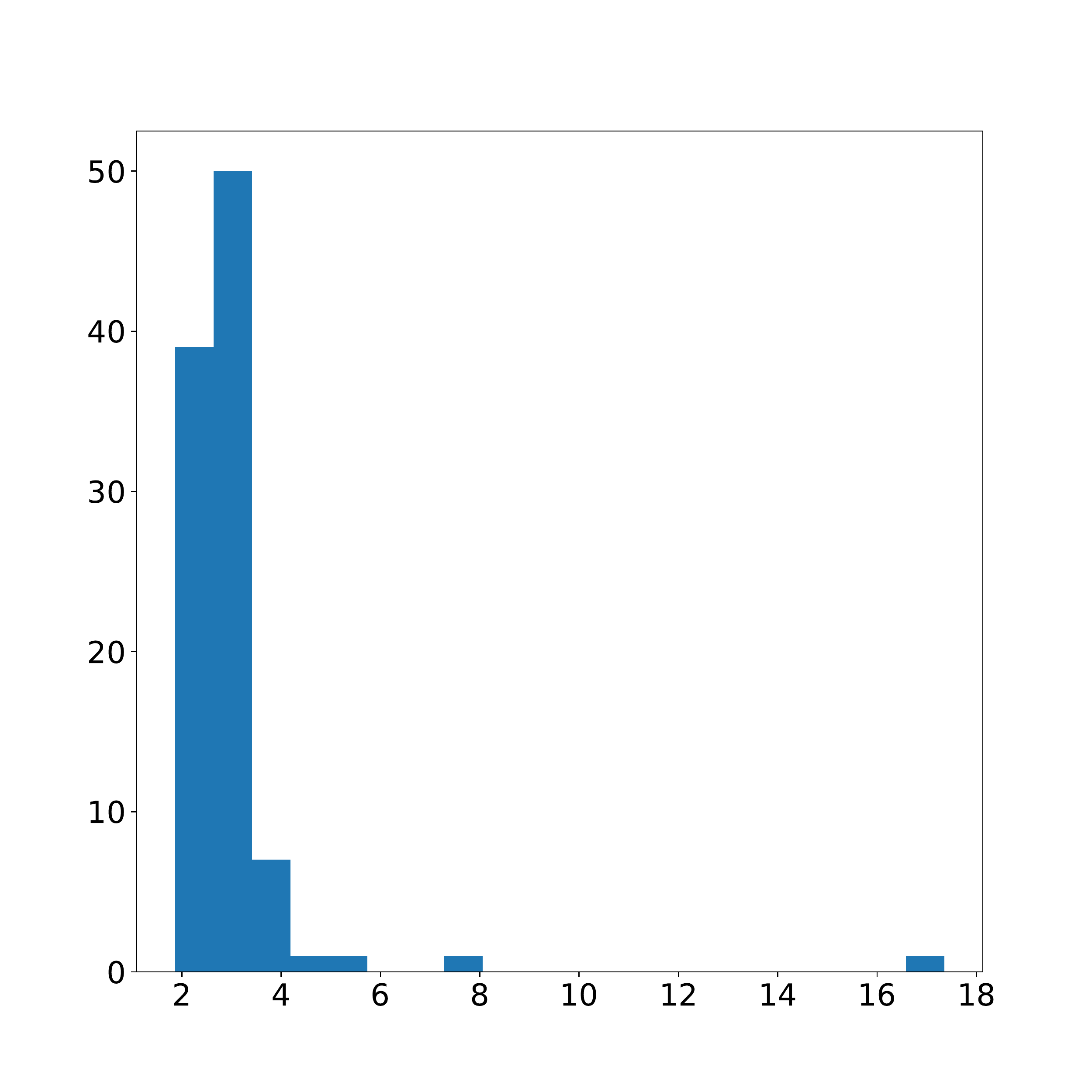}
	\caption{}
	\end{subfigure}
    \begin{subfigure}[b]{.49\textwidth}
	\includegraphics[width=6cm,height=5cm]{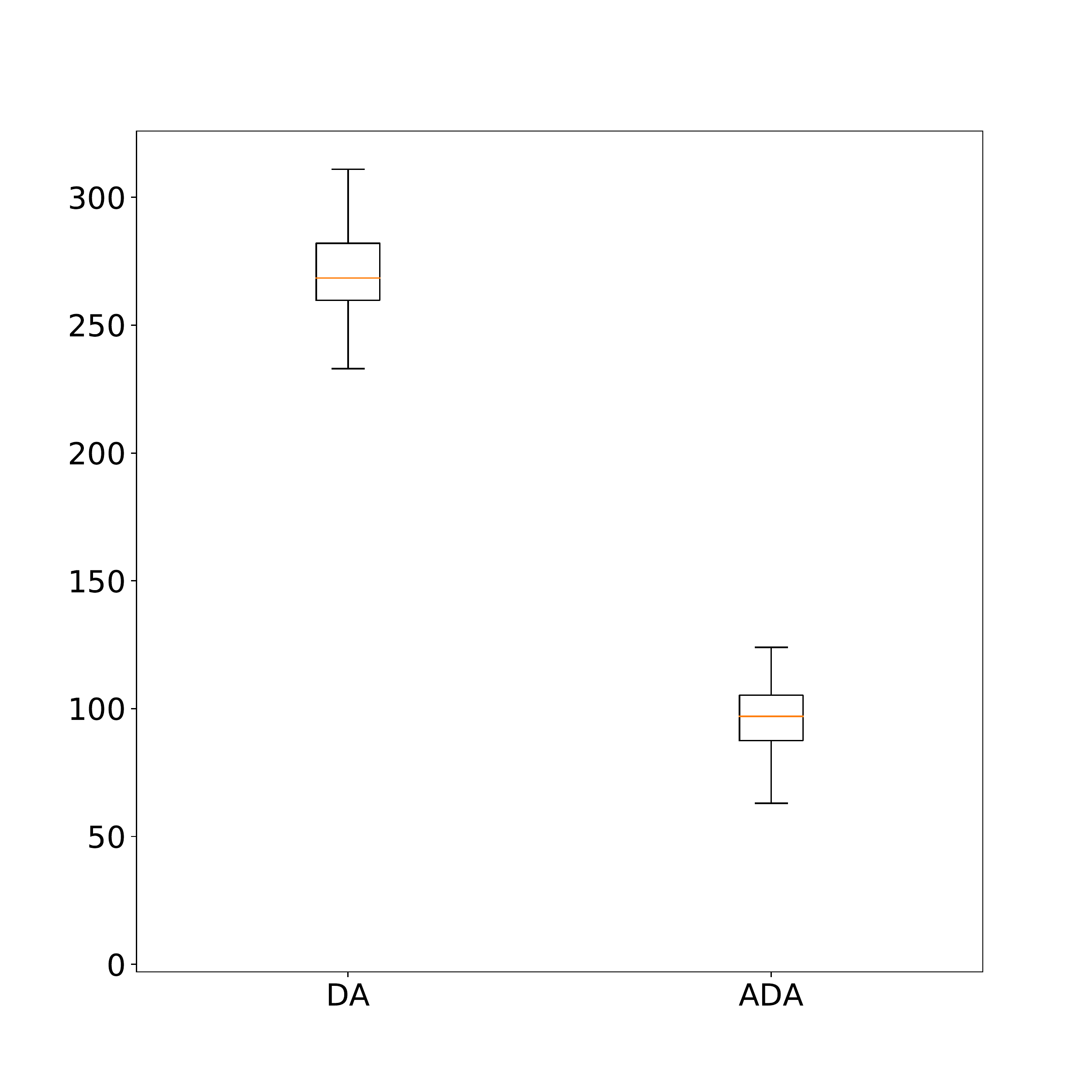}
	\caption{}
	\end{subfigure}
    \begin{subfigure}[b]{.49\textwidth}
      \includegraphics[width=6cm,height=5cm]{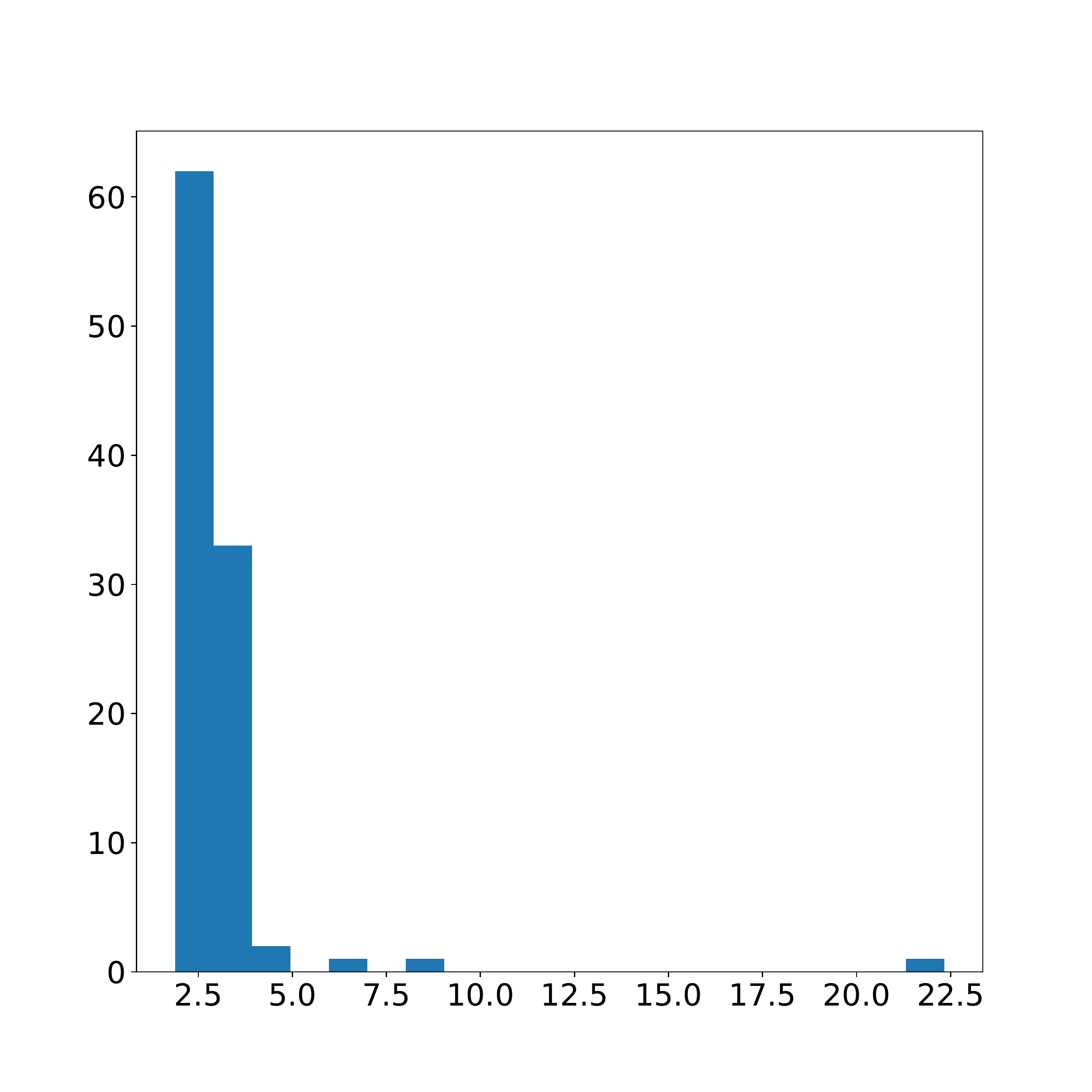}
      \caption{}
      \label{fig:speed_up_ana_dwp_real_data_reduction_factor}
	\end{subfigure}
\end{center}
\caption{\footnotesize{Speed-up analysis for the DWP model across 100 independent simulations, each for 1000 iterations, using the simulated data set. Subfigures: a) Run-times (sec) for DA (left boxplot) and ADA (right boxplot); b) Speed-up of ADA relative to DA; c) Number of particle filter evaluations in the second stage of DA (left) and ADA (right); d) Reduction in number of particle filter evaluations in the second stage for ADA compared to DA. }}
\label{fig:speed_up_ana_dwp_real_data}
\end{figure}

Furthermore, we run our analyses independently on 100 simulated datasets (see the supplementary material) using 1200 particles equally distributed across 4 cores. Results are in Figure \ref{fig:speed_up_ana_dwp_real_data}. We conclude that ADA is about 2 to 4 times as fast as DA. The number of particle filter evaluations for ADA is reduced by a factor of about 3. 

Regarding ADA, it is interesting to study how often each of the four possible cases illustrated in Section \ref{sec:ada-mcmc} are selected, and how likely it is that we run a particle filter conditionally on the selected case. Table \ref{tab:analyses_of_cases} reports our findings for the Ricker model and DWP-SDE. We notice that  proposals are not equally likely to be sent to each of the four cases, and that case 4 is the least likely case for a proposal to be sent to. It is perhaps surprising to observe the marked difference in the percentages of proposals sent to case 3 and case 4, as both cases correspond to likelihood ratios (ratio of GP likelihoods and ratio of particle filter likelihoods) that disagree in sign at the evaluated proposal.
Furthermore, we can also conclude that the probability of running the particle filter varies for the different cases. Not surprisingly, given how the cases are defined, the probability for case 2 is 1 and is 0 for case 4. We also note that the probability of running the particle filter in case 3 is much lower compared to case 1: this means that whenever case 1 is selected for proposal $\theta^{\star}$ it turns that the event $u<\tilde{L}(\theta^{r-1})/\tilde{L}(\theta^{\star})$ is less likely than the event having the opposite inequality. If instead case 3 is selected, event $u>\frac{\tilde{L}(\theta^{r-1})}{\tilde{L}(\theta^{\star})}$ is less likely than the event having the opposite inequality. 

\begin{table}[h!]
\small
\centering
\caption{\footnotesize{Percentage of proposals sent to the different cases (mean over 100 iterations of the ADA-MCMC algorithm), and probability of running the particle filter given the specific selected case (mean over 100 iterations of the ADA-MCMC algorithm). } }
\label{tab:analyses_of_cases}
\begin{tabular}{clllllllll}
& \multicolumn{4}{c}{ \makecell{Percentage of proposals \\ in each case (\%)} } &  & \multicolumn{4}{c}{\makecell{Prob. of running particle \\ filter in each case} } \\ \cline{2-5} \cline{7-10} 
& Case 1    & Case 2    & Case 3   & Case 4   &  & Case 1      & Case 2      & Case 3      & Case 4     \\ \midrule
Ricker model   & 62.59 &  12.61 & 21.31 & 3.59 & & 0.82  &  1 & 0.40  &  0 \\
\makecell{DWP-SDE \\ protein folding data.}  & 18.80 &  6.82 & 73.28 & 1.09 & & 0.98  &  1 & 0.024  &  0 \\ \bottomrule
\end{tabular}
\end{table}

\section{Summary} \label{sec:discussion}

We have provided ways to speed up MCMC sampling by introducing a novel, approximate version of the so-called ``delayed-acceptance'' MCMC introduced in \cite{christen2005markov}. More specifically, our ADA-MCMC algorithm can be used to accelerate MCMC sampling for Bayesian inference by exploiting possibilities to avoid the evaluation of a computationally expensive likelihood function. While the standard DA-MCMC only accepts proposals by evaluating the likelihood function associated to the exact posterior, instead ADA-MCMC in some specific cases can accept proposals even without the evaluation of the likelihood. Clearly, this is particularly relevant in statistical experiments where the likelihood function is not analytically available and is expensive to approximate. This is typical when unbiased approximations of the likelihood are used in pseudo-marginal algorithms for exact Bayesian inference \citep{andrieu2009pseudo}. Another situation where ADA-MCMC comes useful is when the likelihood function turns expensive due to the size of the data. 

Both DA-MCMC and ADA-MCMC depend on the construction of surrogates of the likelihood function. Unfortunately, producing a useful (i.e. informative) surrogate of the likelihood has its own cost. In fact, the construction of the surrogate model is typically the result of a ``learning'' procedure, where the output of a preliminary MCMC run (obtained using the expensive likelihood) is used to understand the relationship between simulated parameters and simulated data (e.g. using neuronal-networks as in \citealp{papamakarios2018sequential}), or between simulated log-likelihoods and parameter proposals (as in \citealp{drovandi2015accelerating}). 

ADA-MCMC samples from an approximate posterior distribution, while the original DA-MCMC algorithm is an exact algorithm. However, our case studies suggest that the approximative posterior inference returned by ADA-MCMC is close to the one obtained with DA-MCMC and Markov-chain-within-Metropolis (MCWM). This result is possibly connected with the quality of the surrogate model. If a poor surrogate model was used, the inference obtained using ADA-MCMC could be biased compared to DA-MCMC. The reason for this is that, in some cases, ADA-MCMC allows us to accept a proposal merely based on the surrogate model.

ADA-MCMC only generates an acceleration in the computations if the evaluation of the likelihood is time-consuming. If this evaluation is relatively fast, ADA-MCMC does not bring any significant gain compared to DA-MCMC (and in this case, any delayed-acceptance procedure should not be considered in the first place). An example of the latter case is shown with the Ricker model case study. However, for the DWP-SDE model, each likelihood evaluation using a particle filter requires about 2-10 seconds, depending on how many particles we use, and the benefits of using our novel approach are clear. Also, for this specific application, the expensive particle filter is invoked 2 to 5 times less often  for ADA-MCMC than for DA-MCMC.

ADA-MCMC is not limited to the Bayesian setting and can be used to sample from a generic distribution, as mentioned in Section \ref{sec:ada-mcmc}. Furthermore, when considering the inference problem in a Bayesian setting, ADA-MCMC can straightforwardly be paired with some other surrogate model than the Gaussian process regression model we employ. Hence, ADA-MCMC is a general algorithm for Monte Carlo sampling that can be exploited in multiple ways, other than the ones we have illustrated.


\begin{center}
{\large\bf SUPPLEMENTARY MATERIAL}
\end{center}

\begin{description}

\item[Further methodological tools:] Details on PMCMC, MCWM, the bootstrap filter, GP regression, diagnostics, further simulation studies and setup for the implementations. (PDF file).

\item[Julia code:] the Julia code used to run the experiments is available at: \\ \href{https://github.com/SamuelWiqvist/adamcmcpaper}{https://github.com/SamuelWiqvist/adamcmcpaper}. 

\end{description}

\bibliographystyle{abbrvnat} 
\bibliography{references}

\newpage


\section*{Supplementary material}

where $P$ is a diagonal matrix, with diagonal entries $[l_1^2,...,l_{dd}^2]$. The parameters of the covariance function are $\phi = \begin{bmatrix} \sigma & \sigma_k & l_1 & \ldots & l_{dd} \end{bmatrix}$, where $\sigma$ is the "nugget", $\sigma_k$ the output standard deviation, and the $l_i$'s the length scales for each dimension. The full set of parameters for the GP model is therefore $\eta = \begin{bmatrix} \phi & \beta \end{bmatrix}$.

We first pre-estimate $\beta$ alone using linear regression, to ease the joint optimization problem described in a moment. When pre-estimating $\beta$ we remove a small number of cases having very low likelihood values. These are considered as outliers and are removed in order to ease the optimization problem. Once this first estimate of $\beta$ is available, the GP model is fitted to $\mathscr{D}$ using maximum likelihood, i.e. both parameters in $\eta = \begin{bmatrix} \phi & \beta \end{bmatrix}$ are jointly estimated (a starting value for $\beta$ is provided by its pre-estimated value) by minimizing the GP negative log-likelihood $g(\eta)$ with respect to $\eta$, where

$$g(\eta) = -\log p(\ell(\theta)|\eta) = \\ (\ell(\theta) - m_\beta(\theta))^\top K_\phi(\Theta,\Theta)^{-1}(\ell(\theta) - m_\beta(\theta)) + \log(\det K_\phi(\Theta,\Theta)) + c. \tag{2}$$

We used $\det(A)$ to denote the determinant of the matrix $A$, while $c$ is a constant not affecting the optimization. Here $\Theta$ denotes the matrix of the $\theta$ proposals that belong to the training data $\mathscr{D}$. The matrix $K_\phi(\Theta,\Theta)$ is the covariance matrix for all the proposals in the matrix $\Theta$. The gradient for the negative log-likelihood (2) is analytically known, and we have that

$$\frac{\partial g}{\partial \beta} = -2 m_\beta(\theta)^\top K_\phi(\Theta,\Theta)^{-1}(\ell(\theta) - m_\beta(\theta)\beta),$$

and

$$\frac{\partial g}{\partial \phi_i} = -(\ell(\theta) - m_\beta(\theta)\beta)^\top K_\phi(\Theta,\Theta)^{-1} \frac{\partial K_\phi(\Theta,\Theta)}{\partial \phi_i} K_\phi(\Theta,\Theta)^{-1}(\ell(\theta) - m_\beta(\theta)\beta) + \\ \operatorname{tr}(K_\phi(\Theta,\Theta)^{-1} \frac{\partial K_\phi(\Theta,\Theta)}{\partial \phi_i}),$$

where $\operatorname{tr}(A)$ denotes the trace of the matrix $A$. We can now use a gradient-based optimization algorithm (and in practice we use the conjugate gradient algorithm) to fit the GP model to the training data $\mathscr{D}$, and we obtain $\hat{\eta} = \begin{bmatrix} \hat{\phi} & \hat{\beta} \end{bmatrix}$ by minimizing (2).

It is simple, and computationally cheap, to generate predictions from the fitted GP model since the predictive distribution is known in closed-form [Rasmussen and Williams, 2006]. This predictive distribution is just the posterior distribution of $\ell(\theta)$ given the training data $\mathscr{D}$ and conditionally to $\hat{\eta}$. That is, for a newly proposed parameter $\theta^\star$

$$\ell(\theta^\star)|\mathscr{D}, \hat{\eta} \sim \mathcal{N}(\bar{\ell}(\theta^\star), \operatorname{Var}(\ell(\theta^\star))), \tag{3}$$

where

$$\bar{\ell}(\theta^\star) = m_{\hat{\beta}}(\theta^\star) + K_{\hat{\phi}}(\theta^\star,\Theta) K_{\hat{\phi}}(\Theta,\Theta)^{-1}(\ell(\Theta) - m_{\hat{\beta}}(\Theta)), \tag{4}$$

and

$$\operatorname{Var}(\ell(\theta^\star)) = K_{\hat{\phi}}(\theta^\star,\theta^\star) - K_{\hat{\phi}}(\theta^\star,\Theta) K_{\hat{\phi}}(\Theta,\Theta)^{-1} K_{\hat{\phi}}(\Theta,\theta^\star). \tag{5}$$

Notice that the (expensive) matrix inversion $K_{\hat{\phi}}(\Theta,\Theta)^{-1}$ in (4)–(5) should only be produced once, since it does not depend on the proposed $\theta^\star$.

The predictive distributions allows for three different types of predictions:

1. *Mean prediction*: The log-likelihood function at a certain $\theta^\star$ is deterministically predicted from its mean value at $\theta^\star$, that is $\bar{\ell}(\theta^\star)$.

2. *Noisy prediction*: Predicting the log-likelihood by sampling from the predictive distribution (3) and including the "nugget" $\sigma$ in $K_\phi(\theta^\star,\theta^\star)$. Hence, $K_\phi(\theta^\star,\theta^\star)$ is computed as $K_\phi(\theta^\star,\theta^\star) = \sigma_k + \sigma$.

3. *Noise-free prediction*: Sample from (3) where the "nugget" $\sigma$ is not included, thereby obtaining a non-noisy prediction. The term $K_\phi(\theta^\star, \theta^\star)$ is therefore computed as $K_\phi(\theta^\star, \theta^\star) = \sigma_k$.

Same as in Drovandi et al. [2018], we are interested in modeling $\ell(\theta)$, and not a noisy estimate of it, and we will therefore use noise-free predictions. In conclusion, in our delayed-acceptance algorithms we will generate proxies to the unknown $\ell(\theta)$ by sampling from the GP predictive (3) using a noise-free approach.

## 2 Particle marginal methods for state-space models

The challenge of approximating the likelihood function for complex models with "intractable likelihoods" has generated a large body of literature in the past fifteen years, most notably approximate Bayesian computation (ABC, see the reviews Sisson and Fan, 2011 and Karabatsos and Leisen, 2017) and pseudo-marginal (particle) MCMC algorithms (Beaumont, 2003, Andrieu and Roberts, 2009, Andrieu et al., 2010). Pseudo-marginal algorithms in particular have found an immediate success in inference for state-space models using sequential Monte Carlo (or particle filters); reviews are Jacob [2015] and Kantas et al. [2015].

Pseudo-marginal algorithms build on the interplay between Markov chain Monte Carlo (MCMC), importance sampling and sequential Monte Carlo (SMC, or particle filters) algorithms. The crucial result is that when the likelihood $p(y|\theta)$ is not available analytically but obtaining a non-negative unbiased estimator $\hat{p}(y|\theta)$ is possible, then a Metropolis-Hastings algorithm using $\hat{p}(y|\theta)$ instead of $p(y|\theta)$ will generate a Markov chain having $p(\theta|y)$ as stationary distribution. This means that it is possible to target the exact posterior even when we deal with an (unbiased) approximation to the likelihood function, rather than the exact likelihood. Andrieu and Roberts [2009] discuss the problem by estimating unbiasedly the unavailable likelihood using $N$ draws from an importance sampler, and the remarkable result is once more that exact Bayesian sampling from $p(\theta|y)$ is possible for any finite value of $N$. Andrieu et al. [2010] frame their particle MCMC (PMCMC) approach for a large class of statistical models, including state-space models (SSM, Cappé et al., 2005). For SSM an unbiased estimator $\hat{p}(y|\theta)$ is given by particle filters using $N$ particles (here and in the following we write $\hat{p}(y|\theta) \equiv \hat{p}_N(y|\theta)$ since the resulting inference for $\theta$ is theoretically unaffected by the value of $N$). In Andrieu et al. [2010] the PMCMC algorithms PMMH (particle marginal Metropolis-Hastings) and PG (particle Gibbs) target the posterior $p(\theta, x_{1:T}|y_{1:T})$ exactly, where $y_{1:T}$ is the sequence of measurements from process $\{y_t\}$ in (6) collected at $T$ discrete times which, to simplify the notation, we assume to be the integers $\{1, 2, ..., T\}$. With $x_{1:T}$ we denote the corresponding latent (unobservable) dynamics, see (6). We employ the following notation for sequences of variables $z_{1:T} \equiv \{z_1, ..., z_T\}$. Therefore PMMH and PG solve simultaneously the parameter inference and the state filtering problem. In the next sections we clarify how these pseudo-marginal methods (PMM) and the delayed-acceptance (DA) framework interact, while emphasizing once more that in order to run a DA algorithm, including our accelerated DA method, the PMM framework is not necessary, nor is our methodology specific for dynamic models such as SSM but can be applied also to "static" models.

A SSM can be written as

$$\begin{cases} y_t \sim p(y_t|x_t; \theta_y) \\ x_t \sim p(x_t|x_s, \theta_x), \quad x_0 \sim p(x_0), \quad t_0 \leq s < t, \end{cases} \tag{6}$$

where $x_0 \equiv x_{t_0}$ is a random initial state with initial distribution $p(x_0)$, observations $y_t \in \mathbb{R}^{d_y}$ depend on a finite dimensional unknown parameter $\theta_y$, and observations are conditionally independent given the latent state $\{x_t\}_{t \geq t_0}$, with $x_t \in \mathbb{R}^{d_x}$, and $d_x, d_y \geq 1$. Here $\{x_t\}$ is a continuous Markov process equipped with a transition density $p(x_t|x_s, \cdot)$ for $s < t$ and depending on another finite dimensional unknown parameter $\theta_x$. Therefore we have that $\theta = (\theta_x, \theta_y)$ is the parameter object of our inference. In this work we consider posterior inference for $\theta$, hence our ideal target is $p(\theta|y_{1:T})$, however, instead of calling the algorithms "pseudo-marginal", we call them PMCMC, since we use particle filters

to approximate the likelihood function. But recall that we are not interested in the filtering problem for $x_{1:T}$.

Despite the existence of these powerful and flexible algorithms, computing an (unbiased) estimator of the likelihood function can be computationally time-consuming for complex models. Computationally cheap surrogate models have therefore been used to accelerate instances of the PMCMC algorithm. As an example, in Drovandi et al. [2018] a surrogate model based on Gaussian processes (GP) is used to replace the time-consuming sequential Monte Carlo estimation of the likelihood function. After an initial, computationally expensive "training phase", a GP regression model is fitted to the output of the training phase (consisting of proposed parameter values and log-likelihoods estimated via particle filters), and the estimated GP is then used as a (cheap) surrogate of the log-likelihood function, allowing for considerable computational acceleration in the MCMC sampling.

Another approach is to not entirely replace the sequential Monte Carlo estimation of the likelihood function, but only compute these estimations for parameter proposals that are not "earlyrejected" by the surrogate model. This is a delayed-acceptance (DA) approach, used for example in Golightly et al. [2015] and Sherlock et al. [2017]. As already mentioned, DA-MCMC has two important properties: the ergodicity of the chain is preserved, and the resulting Markov chain targets the true posterior distribution of $\theta$. In Golightly et al. [2015] the surrogate model is based on Langevin diffusion approximations and linear noise approximations. In Sherlock et al. [2017] the surrogate estimation of the likelihood function is computed using previous estimations via a search-tree. Hence, quite different surrogate models can be employed and still resulting in an valid DA-MCMC for exact Bayesian inference.

## 2.1 Particle Markov chain Monte Carlo

The likelihood function for the SSM (6) can be written as

$$p(y_{1:T}|\theta) = p(y_1|\theta) \prod_{t=2}^{T} p(y_t|y_{1:t-1};\theta)$$

where

$$p(y_t|y_{1:t-1};\theta) = \int p(y_t|x_t;\theta) p(x_t|y_{1:t-1};\theta) dx_t$$

and the latter integral can be efficiently approximated by drawing $N$ "particles" $x_t^n \sim p(x_t|y_{1:t-1};\cdot)$ then taking the sample average $\sum_{n=1}^{N} p(y_t|x_t^n;\cdot)/N$, and similarly to approximate $p(y_1|\cdot)$. This can be accomplished using sequential Monte Carlo methods, such as the bootstrap particle filter [Gordon et al., 1993] given in Algorithm 1. The bootstrap filter returns a non-negative unbiased estimator of the likelihood function $\hat{L}_{PF} \equiv \hat{p}(y_{1:T}|\theta)$, where the expectation of $\hat{L}_{PF}$ is taken with respect to the law underlying the generation of the random variates necessary for the implementation of Algorithm 1. For successful implementations, the number of particles $N$ should be tuned so that the standard deviation of the estimated log-likelihood $\log \hat{L}_{PF}$ does not exceed the value 2 at any given $\theta$, to assure good performance of the PMCMC [Pitt et al., 2012], and avoid problems of sticky chains [Sherlock et al., 2015].

The particle Markov chain Monte Carlo algorithm (PMCMC) in Algorithm 2 uses $\hat{L}_{PF}$ in an otherwise standard Metropolis-Hastings algorithm, to sample from the parameter posterior $p(\theta|y_{1:T})$ exactly, for any value of $N$ (Beaumont, 2003, Andrieu and Roberts, 2009), even though $N$ does have an impact on the mixing properties of the algorithm, as discussed below. An algorithm closely related to PMCMC is Monte Carlo within Metropolis (MCWM), given in Algorithm 3 and due to Beaumont [2003] (but see Medina-Aguayo et al., 2016 for theoretical properties). The only difference between MCWM and PMCMC is that in MCWM the likelihood value at the denominator of the acceptance probability is re-estimated anew as $\hat{L}_{PF}(\theta^{r-1})$. That is, at each iteration of MCWM the estimated likelihood at the denominator of $\alpha$ in step 5 of Algorithm 3 is "refreshed". Notice in particular the double estimations of the likelihood in steps 3–4. Hence, each iteration of the MCWM algorithm requires two

4

**Algorithm 1** Bootstrap particle filter

**Input:** Data $y_{1:T}$, number of particles $N$, and model parameters $\theta$.
**Output:** The likelihood estimation $\hat{L}_{PF}(\theta)$.
1: Initialize particles $\tilde{x}_0^n \sim p(x_0)$.
2: **for** $t = 1, \ldots, T$ **do**
3:    **if** $t = 1$ **then**
4:       For $n = 1, \ldots, N$, propagate particles, $x_1^n \sim p(\cdot | \tilde{x}_0^n)$.
5:       For $n = 1, \ldots, N$, evaluate importance weights, $w_1^n = p(y_1 | x_1^n)$.
6:       Estimate $\hat{p}(y_1 | \theta) = \frac{\sum_{n=1}^N w_1^n}{N}$.
7:       For $n = 1, \ldots, N$, normalize importance weights, $\tilde{w}_1^n = \frac{w_1^n}{\sum_{n=1}^N w_1^n}$.
8:    **else**
9:       Re-sample $N$ times with replacement from $(x_{t-1}^1, \ldots, x_{t-1}^N)$ with associated probabilities $(\tilde{w}_{t-1}^1, \ldots, \tilde{w}_{t-1}^N)$ to obtain a new sample $(\tilde{x}_{t-1}^1, \ldots, \tilde{x}_{t-1}^N)$.
10:      For $n = 1, \ldots, N$, propagate particles, $x_t^n \sim p(\cdot | \tilde{x}_{t-1}^n)$.
11:      For $n = 1, \ldots, N$, evaluate importance weights, $w_t^n = p(y_t | x_t^n)$.
12:      Estimate $\hat{p}(y_t | y_{1:t-1}; \theta) = \frac{\sum_{n=1}^N w_t^n}{N}$.
13:      For $n = 1, \ldots, N$, normalize importance weights, $\tilde{w}_t^n = \frac{w_t^n}{\sum_{n=1}^N w_t^n}$.
14:    **end if**
15: **end for**
16: Estimated likelihood $\hat{L}_{PF} := \hat{p}(y | \theta) = \hat{p}(y_1 | \theta) \prod_{t=2}^T \hat{p}(y_t | y_{1:t-1}; \theta)$.

estimations of the likelihood function, which is a drawback if the estimation is computationally intensive. The mathematical properties of the MCWM algorithm are less well understood than for PM-CMC. The main advantage is, however, that MCWM in many cases generates a chain that mixes better than PMCMC, even when the estimation of the likelihood function is imprecise [Medina-Aguayo et al., 2016]. With MCWM one often avoids problems of stickiness in the simulated Markov chain, a problem that the PMCMC algorithm can suffer from, in particular if the number of particles used in the particle filter is low [Sherlock et al., 2015]. In fact, this causes the estimated likelihoods to have high variability, allowing for the acceptance of the occasional over-estimated $\hat{p}(y_{1:T} | \theta)$ to end-up at the denominator of $\alpha$ in Algorithm 2, hence reducing the chance for newer proposals to be accepted. By "refreshing" the denominator at each iteration, MCWM alleviates this pathology. However, while PMCMC targets the true posterior $p(\theta | y_{1:T})$, this does not hold for MCWM. However, Medina-Aguayo et al. [2016] gives mild conditions on the particle weights such that the stationary distribution targeted by MCWM algorithm will converge to the true posterior distribution as $N \to \infty$. Simulation results show that, for finite $N$, the marginal posteriors obtained from MCWM are often wider than the true marginals implied by the PMCMC algorithm, and MCWM therefore generates a conservative estimation of the posterior distribution [Drovandi et al., 2018].

## 3 Implementation details

Unless else stated, all calculations were carried out on the LUNARC cluster available at Lund University (Sweden), where each node has access to two Intel Xeon E5-2650 v3 (2.3 Ghz, 10-core) CPUs, http://www.lunarc.lu.se. The algorithms are implemented with Julia 0.5.2 [Bezanson et al., 2017], and the code is available at https://github.com/SamuelWiqvist/adamcmcpaper.

    For the considered case studies, the parameters in $\theta$ are all positive, and for convenience we conduct inference on their natural logarithms. The prior distributions will also be set on the log-scale. The weights $w_t^n$ in the particle filter can sometimes take very large and small values, and for numerical stability these are computed on the log-scale. We also make use of standard methods such as subtracting the largest log-weight at time $t$ from the log-weights at time $t$, prior to exponentiate them [Cappé et al., 2007]. Regarding the computation of the sum of the weights, required to compute the denominator of the normalized weights $\tilde{w}_t^n$, the so-called log-sum-exp trick turns useful [Murphy,

**Algorithm 2** PMCMC algorithm

**Input:** Number of iterations $R$, starting parameters $\theta^0$, and corresponding $\hat{L}_{PF}(\theta^0)$.
**Output:** The chain $\theta^{1:R}$.
1: **for** $r = 1, \ldots, R$ **do**
2:     Propose $\theta^\star \sim g(\cdot|\theta^{r-1})$.
3:     Run Algorithm 1 to estimate $\hat{L}_{PF}(\theta^\star)$.
4:     Compute $\alpha = \min(1, \frac{\hat{L}_{PF}(\theta^\star)}{\hat{L}_{PF}^{r-1}(\theta^{r-1})} \cdot \frac{p(\theta^\star)}{p(\theta^{r-1})} \cdot \frac{g(\theta^{r-1}|\theta^\star)}{g(\theta^\star|\theta^{r-1})})$.
5:     Draw $u \sim \mathcal{U}(0,1)$.
6:     **if** $u \leq \alpha$ **then**
7:         Set $\theta^r = \theta^\star$.
8:     **else**
9:         Set $\theta^r = \theta^{r-1}$.
10:     **end if**
11: **end for**

**Algorithm 3** MCWM algorithm

**Input:** Number of iterations $R$, starting parameters $\theta^0$.
**Output:** The chain $\theta^{1:R}$.
1: **for** $r = 1, \ldots, R$ **do**
2:     Propose $\theta^\star \sim g(\cdot|\theta^{r-1})$.
3:     Run Algorithm 1 to estimate $\hat{L}_{PF}(\theta^\star)$.
4:     Run Algorithm 1 to estimate $\hat{L}_{PF}(\theta^{r-1})$.
5:     Compute $\alpha = \min(1, \frac{\hat{L}_{PF}(\theta^\star)}{\hat{L}_{PF}(\theta^{r-1})} \cdot \frac{p(\theta^\star)}{p(\theta^{r-1})} \cdot \frac{g(\theta^{r-1}|\theta^\star)}{g(\theta^\star|\theta^{r-1})})$.
6:     Draw $u \sim \mathcal{U}(0,1)$.
7:     **if** $u \leq \alpha$ **then**
8:         Set $\theta^r = \theta^\star$.
9:     **else**
10:         Set $\theta^r = \theta^{r-1}$.
11:     **end if**
12: **end for**

2012]. In Algorithm 1 particles are resampled using the stratified resampling algorithm [Kitagawa, 1996]. The execution of the bootstrap filter for the Ricker model is relatively cheap, since the model is fairly simple and the data set used is small (it only contains $T = 50$ observations). We can, therefore, easily compute exact Bayesian inference by using the PMCMC algorithm, since it is possible to run the particle filter with sufficiently many particles, so that the standard deviation of the estimated log-likelihood is less than 2.

The DWP-SDE model is a more complex case study, and the particle filter is time-consuming since the data set contains 25,000 observations. On a standard desktop computer it can therefore be computational unfeasible to run the PMCMC algorithm. We assign $N \approx 200 - 1200$ particles to separate cores of the LUNARC cluster (possibly over multiple nodes), and run independent particle filters in parallel (this can also be replicated on a multiprocessor desktop by running several independent estimations of the likelihood). A simple method, exploiting multiple particle filters running in parallel on multiple cores (or multiple CPUs), is in Drovandi [2014], and consists of averaging out likelihood approximations obtained at different cores. Since the likelihood approximations are computed on the log-scale we have to compute the average of the exponential of the log-likelihood approximation, and then take the logarithm of this average. This scheme allows us to obtain an unbiased approximation of the likelihood function with lower variance, compared to the approximation obtained from a single particle filter. The negative log-likelihood function $g$ in (2) is minimized using the function `optimize`, found in the Julia package `Optim.jl`. In particular, we used a conjugate-gradient algorithm. As a measure of efficiency of the different Markov chains produced by the different algorithms, we compute the minimal ESS/(time unit), where ESS is the effective sample size. That is, the ESS for each parameter's chain is obtained via the R-package `mcmcse`, then the minimum ESS value across

all chains is found, and this value is then divided by the run-time. Hence, min ESS/(time unit) tells us how many independent samples the algorithm is generating per time-unit, when we consider the least efficient chain.

## 4 Diagnostics for the GP model and selection methods

For diagnostic purposes of the predictive accuracy of the fitted models (GP and selection methods $s_{13}()$ and $s_{24}()$), we can split the *training* data, to obtain *testing* data. Basically, what we have denoted as $\mathcal{D}$ and $\tilde{\mathcal{D}}$, can be partitioned as $\mathcal{D} = [\mathcal{D}_1, \mathcal{D}_2]$ and $\tilde{\mathcal{D}} = [\tilde{\mathcal{D}}_1, \tilde{\mathcal{D}}_2]$. Then $\mathcal{D}_1$ (and $\tilde{\mathcal{D}}_1$) can be used to fit the GP model, while $\mathcal{D}_2$ (and $\tilde{\mathcal{D}}_2$) is the "test data", which is not used to fit the GP model, nor to fit the selection methods. Instead the test data is merely used used to evaluate the performance of the GP model and the selection methods, as typically done with predictive models. In this case by considering data that is not used to fit the GP model.

To test the fit of the GP model, we predict likelihood values from the GP for each proposal in the test data in $\mathcal{D}_2$, and compare the GP predictions to the corresponding particle filter predictions that are stored in $\mathcal{D}_2$.

Testing the performance of the selection methods is a slightly more involved process. For each proposal in the test data $\mathcal{D}_2$ we compute corresponding GP predictions, and we also compute a new set of particle filter predictions. We then use the GP predictions and check if proposal $r$ belongs to case 1 and 3, or case 2 and 4. Assume that proposal $\theta^{\star,r}$ belongs to case 1 and 3. Then run the selection method $s_{1,3}(\theta^{\star,r})$ for proposal $r$, and check which case proposal $r$ belongs to. After having determined which case proposal $r$ belongs to, according to the selection method, we check if the same case is selected using the new particle filter predictions, where we use the definition of the four cases (see Section 2.1 in the paper) to determine which case we should select, according to the new particle filter predictions. Using this method we can calculate how likely it is that the new particle filter predictions and the selection method are consistent.

## 5 DWP-SDE model: Simulation study

Here we simulate data from model DWP-SDE model, and then produce Bayesian inference for the parameters. Simulated data of length $T = 25,000$ are produced using ground-truth parameters $\theta_{\text{true}}$ set to $\exp(\theta_{\text{true}}) = [0.3, 0.9, 0.01, 28.5, 4, 0.03, 1.5, 1.8, 1.9]$. Similarly to the paper, we consider parameters $A$ and $g$ as known and fixed to $A = 0.01$ and $g = 0.03$. The other parameters are treated as unknown. The simulated data are reported in Figure 2. The parameters were set to produce data resembling data set 1 in Figure 1 which is an additional protein folding dataset.

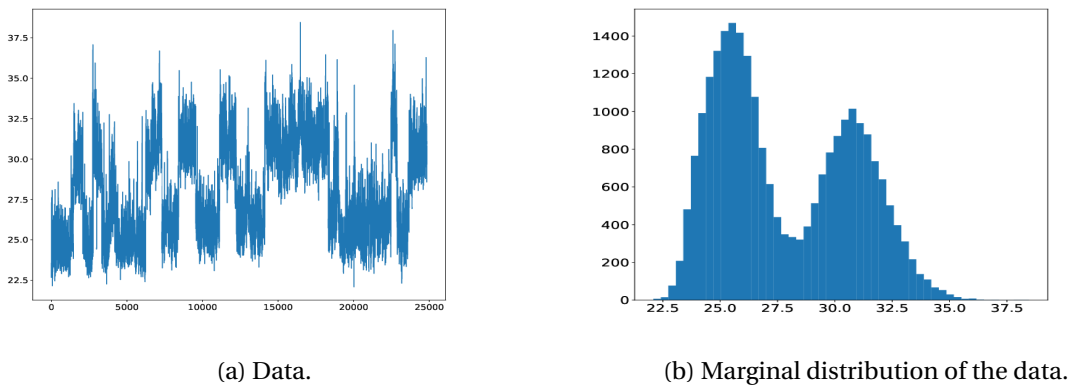

(a) Data.  (b) Marginal distribution of the data.

Figure 1: Additional protein folding dataset.

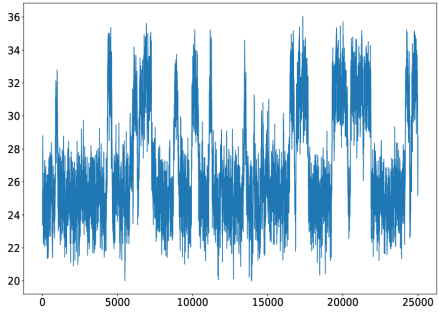
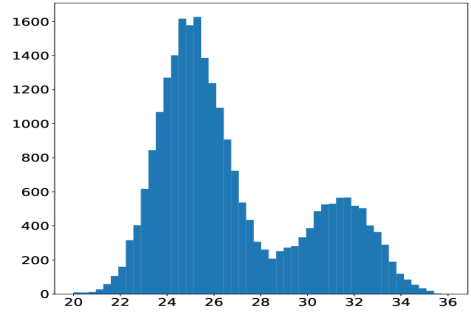

(a) Simulated data.   (b) Marginal distribution of the simulated data.

Figure 2: Data generated from the DWP-SDE model.

Table 1: Posterior means (2.5th and 97.5th quantiles) for MCWM, DA-GP-MCMC, and ADA-GP-MCMC.

|  | True value | MCWM | DA-GP-MCMC | ADA-GP-MCMC |
|---|---|---|---|---|
| $\log \kappa$ | -1.2 | -1.2 [-1.43,-0.97] | -1.21 [-1.43,-0.99] | -1.21 [-1.44,-0.95] |
| $\log \gamma$ | -0.11 | -0.11 [-0.25,0.02] | -0.1 [-0.24,0.01] | -0.11 [-0.25,0.03] |
| $\log c$ | 3.35 | 3.35 [3.34,3.36] | 3.35 [3.34,3.36] | 3.35 [3.34,3.36] |
| $\log d$ | 1.39 | 1.44 [1.17,1.81] | 1.41 [1.18,1.69] | 1.43 [1.16,1.85] |
| $\log p_1$ | 0.41 | 0.43 [0.29,0.63] | 0.42 [0.29,0.57] | 0.43 [0.28,0.65] |
| $\log p_2$ | 0.59 | 0.51 [0.1, 0.82] | 0.54 [0.18,0.88] | 0.52 [0.02,0.92] |
| $\log \sigma$ | 0.64 | 0.65 [0.48,0.81] | 0.64 [0.49,0.77] | 0.65 [0.48,0.79] |

We set Gaussian priors: $p(\log \kappa) \sim \mathcal{N}(-0.7, 0.5^2)$, $p(\log \gamma) \sim \mathcal{N}(-0.7, 0.5^2)$, $p(\log c) \sim \mathcal{N}(3.34, 0.173^2)$, $p(\log d) \sim \mathcal{N}(1.15, 0.2^2)$, $p(\log p_1) \sim \mathcal{N}(0.69, 0.5^2)$, $p(\log p_2) \sim \mathcal{N}(0, 0.5^2)$, and $p(\log \sigma) \sim \mathcal{N}(0, 0.5^2)$. The starting parameter values were set far from the ground truth, as $\exp(\theta_0) = [2, 2, 30, 10, 2, 2, 2]$. The algorithm settings for MCWM, DA-GP-MCMC, and ADA-GP-MCMC are the same as in Section 5.2.1 in the paper.

Notice, before fitting the GP model, we removed the 1% of the observations having the lowest log-likelihood from the training data, in order to obtain a more robust prediction. Marginal posteriors from the two methods are in Figure 3. These results are very similar, given the diffuse priors (also, see the posterior quantile intervals in Table 1). All parameters are well inferred and we manage to capture the true parameter values. From Table 2 we see that the speed-up for ADA-GP-MCMC is larger in this case, compared to the Ricker model, since ADA-GP-MCMC is 4.6 times faster than MCWM, and 1.5 times faster than DA-GP-MCMC. The algorithm efficiency measure min ESS/minute in Table 2 indicates that ADA-GP-MCMC is somewhat more efficient than both MCWM and DA-GP-MCMC. In Table 3 we present the estimated probabilities for the four different cases, and we can conclude that

Table 2: Algorithm properties for the the MCWM, DA-GP-MCMC, and ADA-GP-MCMC algorithm.

|  | Minutes per 1000 iter. | Acceptance rate (%) | min ESS/min | Second stage direct (%) | Early-rejections (%) |
|---|---|---|---|---|---|
| MCWM | 60.29 | 19.84 | 0.57 | NA | NA |
| DA-GP-MCMC | 20.67 | 3.80 | 0.67 | 15.01 | 67.03 |
| ADA-GP-MCMC | 13.39 | 4.02 | 1.04 | 15.10 | 67.01 |

8

Table 3: Estimated probabilities for the different cases and percentage of times the assumption for the different cases in the ADA-GP-MCMC algorithm holds.

|  | Case 1 | Case 2 | Case 3 | Case 4 |
|---|---|---|---|---|
| Est. prob. $(\hat{p}_1,\hat{p}_2,\hat{p}_3,\hat{p}_4)$ | 0.22 | 0.90 | 0.78 | 0.09 |
| Perc. assum. holds | 38.05 | 84.40 | 69.17 | 27.90 |

case 4 is the least likely case. We also notice that the performance of the selection algorithm is much better for case 2 than for case 4: this is due to the unbalance of the two classes, meaning that in our training data case 2 occurrs more frequently than case 4, and therefore it is more difficult to estimate the latter case accurately.

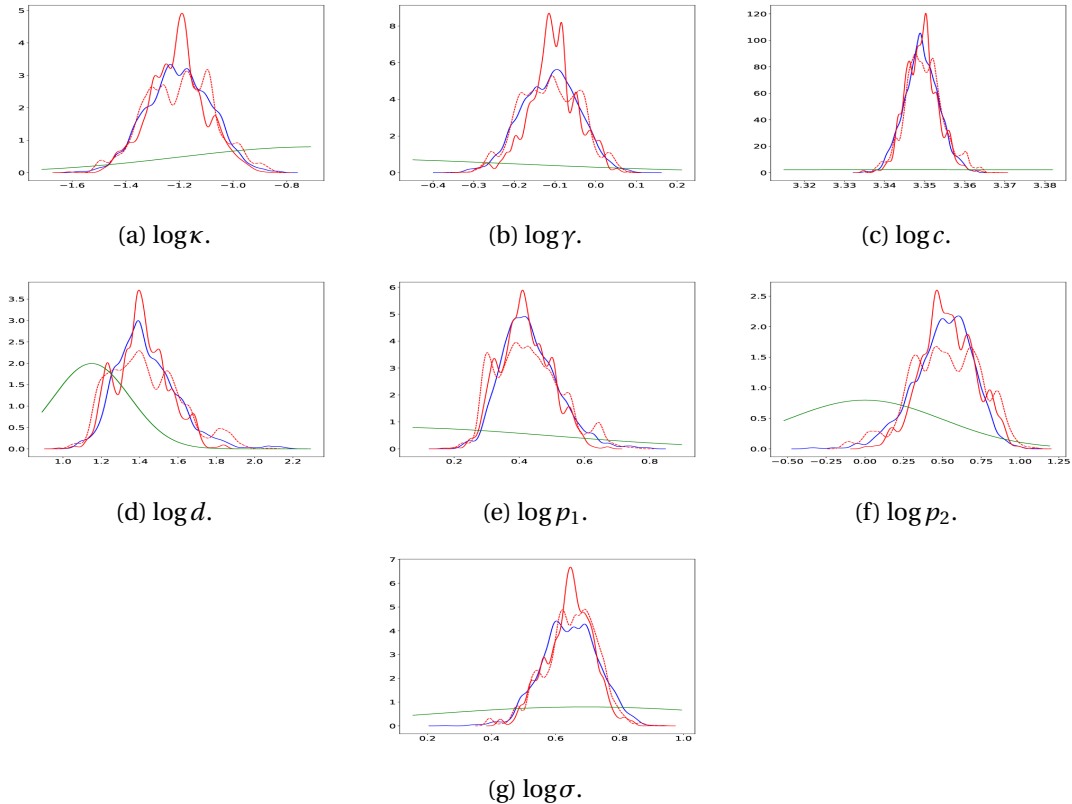

(a) $\log \kappa$.  (b) $\log \gamma$.  (c) $\log c$.

(d) $\log d$.  (e) $\log p_1$.  (f) $\log p_2$.

(g) $\log \sigma$.

Figure 3: Marginal posteriors based on simulated data: MCWM (blue solid line), DA-GP-MCMC (red solid line), and ADA-GP-MCMC (red dashed line). Priors are denoted with green lines (these look "cut" as we zoom on the bulk of the posterior).

We now sample parameters from the high-density region of the posterior distribution and run forward simulations of the DWP-SDE model, similarly to the main paper. In Figure 4 we present three forward simulations, conditionally to parameters from MCMW and ADA-GP-MCMC. The forward simulations in Figure 4 resemble the simulated data better than the forward simulations in the main paper resemble the real data. This seems to point to the fact that the arbitrarily chosen values for $A$ and $g$ in the real-data case study are suboptimal, and (conditionally to those) the inference for the other parameters is probably biased.

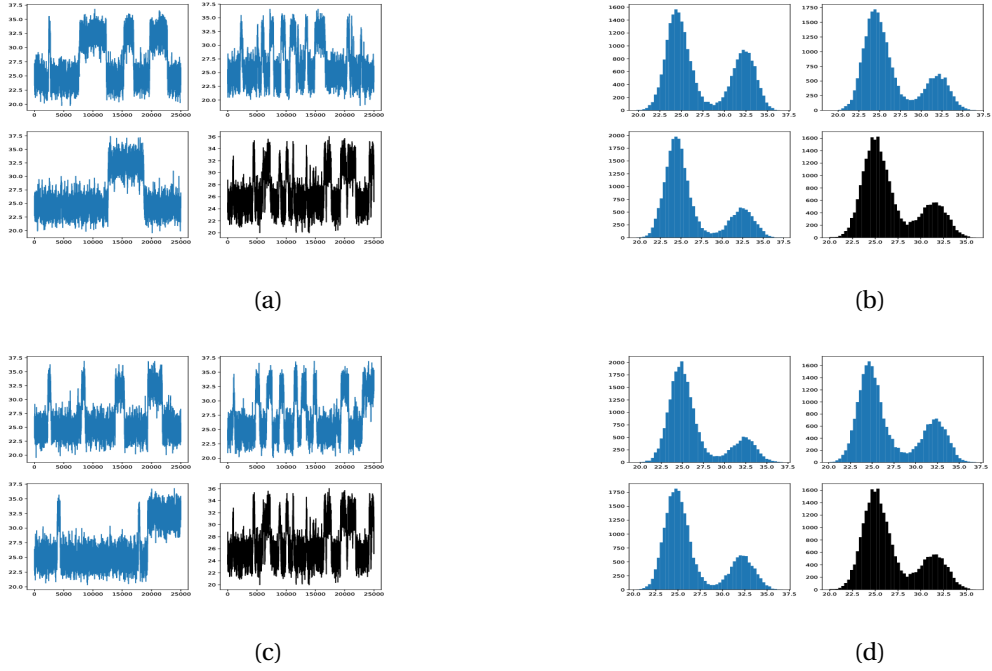

Figure 4: Trajectories obtained by forward simulating the DWP-SDE model based on parameter estimations from MCWM and ADA-GP-MCMC (samples from the high density region of the posterior distribution picked at random). Corresponding marginal distributions. Forward simulations are in blue; real data are in black. Subfigures: a) trajectories from MCWM, b) marginal distributions from MCWM, c) trajectories from ADA-GP-MCMC, and d) marginal distributions from ADA-GP-MCMC.

# 6 Pseudo-code for algorithms

## Algorithm 4 DA-GP-MCMC algorithm

**Input:** Number of iterations $R$, probability to run standard MH update $\beta_{MH}$, a GP model fitted to the training data, a starting value $\theta^0$ and corresponding $\hat{L}_u(\theta^0)$.
**Output:** The chain $\theta^{1:R}$.
1: **for** $r = 1, ..., R$ **do**
2:     Draw $u \sim \mathcal{U}(0,1)$.
3:     **if** $u \leq \beta_{MH}$ **then**     ▷ Skip DA-part
4:         Propose $\theta^\star \sim \tilde{g}(\cdot | \theta^{r-1})$.
5:         Run a single iteration of PMCMC or MCWM for proposal $\theta^\star$.
6:     **else**
7:         Propose $\theta^\star \sim g(\cdot | \theta^{r-1})$.     ▷ Run two stages DA scheme
8:         Sample from (3) to predict independently $\ell_{GP}(\theta^\star)$ and $\ell_{GP}(\theta^{r-1})$. Define $\hat{L}_{GP}(\theta^\star) := \exp(\ell_{GP}(\theta^\star))$ and $\hat{L}_{GP}(\theta^{r-1}) := \exp(\ell_{GP}(\theta^{r-1}))$.
9:         Compute $\alpha_1 = \min\left(1, \frac{\hat{L}_{GP}(\theta^\star)}{\hat{L}_{GP}(\theta^{r-1})} \cdot \frac{g(\theta^{r-1}|\theta^\star)}{g(\theta^\star|\theta^{r-1})} \cdot \frac{p(\theta^\star)}{p(\theta^{r-1})}\right)$.
10:         Draw $u \sim \mathcal{U}(0,1)$.
11:         **if** $u > \alpha_1$ **then**     ▷ Early-reject
12:             Set $\theta^r = \theta^{r-1}$.
13:         **else**
14:             Estimate the likelihood $\hat{L}_u(\theta^\star)$.     ▷ Second stage update scheme
15:             Compute $\alpha_2 = \min(1, \frac{\hat{L}_u(\theta^\star)}{\hat{L}_u(\theta^{r-1})} \cdot \frac{\hat{L}_{GP}(\theta^{r-1})}{\hat{L}_{GP}(\theta^\star)})$.
16:             Draw $u \sim \mathcal{U}(0,1)$.
17:             **if** $u \leq \alpha_2$ **then**     ▷ Accept proposal
18:                 Set $\theta^r = \theta^\star$.
19:             **else**
20:                 Set $\theta^r = \theta^{r-1}$.     ▷ Reject proposal
21:             **end if**
22:         **end if**
23:     **end if**
24: **end for**

**Algorithm 5** ADA-GP-MCMC algorithm

**Input:** Number of iterations $R$, probability to run standard MH update $\beta_{MH}$, a GP model fitted to the training data, model $s_{13}()$ to select between case 1 and 3, model $s_{24}()$ to select between case 2 and 4, a starting value $\theta^0$ and corresponding $\hat{L}_u(\theta^0)$.

1: **for** $r = 1, ..., R$ **do**
2:     Draw $u \sim \mathcal{U}(0, 1)$.
3:     **if** $u \leq \beta_{MH}$ **then**     ▷ Skip DA-part
4:         Propose $\theta^\star \sim \tilde{g}(\cdot|\theta^{r-1})$.
5:         Run a single iteration of PMCMC or MCWM for proposal $\theta^\star$.
6:     **else**
7:         Propose $\theta^\star \sim g(\cdot|\theta^{r-1})$.     ▷ Run A-DA scheme
8:         Sample from the predictive distribution of the GP model to predict independently $\ell_{GP}(\theta^\star)$ and $\ell_{GP}(\theta^{r-1})$. Define $\hat{L}_{GP}(\theta^\star) := \exp(\ell_{GP}(\theta^\star))$ and $\hat{L}_{GP}(\theta^{r-1}) := \exp(\ell_{GP}(\theta^{r-1}))$.
9:         Compute $\alpha_1 = \min\left(1, \frac{\hat{L}_{GP}(\theta^\star)}{\hat{L}_{GP}(\theta^{r-1})} \cdot \frac{g(\theta^{r-1}|\theta^\star)}{g(\theta^\star|\theta^{r-1})} \cdot \frac{p(\theta^\star)}{p(\theta^{r-1})}\right)$.
10:        Draw $u \sim \mathcal{U}(0, 1)$.
11:        **if** $u < \alpha_1$ **then**     ▷ Run second stage of the A-DA scheme
12:             **if** $\hat{L}_{GP}(\theta^\star) > \hat{L}_{GP}(\theta^{r-1})$ **then**
13:                 Select case 1 or 3 according to the model $s_{13}(\theta^\star)$.
14:                 Run the accelerated delayed-acceptance scheme for the selected case.
15:             **else**
16:                 Select case 2 or 4 according to the model $s_{24}(\theta^\star)$.
17:                 Run the accelerated delayed-acceptance scheme for the selected case.
18:             **end if**
19:        **else**     ▷ Early-reject
20:             Set $\theta^r = \theta^{r-1}$.
21:        **end if**
22:     **end if**
23: **end for**

# 7 MCMC trace plots and diagnostics plots for the GP model

Here we show some material pertaining our simulation and data analysis studies. We first report material pertaining the first application (stochastic Ricker model), then the second application (modelling of protein folding data).

Quantities denoted as "residuals" are computed as:

$$r_i = \ell_{PF}(\theta^{\star,i}) - \ell_{GP}(\theta^{\star,i}), \qquad i = 1, \ldots, N_{test}$$

where $N_{test}$ is the number of observations in the test data $\mathcal{D}_2$.

**Ricker model**

Here follow trace plots for MCMC chains obtained under different methods.

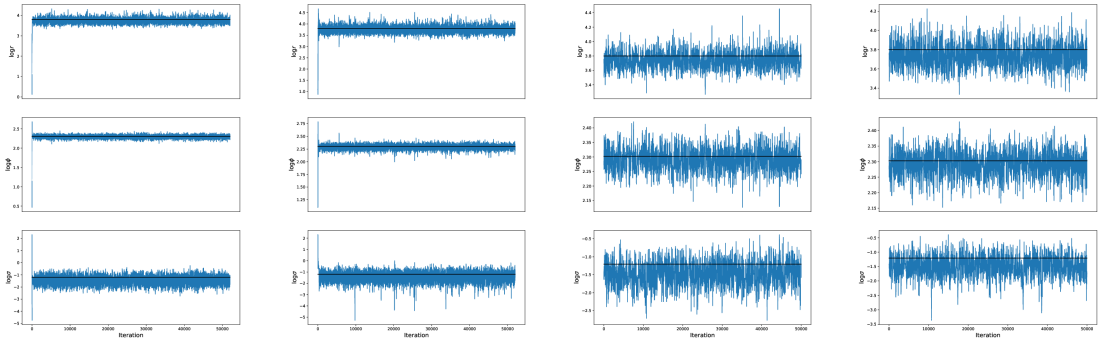

(a) PMCMC.      (b) MCMC.      (c) ADA-GP-MCMC.      (d) ADA-GP-MCMC.

Fit of the GP model.

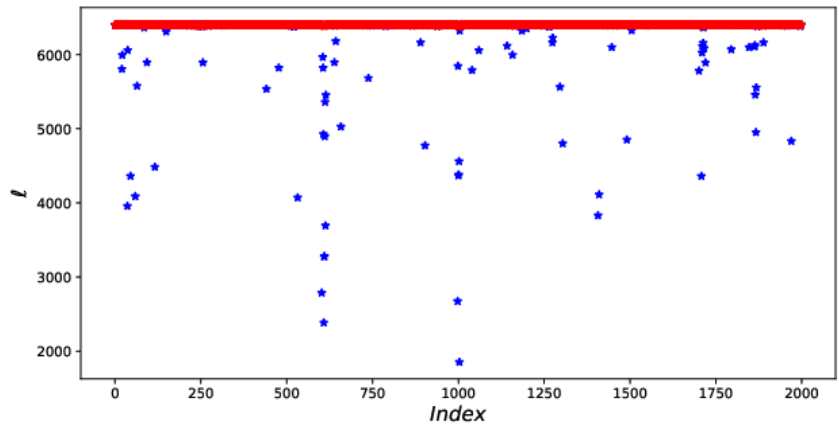

Figure 6: Log-likelihood estimations; particle filter (blue), Gaussian process model (red).

Residual plots.

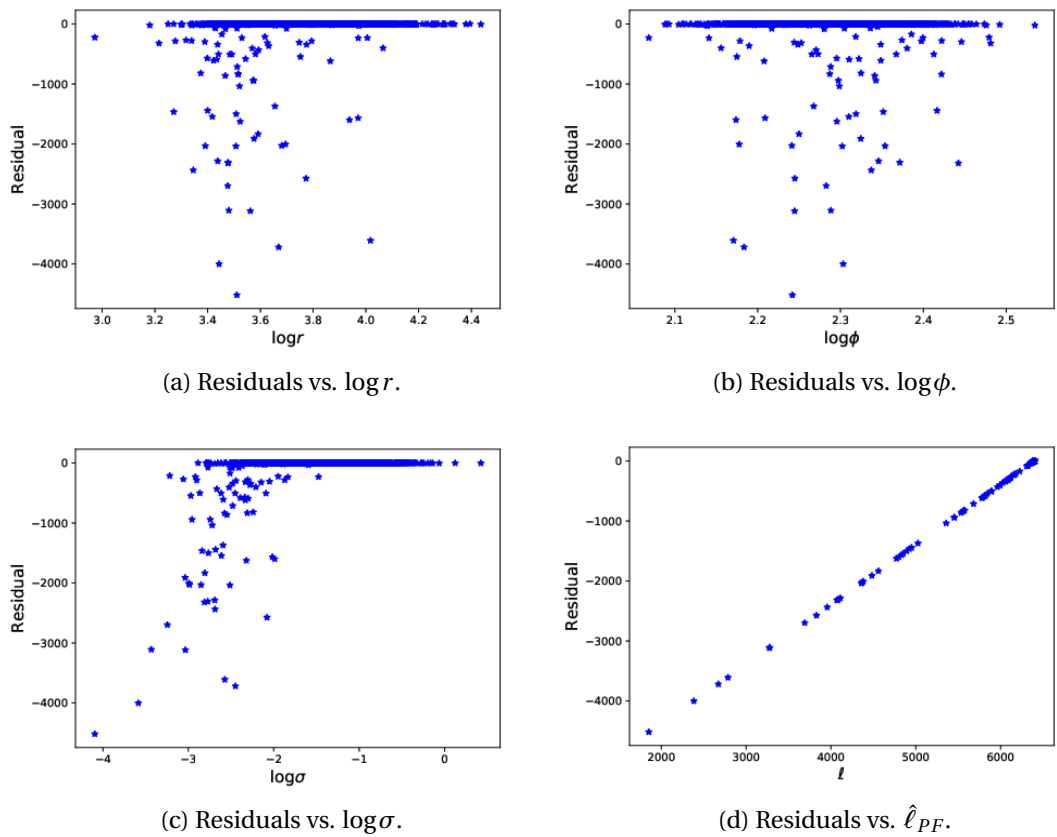

(a) Residuals vs. $\log r$.

(b) Residuals vs. $\log \phi$.

(c) Residuals vs. $\log \sigma$.

(d) Residuals vs. $\hat{\ell}_{PF}$.

Figure 7: Residual plots.

12

Histogram and normal probability plot of the residuals.

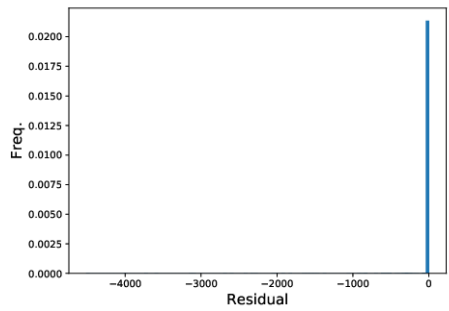 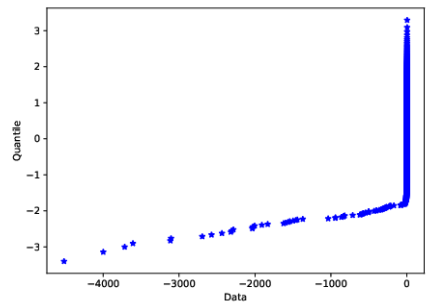

(a) Histogram.                    (b) Normal probability plot.

## DWP-SDE model for simulated data

Here follow trace plots for MCMC chains obtained under different methods.

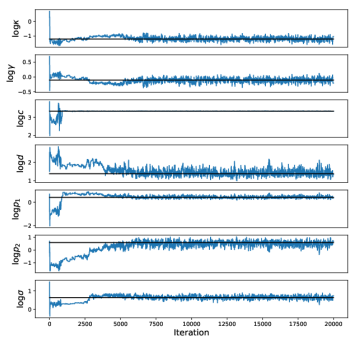 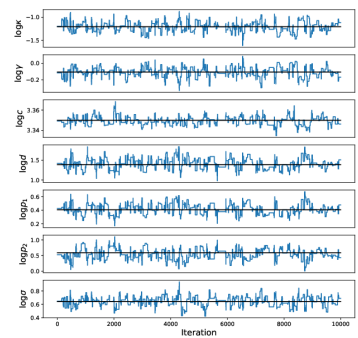 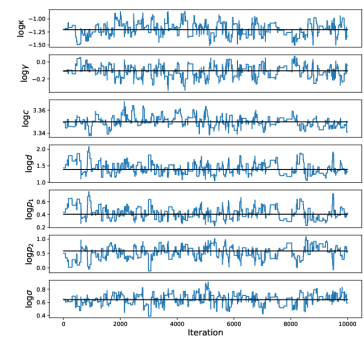

(a) MCWM.              (b) DA-GP-MCMC.              (c) ADA-GP-MCMC.

Fit of the GP model.

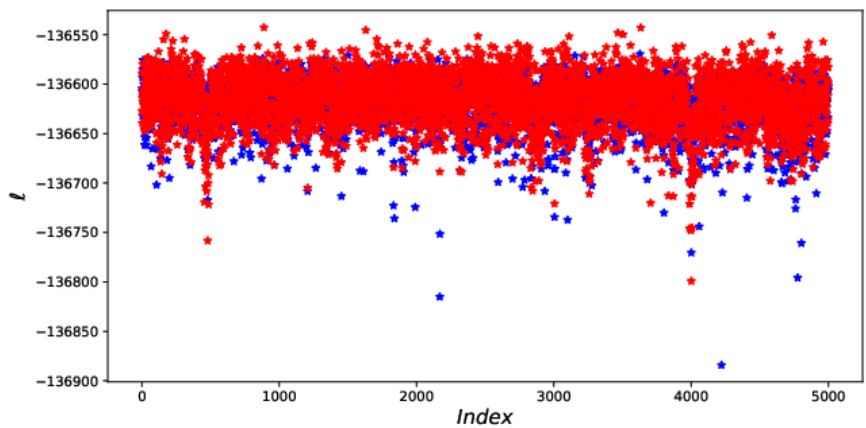

Figure 10: Log-likelihood estimations; particle filter (blue), Gaussian process model (red).

13

Residual plots.

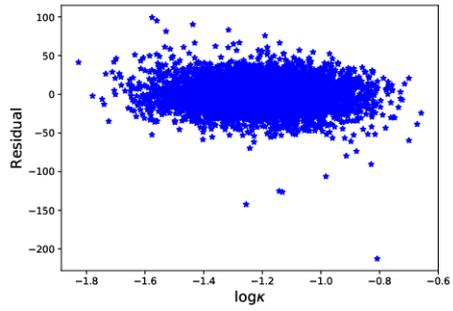
(a) Residuals vs. $\log \kappa$.

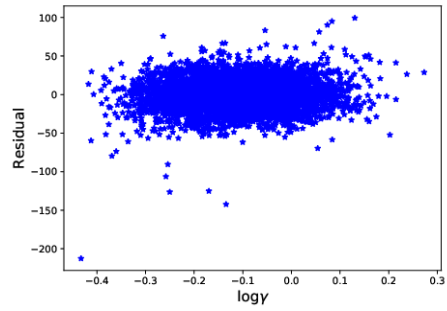
(b) Residuals vs. $\log \gamma$.

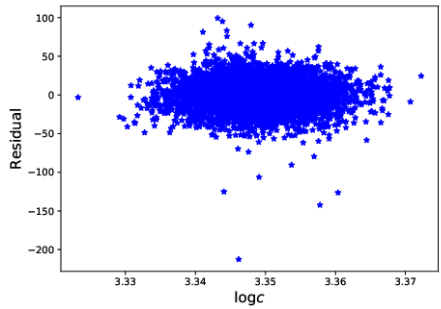
(c) Residuals vs. $\log c$.

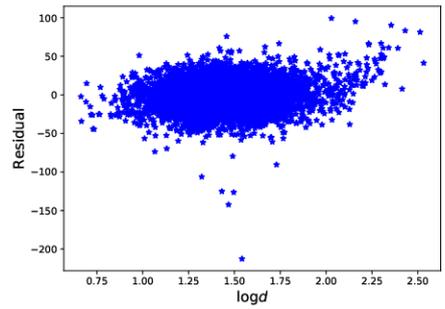
(d) Residuals vs. $\log d$.

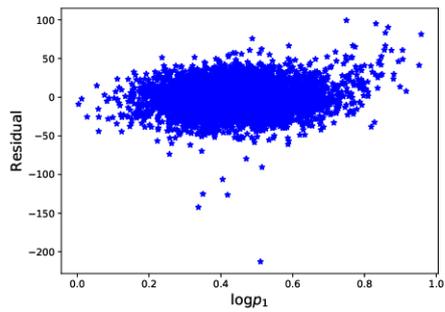
(e) Residuals vs. $\log p_1$.

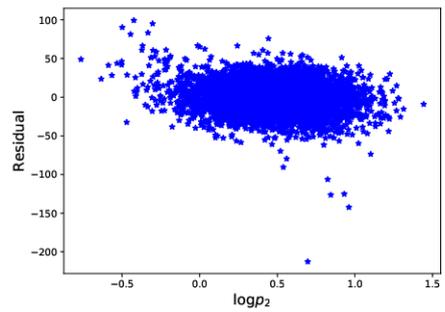
(f) Residuals vs. $\log p_2$.

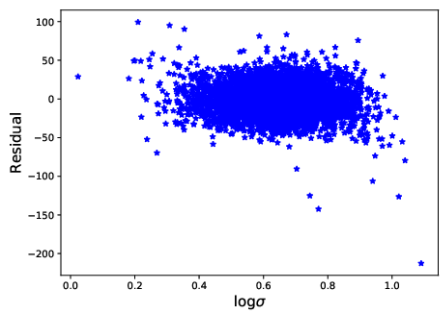
(g) Residuals vs. $\log \sigma$.

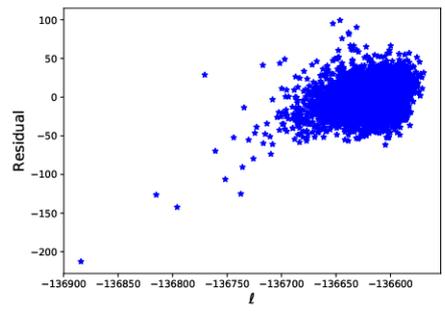
(h) Residuals vs. $\hat{\ell}_{PF}$.

Histogram and normal plot of residuals.

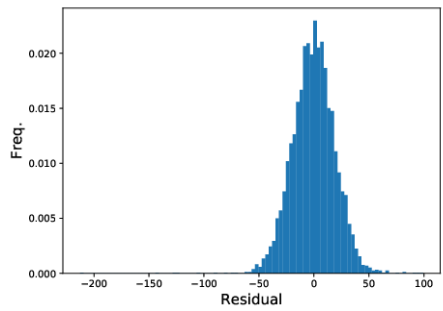 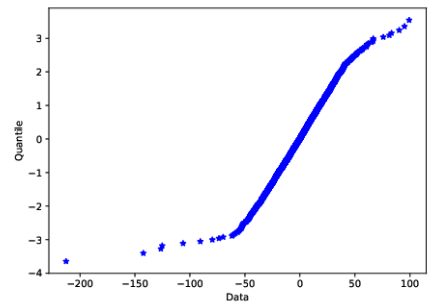

(a) Histogram.  (b) Normal plot.

**DWP-SDE model for protein folding data**

Here follow trace plots for MCMC chains obtained under different methods.

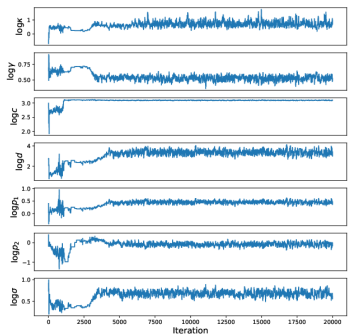 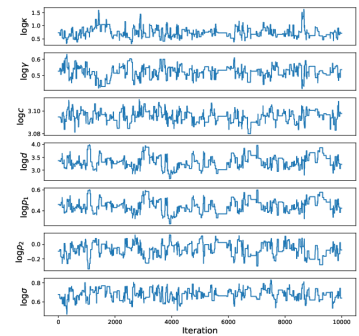 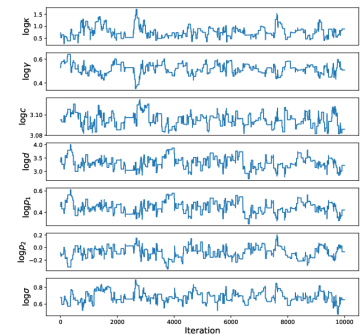

(a) MCWM.  (b) DA-GP-MCMC.  (c) ADA-GP-MCMC.

Fit of the GP model.

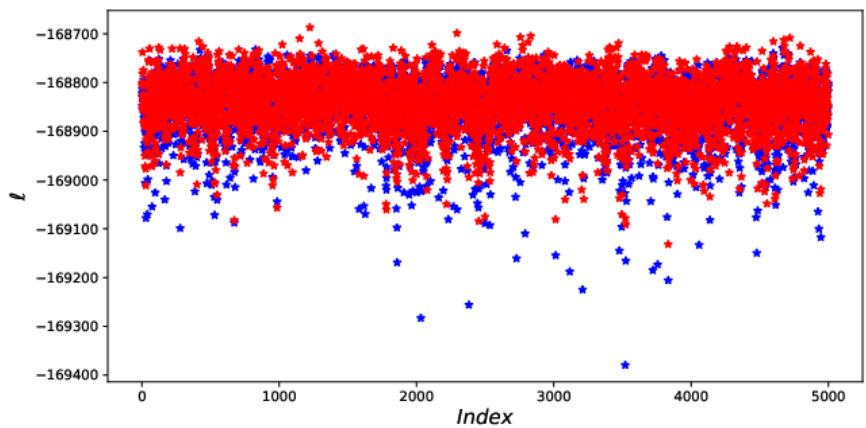

Figure 14: Log-likelihood estimations; particle filter (blue), Gaussian process model (red).

Residual plots.

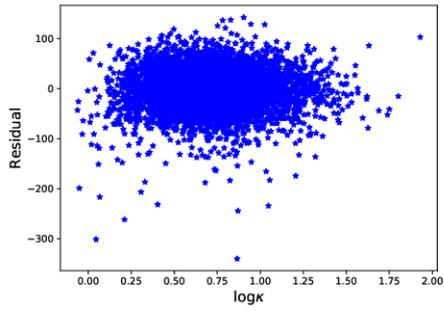

(a) Residuals vs. $\log \kappa$.

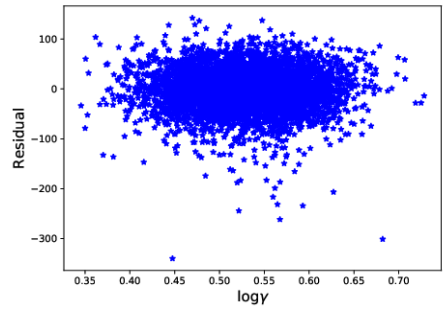

(b) Residuals vs. $\log \gamma$.

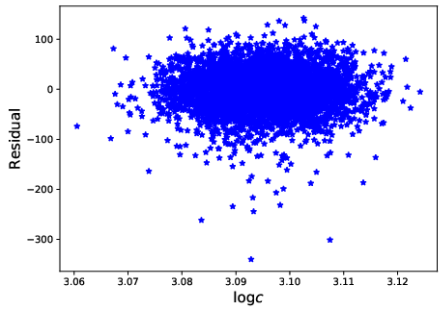

(c) Residuals vs. $\log c$.

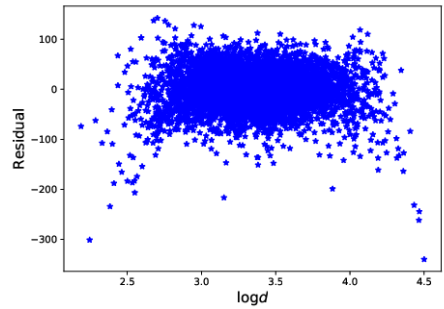

(d) Residuals vs. $\log d$.

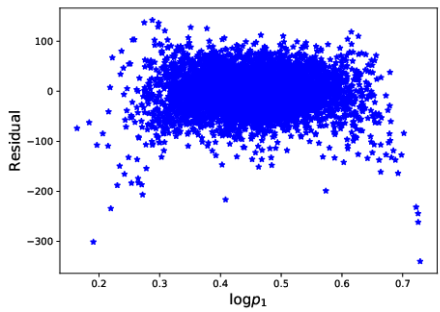

(e) Residuals vs. $\log p_1$.

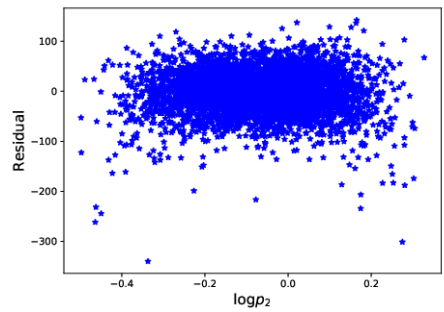

(f) Residuals vs. $\log p_2$.

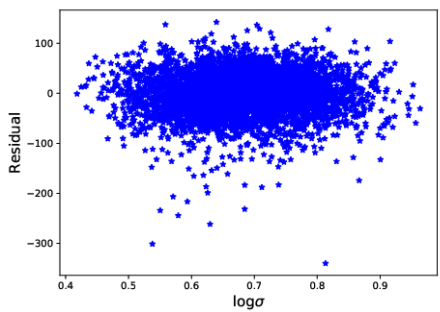

(g) Residuals vs. $\log \sigma$.

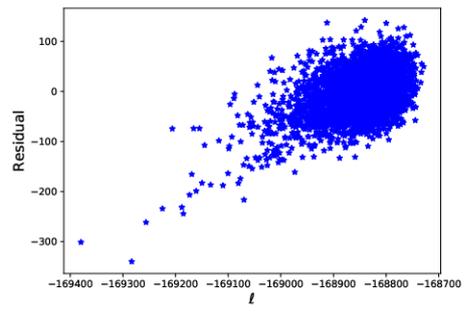

(h) Residuals vs. $\hat{\ell}_{PF}$.

Histogram and normal plot of residuals.

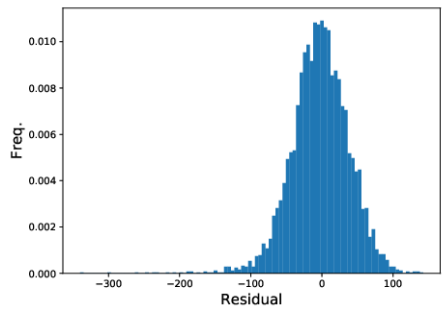 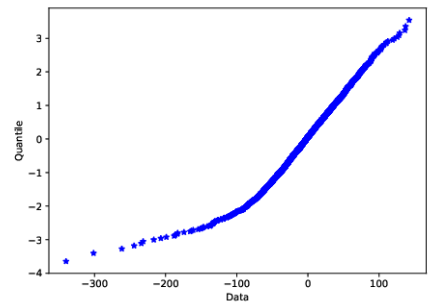

(a) Histogram.   (b) Normal plot.



\end{document}